\documentclass[12pt]{article}
\pdfoutput=1

\usepackage{jheppub}
\usepackage{amsmath}
\usepackage{amssymb}
\usepackage{hyperref}

\usepackage{graphicx}
\usepackage{subcaption}

\usepackage{color}
\definecolor{darkgreen}{rgb}{0,0.5,0}
\definecolor{darkblue}{rgb}{0,0,0.7}
\definecolor{darkred}{rgb}{0.5,0,0.0}
\definecolor{darkorange}{rgb}{0.8,0.4,0.0}

\newcommand{\ie}{i.e. }
\newcommand{\eg}{e.g. }

\newcommand{\kmin}{\ensuremath{k_{\min}}}
\newcommand{\kmax}{\ensuremath{k_{\max}}}

\title{Jet shapes for boosted jet two-prong decays from first-principles}
\author[a]{Mrinal Dasgupta,}
\author[b]{La\'is Schunk}
\author[b]{and Gregory Soyez}
\affiliation[a]{Consortium for
  Fundamental Physics, School of Physics \& Astronomy, University
  of Manchester, Manchester M13 9PL, United Kingdom}
\affiliation[b]{IPhT, CEA Saclay, CNRS UMR 3681, F-91191 Gif-sur-Yvette, France}

\emailAdd{mrinal.dasgupta@manchester.ac.uk}
\emailAdd{lais.sarem-schunk@cea.fr}
\emailAdd{gregory.soyez@cea.fr}


\abstract{
  Several boosted jet techniques use jet shape variables to
  discriminate the multi-pronged signal from Quantum Chromodynamics
  backgrounds.
  In this paper, we provide a first-principles study of an important
  class of jet shapes all of which put a constraint on the subjet
  mass: the mass-drop parameter ($\mu^2$), the $N$-subjettiness ratio
  ($\tau_{21}^{(\beta=2)}$) and energy correlation functions
  ($C_2^{(\beta=2)}$ or $D_2^{(\beta=2)}$). We provide analytic
  results both for QCD background jets as well as for signal
  processes.
  We further study the situation where cuts on these variables are
  applied recursively with Cambridge-Aachen de-clustering of the
  original jet. We also explore the effect of the choice of axis for
  $N$-subjettiness and jet de-clustering.
  Our results bring substantial new insight into the nature, gain and
  relative performance of each of these methods, which we expect will
  influence their future application for boosted object searches.  }

\begin{document}

\maketitle

\section{Introduction}\label{sec:intro}

In recent years jet substructure  studies have received unprecedented
attention and have been the focus of many theoretical and experimental
studies. Most of this research has been carried out in the direct
context of boosted new particle searches at the LHC. For reviews and
detailed studies we refer the reader to Refs.~\cite{Abdesselam:2010pt,Altheimer:2012mn,Altheimer:2013yza,Adams:2015hiv} and references therein. 

The basic ideas that underpin such studies are simple to understand. A
high $p_T$ resonance with a mass $m \ll p_T$ will exhibit
collimated decays where in a significant fraction of events the decay
products would be reconstructed in a single ``fat'' jet.  Tagging
signal jets and removing jets arising from QCD background will thus rely
crucially on detailed information about the jets themselves. In this
context it is clear that valuable information will be obtained by
studying the internal structure of jets in some detail. 

Let us for example contrast the two-pronged hadronic decays of an electroweak
boson (W/Z/H) with  $1 \to 2$ QCD splittings. QCD emission
probabilities are infrared enhanced, favouring soft splittings, and
hence a QCD jet would typically consist of a single hard prong. On the
other hand decays of electroweak bosons show no preference for soft
splittings and this results in a more symmetric energy sharing which
gives rise to jets with a characteristic two-pronged internal
structure. Another important difference results from the colour
neutral nature of electroweak bosons which results in a strong
suppression of radiation at angles that are large compared to the
opening angle between the hard decay products. Soft large-angle radiation
in a signal jet would thus typically arise from emissions that are uncorrelated
with the decay of the electroweak boson in question \ie from
initial state radiation (ISR) and underlying event (UE) as well as
from pile-up. Such radiation serves to degrade signal peaks making
them less visible and also pushes up the masses of background jets. It
is therefore also desirable to eliminate this radiation. In the above
context the two principal aims of a substructure analysis therefore
emerge as  identification of two hard prongs (tagging) and removal
of uncorrelated soft radiation (grooming).

In recent years there have been many tools developed to achieve the
above aims of tagging and grooming jets. These include the
mass-drop+filtering methods \cite{Butterworth:2008iy}, trimming
\cite{Thalertrim} and pruning \cite{Ellis:2009su, Ellis:2009me}
amongst a whole host of other techniques. Monte Carlo event generator
studies involving several of these techniques can be found in
Refs.~\cite{Abdesselam:2010pt,Altheimer:2012mn,Altheimer:2013yza,Adams:2015hiv}
and the original references.

Somewhat more recently there has also been the emergence of jet shape
variables that directly attempt to quantify the $N$-pronged nature of a
fat jet. Examples include the $N$-subjettiness variables
\cite{Thaler:2010tr,Kim:2010uj,Thaler:2011gf} and the $N$-point energy
correlation functions (ECFs) \cite{Larkoski:2013eya,Larkoski:2014gra},
both of which are designed to take on small values for particle
configurations corresponding to $N$ collimated subjets of a fat jet,
which one can naturally associate to an $N$-pronged decay.
These techniques typically put constraints on the gluon radiation
patterns in a jet. We expect this to have a good discriminating power
both at small and large angles because gluon radiation is different
for colour-neutral bosons compared to coloured QCD jets.
At small angles, gluon radiation tends to be larger in QCD jets, made
of a mixture of quarks and gluons, than in resonances, which decay
mostly into quarks.
At large angles, this is an even bigger effect since one expects a
strong suppression of the radiation from collimated colour-neutral
resonance decays compared to QCD jets.
It is interesting to notice at this stage that the large-angle region,
which shape variables try to constrain, is also the region that is
sensitive to initial-state radiation and the underlying event. One
typically uses grooming techniques to mitigate these effects and,
therefore, one may wonder about the effectiveness of shape variable
constraints when combined with grooming.

For studies involving two-pronged (W/Z/H) signal jets the
$N$-subjet\-tiness ratio
$\tau_{21}^{(\beta)}=\tau_2^{(\beta)}/\tau_1^{(\beta)}$ and the ECF
$C_2^{(\beta)}$ are known to provide good discrimination between
signal and background, where $\beta$ is a parameter (angular exponent)
that enters the definition of both variables. We shall provide precise
definitions of these variables in the following section.\footnote{Note
  that to satisfy infrared and collinear (IRC) safety one has the
  requirement $\beta>0$.}

There have also been several detailed studies carried out for both
$\tau_{21}$ and $C_2$ in the literature. Again, nearly all of these
studies have been done using Monte Carlo event generator tools. As
examples we refer the reader to the work carried out in the original
references \cite{Thaler:2010tr,Thaler:2011gf} while for more recent
studies also including the implementation of these variables in
multivariate combinations we refer to Ref.~\cite{Adams:2015hiv}.

In contrast our principal aim here is to carry out {\it{analytical
    calculations}} for the above variables, based on the first
principles of QCD. Such calculations have, for instance, been carried
out for the mass-drop, pruning and trimming methods
\cite{Dasgupta:2013ihk} and provided considerable new insight into the
performance of those tools over and above what could be gained purely
from Monte Carlo methods. 
We would therefore expect a similar level of information from
analytical studies of the shape variables considered here.
For our calculations in this paper we shall
make the choice of $\beta=2$, \ie focus on $\tau_{21}^{(\beta=2)}$ and
$C_2^{(\beta=2)}$ for which calculations are relatively
straightforward to perform. Detailed numerical studies of the
dependence on $\beta$ have been carried out in particular for
$C_2^{(\beta)}$, in Ref.~\cite{Larkoski:2013eya}. These studies found
that in the transverse momentum range $p_T \in \left[400,500\right]$
GeV for jet masses relevant to W/Z/H tagging, optimal $\beta$ values
ranged between 1.5 and 2.  For larger masses the optimal $\beta$
values were found to be smaller. An analytical understanding of the
$\beta$ dependence of discrimination power would also be desirable but
is left to future work.

As we shall show explicitly later in the article, cuts on
$\tau_{21}^{(\beta=2)}$ and $C_2^{(\beta=2)}$ effectively serve to
constrain subjet masses. Another similar variable, that has been far
less investigated in the literature, is the parameter $\mu^2$ of the
mass-drop tagger (MDT) \cite{Butterworth:2008iy}. This is obtained by
declustering a jet into two subjets and taking the ratio of the
squared jet mass for the heavier subjet to that for the original
jet. The original mass drop tagger uses a cut on $\mu^2$ along with an
energy cut designed to discriminate against soft splittings \ie the
$y_{cut}$ parameter of the MDT. It was shown in
Ref.~\cite{Dasgupta:2013ihk} that in fact in the presence of the
$y_{cut}$ condition the dependence on $\mu^2$ could essentially be
neglected. In the present article we instead study the dependence on
$\mu^2$ without any $y_{cut}$ requirement and compare the
discriminating power it provides, to that from similar variables
\ie $\tau_{21}^{(\beta=2)}$ and $C_2^{(\beta=2)}$. Note that while
the standard mass-drop tagger recurses, successively undoing the last
step of a Cambridge/Aachen clustering, until the cut on $\mu^2$ (and
the $y_{cut}$ condition) is satisfied here we study both recursive and
non-recursive variants for each of the shape variables.

We carry out analytical studies for the jet mass distributions of QCD
background jets with cuts on shape variables $v <v_{\rm max}$, with
$v=\tau_{21},C_2 $ and $\mu^2$. We also study the probability for
signal jets to pass the same cuts. We define $\rho= m^2/(p_T^2 R^2)$,
with $m$ being the jet mass and work in the limit $\rho \ll 1$
(relevant for boosted object studies) and $v_{max} \ll 1$ which is
desirable to separate two-pronged structures from QCD background. Our
analytical results aim only to capture leading-logarithmic accuracy
although we also retain several sources of next-to--leading
logarithmic corrections.  We test our analytical results by comparing
to fixed-order results from EVENT2 \cite{Catani:1996vz,event2} to
results from parton shower Monte Carlos and additionally carry out
pure Monte Carlo studies of the impact of non-perturbative
corrections.
Since non-perturbative corrections are found to be large, we further
examine with Monte Carlo studies the impact of grooming with SoftDrop
\cite{Larkoski:2014wba}. This shows an important reduction of the
non-perturbative effects. To avoid diluting the main message of this
paper with additional technical considerations, we defer the study of
groomed jet shapes to a forthcoming work.

Note that some level of analytic understanding for jet shapes already
exists. For example, studies of the lowest-order Energy-Correlation
Functions, $C_1^\beta$, have been carried out in
Ref.~\cite{Larkoski:2013eya}. Also, in the framework of Soft-Collinear
Effective Theory (SCET)~\cite{Bauer:2000yr,Bauer:2001ct,Bauer:2001yt}
and its extension SCET$_+$~\cite{Bauer:2011uc}, results for
$N$-subjettiness have been obtained at the N$^3$LL accuracy for signal
jets \cite{Abbate:2010xh} and studies of the Energy-Correlation
Functions $C_2^\beta$ and $D_2^\beta$ \cite{Larkoski:2015kga} appeared
as the present paper was being finalised.
In contrast, rather than providing a high-accuracy calculation of a
given method, the main aim of our work is a transparent comparison of
different shapes for both signal and background jets with
phenomenological applications in mind.

With that in mind, it is however interesting to compare our approach
and results to what is obtained for $D_2$ in Ref.~\cite{Larkoski:2015kga}. 
Besides using different approaches (SCET-based v. more standard pQCD
language), the main difference between this work and
Ref.~\cite{Larkoski:2015kga} is that, to the best of our understanding
in terms of the variable $\rho$ and $D_2$, the latter provides a NLL
resummation\footnote{The treatment of the non-global logarithms and of
  their resumamtion is not totally clear to us.} in $\rho$, regardless
of the value of $D_2$ while our approach assumes small $D_2$ and
treats $\log(D_2)$ and $\log(\rho)$ on an equal
footing.\footnote{Strictly speaking, we reach (modified) LL accuracy
  but we include a series of NLL effects, see
  Section.~\ref{sec:bkg-towards-nll}.}
Therefore, the calculation in Ref.~\cite{Larkoski:2015kga} has likely
a higher accuracy, at least in the region used in many
phenomenological applications.
However, it is limited to $D_2$ while our main goal here is to
discover the source of and address the main diffferences between
various shapes. The results of Ref.~\cite{Larkoski:2015kga} require at
least four numerical integration (compared to a single one for our
results), which, keeping in mind our purposes, makes a physical
interpretation more involved.

This article is organised as follows: In the next section we provide
detailed definitions of the shapes mentioned above. Following this, in
section \ref{sec:generic}, we discuss the general form of the results
obtained for all the shapes under consideration, both for signal and
background jets. In section \ref{sec:bkg} we perform the detailed
calculations for background jets for both non-recursive and recursive
variants for each shape variable. In the same section we compare the
expansion of our results to fixed-order results from EVENT2, as a
check on our calculations. We also carry out comparisons to results
from Pythia with only final state radiation (FSR) turned on, to give a
direct comparison against our calculations. In section \ref{sec:sig}
we perform the calculations, checks and comparisons to Monte Carlo for
signal jets. Following this, in section \ref{sec:grooming} we study
the impact of non-perturbative corrections where we note the
significant contributions from initial state radiation and the
underlying event in particular. In order to obtain better control over
such effects we combine shape variable studies with grooming using
SoftDrop and study the impact on both signal and
background efficiencies. In section \ref{sec:discussion} we discuss
our findings in detail including an assessment of the comparative
performance of all the shapes studied here. Finally we present our
conclusions.

\section{Radiation-constraining jet shapes}\label{sec:shapes}

Among a large family of jet shapes, this paper will identify and focus
on a series of variables all of which place constraints on the subjet
mass.
In this category, we will study the following three variables:
\begin{itemize}
\item $N$-subjettiness computed with $\beta=2$,
  $\tau_{21}^{(\beta=2)}=\tau_2^{(\beta=2)}/\tau_1^{(\beta=2)}$ with
  $\tau_N^{(\beta=2)}$ defined as \cite{Thaler:2010tr,Stewart:2010tn}
  \begin{equation}\label{eq:tau21def}
    \tau_N^{(\beta=2)} = \frac{1}{p_{t,\rm jet}R^2} \sum_{i\in\rm
      jet}p_{t,i}{\rm min}_{a_1\dots a_N}(\theta_{ia_1}^2,\dots,\theta_{ia_N}^2),
  \end{equation}
  where the sum runs over all the constituents of a given jet and
  $a_1,\dots,a_N$ denote the partition axes.
  While the choice $\beta=1$ is more common in experimental studies at
  the LHC --- likely because of an expected smaller sensitivity to
  non-perturbative effects ---, analytic studies have thus far mostly
  focused on $\beta=2$. As argued eariler, the latter is expected to
  give better discriminative power. We decided to choose $\beta=2$ for
  the present study because in that case, $\tau_N$ acts like a measure
  of the subjet mass which allows for a direct comparison with the
  mass-drop $\mu^2$ cut.\footnote{The choice $\beta=1$ would fall in
    another category of observables, together with energy-correlation
    functions with $\beta=1$ and Y-splitter
    \cite{Butterworth:2002tt}. A calculation similar to the one in
    this paper can be performed, although the situation is often more
    complicated. We leave the study of these variables for future work
    together with a comparison of the performance of the ``$\beta=1$''
    and ``$\beta=2$'' shapes.}  w
  To fully define $\tau_{21}$, we still need to specify our choice for
  the partition axes $a_1,\dots,a_N$ in (\ref{eq:tau21def}). We shall
  consider the following three options:\footnote{See also
    Refs.~\cite{Stewart:2015waa,Thaler:2015xaa,Larkoski:2015uaa} for recent
    studies of axis choice for $N$-subjettiness.}
  \begin{itemize}
  \item the {\it optimal axes} which should minimise $\tau_N$;
  \item the {\it $k_t$ axes} obtained by clustering the jet with the
    $k_t$ algorithm \cite{Catani:1991hj,Catani:1993hr,Ellis:1993tq}
    and taking the $N$ exclusive subjets;
  \item the {\it generalised-$k_t$ axes with $p=1/2$} (gen-$k_t(1/2)$)
    obtained by clustering the jet with the generalised-$k_t$
    algorithm (see Section~4.4 of \cite{fastjet-manual}), with its
    extra parameter $p$ set to $1/2$, and taking the $N$ exclusive
    subjets.
  \end{itemize}
  The third option is new and leads to similar performance to the
  optimal axes at much smaller computational cost.
  The motivation to look into gen-$k_t(1/2)$ axes is that its distance
  measure behaves again like a mass, as does $\tau_{21}^{\beta=2}$,
  and we can expect the resulting axes to be very close to the optimal
  axes. More generally, for $\tau_{21}^{\beta}$ with a generic
  $\beta$, we would expect the generalised-$k_t$ axes with $p=1/\beta$
  to give a close-to-optimal result.
\item a version of the mass-drop parameter \cite{Butterworth:2008iy},
  $\mu^2$ which, given two subjets $j_1$, $j_2$ in a given jet $j$ is
  defined as $\mu^2={\rm max}(m_{j_1}^2,m_{j_2}^2)/m_j^2$.
  In its original formulation, the cut on $\mu^2$ was applied in a
  recursive de-clustering of a jet obtained with the Cambridge/Aachen
  (C/A) algorithm~\cite{Dokshitzer:1997in,Wobisch:1998wt}. The present
  definition of $\mu^2$ is however defined non-recursively, \ie as a
  cut that the jet $j$ satisfies, or not, without any further
  de-clustering if it does not.
  Similarly to the definition of the $N$-subjettiness axes, we need to
  specify the procedure to separate the jet $j$ into two subjets
  $j_1$, $j_2$.
  We will denote by $\mu_p^2$ the result obtained by undoing the last
  step of a generalised-$k_t$ clustering, with extra parameter $p$, of
  the jet $j$.
  We shall concentrate on $\mu_{1/2}^2$, since it follows the ordering
  in mass, and $\mu_0^2$ since it corresponds to the historical
  choice.\footnote{We shall see that, unless it is completed by a
    recursive declustering (as it is the case in the original
    formulation) or a pre-grooming of the jet \eg using the
    SoftDrop procedure, $\mu_0^2$ is infrared unsafe.}
\item the energy correlation function double ratio. Here we again use
  $\beta=2$, which will be kept fixed here, and define
  \cite{Larkoski:2013eya},
  \begin{align}
    e_2 & = \frac{1}{p_T^2R^2}\sum_{i<j\in \rm jet}p_{t,i}p_{t,j}\theta_{ij}^2,\\
    e_3 & = \frac{1}{p_T^3R^6}\sum_{i<j<k\in \rm jet}p_{t,i}p_{t,j}p_{t,k}\theta_{ij}^2\theta_{ik}^2\theta_{jk}^2,
  \end{align}
  and work with $C_2=e_3/e_2^2$. 
  Note that, at the order of accuracy targeted in this paper, we can
  alternatively use the recently-proposed $D_2=e_3/e_2^3$,
  \cite{Larkoski:2014gra}, since, up to our accuracy, they only differ
  by a rescaling by the total jet mass.
\end{itemize}

For any of these three shapes, $v$, a cut of the form $v<v_{\rm cut}$
is expected to show good performance in discriminating two-pronged boosted
objects from standard QCD jets.
Note also that, if the cut is not satisfied, the jet is discarded.

Additionally, we shall also consider the cases where one of the three
shape constraints introduced above is applied recursively.
By this we mean that, for a shape $v$, we apply the following
procedure:
\begin{enumerate}
\item recluster the jet $j$ with the C/A algorithm,
\item compute $v$ from $j$; if $v<v_{\rm cut}$, $j$ is the result of
  the procedure and exit the loop,
\item undo the last step of the clustering to get two subjets $j_1$
  and $j_2$, define the hardest of $j_1$ and $j_2$ (in terms of their
  $p_t$) as the new $j$ and go back to 2.
\end{enumerate}
This is of course motivated by the original mass-drop tagger proposal
\cite{Butterworth:2008iy}, where a cut was placed on the $\mu^2$
parameter. We have to note that, here, the recursion follows the
hardest branch, as suggested in the modified version of the mass-drop
tagger \cite{Dasgupta:2013ihk}, rather than the most massive one, as
in the original proposal.

\section{Generic structure of the results}\label{sec:generic}

For QCD jets, there are two basic physical quantities that we will be
interested in: the jet mass distribution after applying a given fixed,
recursive or not, cut on one of the shapes described in the previous
section; or the distribution of a jet shape for a given fixed value of
the jet mass. The latter situation only applies to the non-recursive
cases.

For signal jets, we are interested in jets of a fixed mass so the
calculation will mostly focus on what fraction of these jets satisfy
the constraint on the jet shape $v$, hence on the distribution of $v$
for an object of a given mass. Jets which fail the constraint on $v$
will be discarded.

Our calculations apply to the boosted regime, where the jet transverse
momentum is much larger than its mass. In that context, it is
convenient to introduce $\rho=m^2/(p_tR)^2$, with $R$ the radius of
the jet. The boosted regime means that we can take the limit $\rho\ll
1$.
Furthermore, in this work, we shall focus on two-pronged decays, where
we expect that the radiation-constraining shapes introduced above
would be smaller for signal jets than for the QCD background. It is
therefore natural to start the study of these shapes in the limit
where they are small. In the following we shall thus also assume that
the cut on the shape is small compared to 1.
In this limit, we focus on the leading double logarithm\footnote{We
  will also include the hard-splitting corrections and discuss a
  series of NLL corrections in Section~\ref{sec:bkg-towards-nll}.} for
which soft and collinear emissions can be considered as strongly
ordered and the mass of the jet is dominated by the strongest of these
emissions.
Throughout the paper, we will therefore assume that this emission,
dominating the mass of the jet, occurs at an
angle\footnote{Practically, it is easier to normalise all angles to
  the jet radius $R$.} $R\theta_1$ and with a fraction $z_1$ of the
jet transverse momentum $p_t$. This has to satisfy the constraint
$z_1(1-z_1)\theta_1^2=\rho$, where, for QCD jets we can neglect the
$(1-z_1)$ factor which would only lead to subleading power corrections
in $\rho$.

All the shapes, $v$, that we consider put constraints on additional
emissions. This means that we can always consider, as a starting
point, a system made of two partons --- the ``leading parton $p_0$''
initiating the jet and the ``first, leading, emission $p_1$'' which
sets the jet mass for QCD jets, or the two prongs of a massive boson
decay for signal jets --- and study additional radiation from this
system.

In the leading-logarithmic approximation, the constraint on radiation
will always take the form of a Sudakov suppression coming on top of
the mass requirement. For QCD jets, the mass distribution with a cut on
$v$ can always be written as
\begin{align}\label{eq:mass_generic_bkg_full}
\left.\frac{\rho}{\sigma}\frac{{\rm d}\sigma}{{\rm d}\rho}\right|_{<v}
& = \int_\rho^1 \frac{d\theta_1^2}{\theta_1^2} \int_\rho^1
dz_1\,P(z_1)\,\rho\,\delta(z_1\theta_1^2-\rho)
\frac{\alpha_s(z_1\theta_1p_tR)}{2\pi}e^{-R_{\rm mass}(\rho)-R_v(z_1,\rho)}\nonumber\\
& = \int_\rho^1 dz_1\, P(z_1)\,
\frac{\alpha_s(\sqrt{z_1\rho}\,p_tR)}{2\pi}e^{-R_{\rm mass}(\rho)-R_v(z_1,\rho)}.
\end{align}
In the above $R_{\rm mass}(\rho)$ is the Sudakov resumming the leading
$\log(1/\rho)$ contributions to the plain jet mass and $R_v(z_1,\rho)$ the extra
contribution coming from the additional cut on $v$.

In the approximation we shall be working at, instead of $P(z_1)$, it is
sufficient to consider its leading logarithmic contribution from its
$2C_R/z_1$ term and a subleading hard collinear contribution
$2C_RB_i\delta(z_1-1)$, where $C_R$ is the colour charge of a jet
initiated by a parton of flavour $i$ and $B_i$ is the integral of the
non-singular part of the splitting function:
\begin{align}
B_q & = \int_0^1 dz\,\left(\frac{1}{2C_F}P_{qq}(z)-\frac{1}{z}\right) = -\frac{3}{4},\\
B_g & = \int_0^1 dz\,\left(\frac{P_{gg}(z)+2n_fP_{qg}(z)}{2C_A}-\frac{1}{z}\right) = -\frac{11C_A-4n_fT_R}{12C_A}.
\end{align}
Eq.~(\ref{eq:mass_generic_bkg_full}) can therefore be replaced by
\begin{align}\label{eq:mass_generic_bkg}
\left.\frac{\rho}{\sigma}\frac{{\rm d}\sigma}{{\rm d}\rho}\right|_{<v}
& = \int_\rho^1 \frac{dz_1}{z_1}
\frac{\alpha_s(\sqrt{z_1\rho}\,p_tR)C_R}{\pi}e^{-R_{\rm mass}(\rho)-R_v(z_1,\rho)}\nonumber\\
& + \frac{\alpha_s(\sqrt{\rho}\,p_tR)C_R}{\pi}\,B_i\,e^{-R_{\rm mass}(\rho)-R_v(z_1=1,\rho)}.
\end{align}

Note however that keeping the full integration over the splitting
function is sometimes useful in comparing background and signal
efficiencies and can lead to potentially large subleading
corrections.\footnote{See also the discussion in
  Section~\ref{sec:bkg-towards-nll}.} For all the analytic plots in
this paper, where the integration over $z_1$ is done numerically, we
have decided to keep the exact $P(z_1)$ splitting function and use
Eq.~(\ref{eq:mass_generic_bkg_full}).

If instead we want to obtain the probability to satisfy the cut on the
shape $v$ for a jet of a given mass one get (for the non-recursive
versions):
\begin{equation}\label{eq:distrib_generic_bkg}
\Sigma(v)=\left[R'_{\rm mass}(\rho)e^{-R_{\rm mass}}\right]^{-1}\left.\frac{\rho}{\sigma}\frac{{\rm d}\sigma}{{\rm d}\rho}\right|_{<v},
\end{equation}
with $R'_{\rm mass}$ being the derivative of $R_{\rm mass}$ wrt
$\log(1/\rho)$. Note that the shapes we consider all require at least
three particles in the jet to be non-zero, meaning that the
distribution $\left. {\rm d}\sigma/{\rm d}\rho\right|_{<v}$ --- or,
equivalently, the double-differential distribution in both the mass
and the shape, ${\rm d}^2\sigma/{\rm d}\rho{\rm d}v$ --- starts at
order $\alpha_s^2$. Conversely, $\Sigma(v)$ will start at order
$\alpha_s$, since it is normalised to the jet mass which itself starts
at order $\alpha_s$.

At fixed coupling, the integration over $z_1$ can usually be carried
out analytically. This however does not bring any additional insight
on the underlying physics mechanisms and so will not be done
explicitly.
For the sake of clarity, we will give fixed-coupling results in the
main body of the text, see Section~\ref{sec:bkg}, and defer the full
results, including running-coupling corrections, to
Appendix~\ref{app:res-rc} (more precisely, Appendix~\ref{app:bkg-rc}
for QCD jets).
The analytic results presented for the radiator function $R_v$ in the
main text therefore correspond to a fixed-coupling (modified) LL
accuracy, i.e. they include the leading logarithms as well as the
corrections due to the hard collinear splittings (the ``$B$ terms'' in
the forthcoming equations). Note that we treat logarithms of the shape
and the jet mass on an equal footing. Hence, by leading logarithms, we
mean, for fixed coupling, double logarithms of any kind, i.e. in
either the shape or the jet mass or both. For the figures and the
comparisons to Monte-Carlo simulations, we will also include the
(leading order) running-coupling contributions as well as a few
relevant NLL effects, discussed in Section~\ref{sec:bkg-towards-nll}
and Appendix~\ref{app:res-rc}.

For signal jets, we will directly be interested in the efficiency,
\ie in the fraction of jets (of the original jet mass) that will
satisfy the constraint on $v$. This can be written as 
\begin{equation}\label{eq:distrib_generic_sig}
\Sigma_{\rm sig}(v) = \int_\rho^1 dz_1\,P_{\rm sig}(z_1) e^{-R_{v,\rm sig}(z_1,\rho)}
\end{equation}
where the signal ``splitting function'' $P_{\rm sig}(z_1)$ is assumed
to be normalised to unity. Again, we can either decide to keep the
full integration over $z_1$ or, at our level of accuracy, keep only
the dominant part without any $z_1$ dependence and the first
$\log(1/z_1)$ and $\log(1/(1-z_1))$ corrections. Note that here $z_1$
can no longer be neglected in the constraint on the jet mass,
$\rho=z_1(1-z_1)\theta_1^2$.
For the illustrative fixed-coupling results given in
Section~\ref{sec:sig}, we will only keep the first corrections in
$\log(1/z_1)$ and $\log(1/(1-z_1))$, while for the full results
including running-coupling corrections given in
Appendix~\ref{app:sig-rc}, we will include these factors in the
resummation, mainly for simplicity reasons.

Given these basic expressions, our main task is to compute the Sudakov
factors $R_v$ for all the shapes under consideration. We do that in
the next two sections.

\section{Calculations for the QCD background}\label{sec:bkg}

The results below give the generic expression for the Sudakov form
factor assuming one works in the (modified) leading-log
approximation. It is helpful to clarify the notations once and for all:
\begin{align}
&L_\rho = \log(1/\rho)=\log(p_t^2R^2/m^2),&&
L_\tau = \log(1/\tau_{21}),\nonumber\\
&L_1 = \log(1/z_1),&&
L_\mu = \log(1/\mu^2),\label{eq:deflogs}\\
&L_v = \log(1/[\tau_{21},\,\mu^2\text{ or }C_2]),&&
L_e = \log(1/C_2).\nonumber
\end{align}

We assume, as stated before, that the angles are normalised to the jet
radius $R$ and we work with a jet initiated by a parton of flavour
$i$.
For a fixed mass $\rho$ and momentum fraction $z_1$, we have
$\theta_1^2=\rho/z_1$.

\subsection[$\tau_{21}$ cut (pure $N$-subjettiness cut)]{\boldmath $\tau_{21}$ cut (pure $N$-subjettiness cut)}
\label{sec:bkg-tau}

We first consider the case where we impose a cut
$\tau_{21}<\tau_{\rm cut}$ on the $N$-subjettiness of a jet of a
given mass $\rho$. We are interested in the limit
$\tau_{\rm cut}\ll 1$.\footnote{In order to keep the notation as
  light as possible, we shall drop the ``${\rm cut}$'' subscript when no
  confusions are possible.}

The first step is to find an expression for $\tau_{21}$ in the limit
where emissions are strongly ordered in angle and transverse momentum
fraction. For this, let us assume that the second leading emission
occurs at an angle $\theta_2$, wrt the leading parton $p_0$,
(initiating the jet) and carries a transverse momentum fraction $z_2$
of the leading parton.

The expression obtained for $\tau_{21}$ in this limit depends on the
choice of axes. It is useful to consider three specific options:
\begin{itemize}
\item the {\it optimal axes} \cite{Thaler:2011gf} which minimise
  $\tau_2$,
\item the {\it $k_t$ axes}, which take the 2 exclusive $k_t$
  subjets as axes,
\item the {\it gen-$k_t(1/2)$ axes}, which also takes exclusive
  subjets as axes, except that this time, we use the generalised $k_t$
  algorithm with $p=1/2$.
\end{itemize}

We defer most of the technical discussions regarding how to obtain
$\tau_{21}$ for the above choices to
Appendix~\ref{app:tau21-details}. In the end, the $k_t$ axes choice
leads to a more complex phase-space, while the optimal and
gen-$k_t(1/2)$ options are equivalent to taking the leading parton and
the emission setting the mass (emission $p_1$) as axes, clustering
emission $p_2$ with whichever axis is closest, and both lead to
\begin{equation}\label{eq:tau21-value-primary}
\tau_{21} = \frac{z_2\theta_2^2}{z_1\theta_1^2},
\end{equation}
up to corrections which are beyond the LL accuracy we aim for
here.\footnote{Note however that there is a bug in {\tt
    MultiPass\_Axes} in version 2.1.0 of the $N$-subjettiness
  implementation \cite{Nsubjettiness-implementation} available from
  FastJet contrib \cite{fjcontrib} which makes the minimisation step
  ineffective. Optimal axes obtained with that version of the
  $N$-subjettiness implementation will therefore return the $k_t$
  axes.}  In what follows, we shall concentrate on the generalised
$k_t$ axes choice since they are simpler than the optimal axes.

Furthermore, we also have to consider secondary emissions, where the
radiation is emitted from the gluon $(z_1, \theta_1^2)$ itself. If
$z_2$ denotes the fraction of the (first emitted) gluon energy carried
by the extra emission at an angle $\theta_{12}$, with
$\theta_{12}<\theta_1$ due to angular ordering, we find
\begin{equation}
\tau_{21}^{\text{secondary}} = z_2 \frac{\theta_{12}^2}{\theta_1^2},
\end{equation}
where the different normalisation wrt
Eq.~\eqref{eq:tau21-value-primary} is purely due to $z_2$ being
normalised to the gluon energy fraction $z_1$.

\begin{figure}[t]
  \begin{subfigure}[b]{0.33\textwidth}
    \centerline{\includegraphics[width=0.9\textwidth]{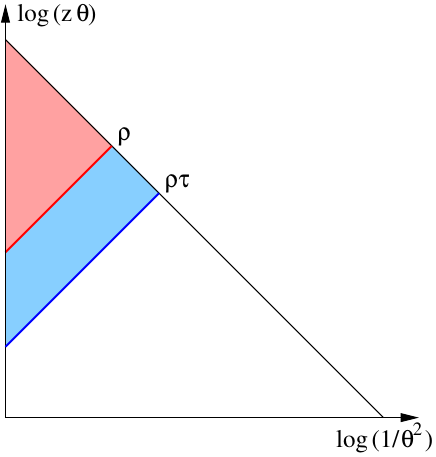}}
    \caption{$N$-subjettiness}\label{fig:phasespace-tau}
  \end{subfigure}%
  \hfill
  \begin{subfigure}[b]{0.33\textwidth}
    \centerline{\includegraphics[width=0.9\textwidth]{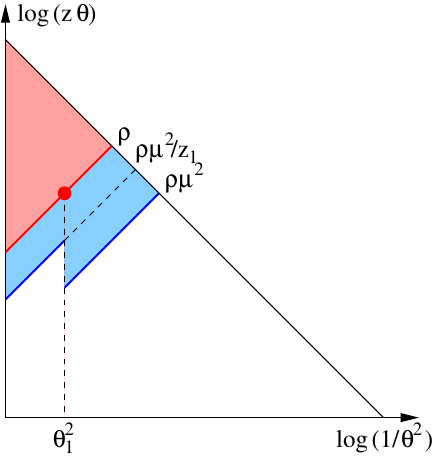}}
    \caption{Mass-drop}\label{fig:phasespace-mu2}
  \end{subfigure}%
  \hfill
  \begin{subfigure}[b]{0.33\textwidth}
    \centerline{\includegraphics[width=0.9\textwidth]{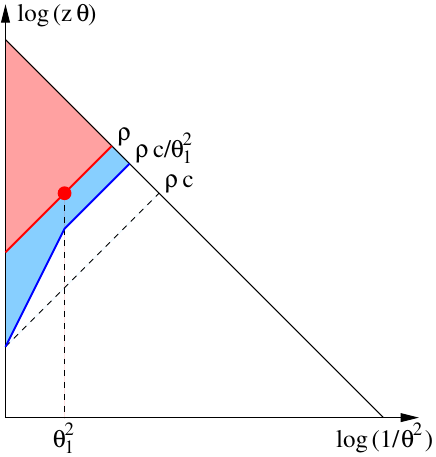}}
    \caption{Energy correlation function}\label{fig:phasespace-ecf}
  \end{subfigure}
  \caption{Plots of the phase-space constraints on emissions setting
    the mass (in red) and the jet shape (in blue).}
  \label{fig:phasespace}
\end{figure}

In the limit of small $\tau_{21}$, additional emissions at smaller
mass do not affect the result. The one-gluon emission will
thus exponentiate according to eq.~(\ref{eq:mass_generic_bkg_full})
and we get
\begin{align}\label{eq:tau21}
R_{\tau}(z_1)
 & = \int_0^1\frac{d\theta_2^2}{\theta_2^2}\int_0^1dz_2 \,\frac{\alpha_s(z_2\theta_2)}{2\pi}\,P_i(z_2)
  \,\Theta(\rho>z_2\theta_2^2>\rho\tau)\nonumber\\
 & + \int_0^{\theta_1^2}\frac{d\theta_{12}^2}{\theta_{12}^2}\int_0^1dz_2 \,\frac{\alpha_s(z_1z_2\theta_{12})}{2\pi}\,P_g(z_2)
  \,\Theta(z_2\theta_{12}^2/\theta_1^2>\tau),
\end{align}
where the first line takes into account emissions from the leading
parton $p_0$ while the second accounts for secondary gluon emissions
from the first emitted gluon $p_1$. 
The arguments of the strong coupling are given as factors multiplying
the ``natural'' scale of the problem, $p_tR$.
The phase-space corresponding to the primary emissions is represented
in Fig.~\ref{fig:phasespace-tau}.

For simplicity, we shall only quote results with a fixed coupling
approximation in the main body of the paper. Results with a proper
treatment of the running-coupling corrections are presented in
the Appendices. 
In this case, the final exponent does not depend\footnote{This is no
  longer valid if we include running-coupling corrections due to the
  scale entering the secondary emissions.} on $z_1$ and we find
\begin{equation}\label{eq:tau21-fc}
R_{\tau}^{{\rm (fixed)}}(z_1)
 = \frac{\alpha_sC_R}{\pi}\left[L_\tau^2/2+L_\rho L_\tau+B_iL_\tau\right]
 + \frac{\alpha_sC_A}{\pi}\left[L_\tau^2/2+B_gL_\tau \right],
\end{equation}
where, for quark jets, we have $C_R=C_F$ and $B_i=B_q=-3/4$ while for
gluon jets we have $C_R=C_A$ and $B_i=B_g=-(11C_A-4n_fT_R)/(12C_A)$.

\subsection[$\mu^2$ cut]{\boldmath $\mu^2$ cut}
\label{sec:bkg-mu2}

As for the case of $N$-subjettiness, we first have to find, given the
emissions $p_1$ and $p_2$ with $p_1$ giving the dominant contribution
to the mass, what is the value of the mass-drop parameter $\mu^2$. 
Since $\mu^2$ is defined by undoing the last clustering step, it will
depend on the jet algorithm we use to (re-)cluster the jet. The
Cambridge/Aachen algorithm is a common choice but does not work
here. Indeed, undoing the last step of a Cambridge/Aachen clustering
would separate the emission at the largest angle from the rest of the
jet, regardless of the transverse momentum of that emission. This is
not infrared safe. We further discuss infrared-safety issues in
Appendix~\ref{app:irc-safety}.

Instead, we shall define $\mu^2$ by undoing the last step of a
generalised-$k_t$ clustering with $p=1/2$. The motivation for this is
the same as the motivation for the axes choice in the previous
section: the generalised-$k_t$ algorithm with $p=1/2$ follows closely
the ordering in mass.
To keep things unambiguous, we shall denote by $\mu_p^2$ the mass-drop
parameter obtained by undoing the last step of a generalised-$k_t$
clustering with parameter $p$. The (infrared-unsafe) case of a C/A
clustering would correspond to $\mu_0^2$ while we will be interested
in $\mu_{1/2}^2$, although the calculation can be performed for any
positive $p$.

Again, we leave the technical details of the calculation for Appendix
\ref{app:mu2-details}. In a nutshell, the hard parton and the first
emission (setting the mass) will form two subjets, and the second
emission, setting the subjet mass, will be clustered with whichever of
these two subjets is closest. 
In the end, keeping in mind that, to our leading-logarithmic accuracy
we can assume strong ordering in angle ($\theta_2\ll\theta_1$ or
$\theta_2\gg\theta_1$), we find
\begin{equation}\label{eq:mu2value}
(z_1\theta_1^2) \mu_{1/2}^2 \approx \begin{cases}
z_2\theta_2^2 & \text{for }\theta_2<\theta_1\text{ or } 
 (\theta_2>\theta_1\text{ and }\theta_2<\theta_{12}),\\
z_1 z_2\theta_2^2 & \text{for }(\theta_2>\theta_1\text{ and }\theta_2>\theta_{12}),\\
z_1^2z_2\theta_{12}^2 & \text{for secondary emissions.}
\end{cases}
\end{equation}

There is a crucial difference between mass-drop and $N$-subjettiness:
the latter can be seen as $(1/p_t) \sum_{\rm j\in subjets}
m_j^2/p_{t,j}$ which has an extra $1/p_{t,j}$ compared to
$\mu_{1/2}^2$. This leads to different expressions whenever the jet
with the largest mass is not the one with the largest $p_t$. The
secondary emissions and large-angle radiations will therefore give
additional suppressions for $N$-subjettiness compared to the mass-drop.

With similar arguments, it is easy to realise that additional
emissions with smaller masses will not affect this calculation, so
that, at leading-logarithmic accuracy, the lowest order simply
exponentiates according to eq.~(\ref{eq:mass_generic_bkg_full}). The
vetoed phase-space for emissions is represented in
Fig.~\ref{fig:phasespace-mu2} and we get
\begin{align}\label{eq:mu2}
R_{\mu_{1/2}^2}(z_1)
 & = \int_0^1\frac{d\theta_2^2}{\theta_2^2}\int_0^1dz_2 \,\frac{\alpha_s(z_2\theta_2)}{2\pi}\,P_i(z_2)
  \,\bigg\{ \Theta(\theta_2^2<\theta_1^2) \,\Theta(\rho>z_2\theta_2^2>\rho\mu^2)\nonumber\\
 & \qquad\qquad + \Theta(\theta_2^2>\theta_1^2) \Big[\frac{1}{2}\Theta(\rho>z_2\theta_2^2>\rho\mu^2)
                                              +\frac{1}{2}\Theta(\rho>z_2\theta_2^2>\theta_1^2\mu^2)
    \Big]\bigg\}
\nonumber\\
 & + \int_0^{\theta_1^2}\frac{d\theta_{12}^2}{\theta_{12}^2}\int_0^1dz_2 \,\frac{\alpha_s(z_1z_2\theta_{12})}{2\pi}\,P_g(z_2)
  \,\Theta(z_1z_2\theta_{12}^2/\theta_1^2>\mu^2).
\end{align}

For a fixed coupling approximation, we find
\begin{align}\label{eq:mu2-fc}
R_{\mu_{1/2}^2}^{{\rm (fixed)}}(z_1)
& = \frac{\alpha_sC_R}{\pi}\big[(L_\rho+L_1+L_\mu) L_\mu/2+\frac{1}{2}(L_\rho-L_1)(L_\mu-L_1)\Theta(L_\mu>L_1)+B_iL_\mu\big]\nonumber\\
& + \frac{\alpha_sC_A}{\pi}\left[(L_\mu-L_1)^2/2+B_g (L_\mu-L_1)\right]\,\Theta(L_\mu>L_1).
\end{align}

\subsection[$C_2$ cut]{\boldmath $C_2$ cut}\label{sec:bkg-ecf}

For two strongly-ordered emissions $p_1(z_1,\theta_1)$ and
$p_2(z_2,\theta_2)$, such that $z_1\theta_1^2\gg z_2\theta_2^2$, one finds, for
primary emissions,
\begin{equation}\label{eq:ecf-value-primary}
C_2 = \frac{1}{z_1^2\theta_1^4} z_1 z_2
       (1-z_1-z_2)\theta_1^2\theta_2^2\theta_{12}^2
 \simeq  \frac{z_2\theta_2^2}{z_1\theta_1^2}\,{\rm max}(\theta_1^2,\theta_2^2)
\end{equation}
which is the same result as the one we obtained in the
$N$-subjettiness case with an extra factor
${\rm max}(\theta_1^2,\theta_2^2)$.\footnote{Contrary to what we have
  for $\mu_{1/2}^2$ (see Appendix.~\ref{app:soft-large-angle}),
  Eq.~(\ref{eq:ecf-value-primary}) is continuous for
  $\theta_1=\theta_2$. Using the exact expression for $\theta_{12}$ in
  the region $\theta_2\approx\theta_1$ will therefore not lead to
  (single) logarithmically enhanced terms.} For secondary emissions,
$\theta_{12}\ll\theta_1$, hence $\theta_2\simeq\theta_1$ and we have
(with $z_2$ measuring the momentum fraction wrt emission 1)
\begin{equation}\label{eq:ecf-value-secondary}
C_2\simeq  z_2\frac{\theta_{12}^2}{\theta_1^2}\theta_1^2 = z_2\theta_{12}^2.
\end{equation}

The corresponding phase-space is represented in
Fig.~\ref{fig:phasespace-ecf} and gives
\begin{align}\label{eq:ecf}
R_{C_2}(z_1)
 & = \int_0^1\frac{d\theta_2^2}{\theta_2^2}\int_0^1dz_2 \,\frac{\alpha_s(z_2\theta_2)}{2\pi}\,P_i(z_2)
  \,\Theta(\rho>z_2\theta_2^2)\nonumber \\
& \qquad\qquad\Big[ \Theta(\theta_2^2<\theta_1^2) \,\Theta(z_2\theta_2^2\theta_1^2>\rho C)
 + \Theta(\theta_2^2>\theta_1^2)\,\Theta(z_2\theta_2^4>\rho C)\Big]
\nonumber\\
 & + \int_0^{\theta_1^2}\frac{d\theta_{12}^2}{\theta_{12}^2}\int_0^1dz_2 \,\frac{\alpha_s(z_1z_2\theta_{12})}{2\pi}\,P_g(z_2)
  \,\Theta(z_2\theta_{12}^2>C).
\end{align}

For a fixed coupling approximation, one finds
\begin{align}\label{eq:ecf-fc}
R_{C_2}^{{\rm (fixed)}}(z_1)
& =
\frac{\alpha_sC_R}{\pi}\big[L_e^2/2+(L_e-L_\rho+L_1)(L_1+B_i)\Theta(L_e>L_\rho-L_1)\big]\nonumber\\
& + \frac{\alpha_sC_A}{\pi}\big[(L_e-L_\rho+L_1)^2/2+ B_g(L_e-L_\rho+L_1)\big]\Theta(L_e>L_\rho-L_1).
\end{align}

If we decide to work with $D_2=C_2/\rho$ rather than $C_2$, and define
$L_d=\log(1/D_2)=L_e-L_\rho$, we get, assuming $L_d>0$, 
\begin{align}\label{eq:d2-fc}
R_{D_2}^{{\rm (fixed)}}(z_1)
& =
\frac{\alpha_sC_R}{\pi}\big[(L_d+L_\rho)^2/2+(L_1+L_d)(L_1+B_i)
\big]\nonumber\\
& + \frac{\alpha_sC_A}{\pi}\big[(L_d+L_1)^2/2+ (L_d+L_1) B_g\big].
\end{align}

\subsection[Recursive $\tau_{21}$ cut]{Recursive \boldmath $\tau_{21}$ cut}\label{sec:bkg-taurec}
We now move to the same calculations as above but apply the cut
recursively declustering a C/A jet until the cut is met (see
Sec.~\ref{sec:shapes}).

The calculation of the shapes mostly remains unchanged but the
recursion will affect the allowed phase-space for emissions.
As before, let us assume that $p_1(\theta_1,z_1)$ is the emission that
dominates the mass {\it after} the recursion procedure has been
applied and see what constraints on the phase-space the cut imposes on
additional emissions $p_2(\theta_2,z_2)$.

For emissions at angles $\theta_2$ smaller than $\theta_1$, the
de-clustering will reach $p_1$ before $p_2$, which corresponds to the same
situation as for the non-recursive case. In fact it remains true for all
shape variables under consideration in this paper that for such
angular configurations the results from the recursive and
non-recursive variants coincide.

Differences occur for emissions at angles larger than $\theta_1$. The
physical reason for that comes from emissions at angles larger than
$\theta_1$ and which would dominate the mass, \ie for which
$z_2\theta_2^2>z_1\theta_1^2$. In the non-recursive case, these
emissions are forbidden by our constraint on the jet mass and this is
included in the Sudakov suppression for the jet mass
$R_{\rm mass}(\rho)$ in Eq.~(\ref{eq:mass_generic_bkg_full}), which
imposes that the mass of the jet is truly dominated by the
$(z_1,\theta_1^2)$ emission.
In the situation where the cut on the shape is applied recursively,
some extra care is needed since some of these emissions --- that are
vetoed in the non-recursive case because they would lead to a larger
jet mass --- can be simply discarded by the recursive procedure. In
such a case, they should no longer be forbidden.

For the large-angle region, $\theta_2>\theta_1$ we therefore have to
separate 4 different regions:
\begin{itemize}
\item for $z_2\theta_2^2<\rho\tau$, we have $\tau_{21}\approx
  z_2\theta_2^2/z_1\theta_1^2=z_2\theta_2^2/\rho<\tau$, meaning that
  the constraint is satisfied. That region is therefore allowed,
\item for $\rho\tau<z_2\theta_2^2<\rho$, we have
  $\tau_{21}\approx z_2\theta_2^2/z_1\theta_1^2=z_2\theta_2^2/\rho$ as
  in the previous case, but this time it does not satisfy the
  condition $\tau_{21}<\tau$. The emission $(z_2,\theta_2^2)$ will
  thus be discarded, meaning that this region is again allowed,
\item for $\rho<z_2\theta_2^2<\rho/\tau$, we now have
  $\tau_{21}\approx z_1\theta_1^2/z_2\theta_2^2=\rho/z_2\theta_2^2$,
  \ie $\tau_{21}>\tau$. The condition is once again not satisfied and
  the region is allowed. 
\item for $z_2\theta_2^2>\rho/\tau$, we find similarly $\tau_{21}\approx
  z_1\theta_1^2/z_2\theta_2^2=\rho/z_2\theta_2^2<\tau$. The condition
  on $\tau_{21}$ would be met, leaving a jet with a mass
  $z_2\theta_2^2>\rho$. This region is therefore forbidden.
\end{itemize}
Compared to the non-recursive case, the vetoed region at large angle
is therefore reduced.

In the above discussion, we tacitly assumed that we were working with
the gen-$k_t(1/2)$ axes or with the optimal axes, but the argument is more
general. We could also define $\tau_{21}$ using the exclusive C/A
axes, automatically available from the declustering procedure. Indeed,
in that case, all emissions with $z_2\theta_2^2<\rho/\tau$ would
fail the cut on $\tau_{21}$ and be discarded.
We will come back to that point later on.

\begin{figure}
  \begin{subfigure}[b]{0.5\textwidth}
    \centering{\includegraphics[width=0.6\textwidth]{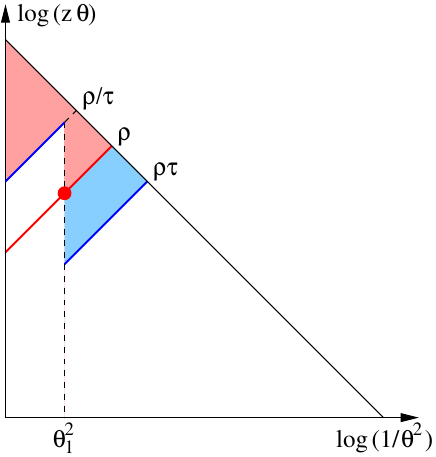}}
    \caption{$N$-subjettiness or mass-drop}\label{fig:phasespace-taurec}
  \end{subfigure}%
  \begin{subfigure}[b]{0.5\textwidth}
    \centering{\includegraphics[width=0.6\textwidth]{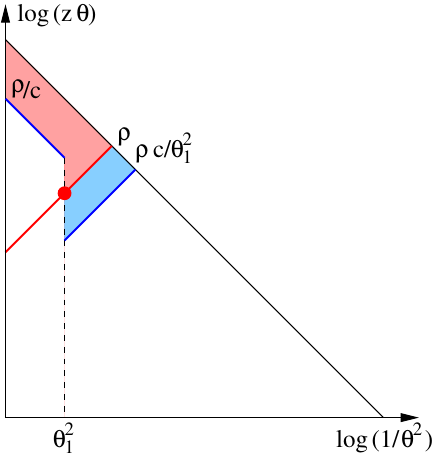}}
    \caption{Energy-correlation function}\label{fig:phasespace-ecfrec}
  \end{subfigure}
  \caption{Same as Fig.~\ref{fig:phasespace} but this time for cases
    where the cut is applied recursively.}
  \label{fig:phasespace-rec}
\end{figure}

Again, the lowest order result simply exponentiates and the Sudakov
suppression, depicted in Fig.~\ref{fig:phasespace-taurec} is
\begin{align}\label{eq:tau21rec}
R_{\tau,\rm rec}(z_1)
 & = \int_0^1\frac{d\theta_2^2}{\theta_2^2}\int_0^1dz_2 
  \,\frac{\alpha_s(z_2\theta_2)}{2\pi}\,P_i(z_2)
  \,\big[\Theta(\theta_2^2>\theta_1^2)\,\Theta(z_2\theta_2^2>\rho/\tau)
   \nonumber\\
 & \phantom{=\int_0^1\frac{d\theta_2^2}{\theta_2^2}\int_0^1dz_2
   \,\frac{\alpha_s(z_2\theta_2)}{2\pi}\,P_i(z)}
   + \Theta(\theta_2^2<\theta_1^2)\, \Theta(z_2\theta_2^2>\rho \tau) \big]
\nonumber\\
 & + \int_0^{\theta_1^2}\frac{d\theta_{12}^2}{\theta_{12}^2}\int_0^1dz_2 
  \,\frac{\alpha_s(z_1z_2\theta_2)}{2\pi}\,P_g(z_2)
  \, \Theta(z_2\theta_{12}^2/\theta_1^2>\tau)-R_{\rm mass}(\rho),
\end{align}
where we have subtracted $R_{\rm mass}(\rho)$ which has already been
included in (\ref{eq:mass_generic_bkg_full}).

For a fixed coupling approximation, this gives
\begin{align}\label{tau21rec-fc}
R_{\tau,\rm rec}^{{\rm (fixed)}}(z_1)
& = \frac{\alpha_sC_R}{\pi}\Big\{
  \left[L_\tau^2/2-L_\rho L_\tau+2 L_1 L_\tau+B_iL_\tau\right]\,\Theta(L_\tau<L_1)\nonumber\\
& \phantom{= \frac{\alpha_sC_F}{\pi}}
 +
 \left[L_\tau^2 -L_\rho L_\tau+L_1 L_\tau+L_1^2/2+B_iL_1\right]\,\Theta(L_1<L_\tau<L_\rho)\nonumber\\
& \phantom{= \frac{\alpha_sC_F}{\pi}}
 +
 \Big[\frac{1}{2}(L_\rho+L_1+L_\tau+2B_i)(L_\tau+L_1-L_\rho)\Big]\,\Theta(L_\rho<L_\tau)\Big\}\nonumber\\
& + \frac{\alpha_sC_A}{\pi}\left[L_\tau^2/2+B_gL_\tau \right].
\end{align}

\subsection[Recursive $\mu^2$ cut (pure mass-drop tagger)]{Recursive \boldmath $\mu^2$ cut (pure mass-drop tagger)}
\label{sec:bkg-murec}

The situation is mostly the same as for the recursive $\tau_{21}$
cut. Here, the use of a recursive criterion allows to use either the
subjets naturally given by the C/A declustering or the gen-$k_t(1/2)$
subjets. The results presented in this section are valid for both
$\mu_0^2$ and $\mu_{1/2}^2$, although, as we will see in the next
paragraph, different axes choice yield the same answer for the mass
distribution in different ways, and would give different answers for
other observables.

As before, for $\theta_2$ smaller than $\theta_1$, the declustering has
no effect and the results are as obtained in Sec.~\ref{sec:bkg-mu2}.
The complication related to the clustering distance for
$\theta_2\gg\theta_1$ is absent here because of the declustering, and
only emissions with $z_2\theta_2^2>\rho/\mu^2$ have to be
vetoed. In all other cases, either the mass-drop condition fails and
the emission is simply discarded, or the mass-drop condition is
satisfied but the mass of the jet remains $z_1\theta_1^2$.\footnote{As
  for the axes choice in $N$-subjettiness, these regions will differ
  for $\mu_0^2$ and $\mu_{1/2}^2$.} E.g., for the natural choice,
$\mu_0^2$, all emissions in the region $z_2\theta_2^2<\rho/\mu_0^2$
will fail the condition and be discarded before the recursion
continues.
That said, the only remaining difference between a recursive $\mu^2$
cut and a recursive $\tau_{21}$ cut will be in the extra factor $z_1$
in the secondary emissions (see, \eg Sec.~\ref{sec:bkg-mu2})
and we find
\begin{align}\label{eq:mu2rec}
R_{\mu^2,\rm rec}(z_1)
 & = \int_0^1\frac{d\theta_2^2}{\theta_2^2}\int_0^1dz_2 
  \,\frac{\alpha_s(z_2\theta_2)}{2\pi}\,P_i(z_2)
  \,\big[ \Theta(\theta_2^2>\theta_1^2)\,\Theta(z_2\theta_2^2>\rho/\mu^2)
  \nonumber\\
 & \phantom{=\int_0^1\frac{d\theta_2^2}{\theta_2^2}\int_0^1dz_2 
  \,\frac{\alpha_s(z_2\theta_2)}{2\pi}\,P_i(z)}
 + \Theta(\theta_2^2<\theta_1^2)\, \Theta(z_2\theta_2^2>\rho \mu^2) \big]
  \nonumber\\
 & + \int_0^{\theta_1^2}\frac{d\theta_{12}^2}{\theta_{12}^2}\int_0^1dz_2 
  \,\frac{\alpha_s(z_1z_2\theta_2)}{2\pi}\,P_g(z_2)
  \, \Theta(z_1z_2\theta_{12}^2/\theta_1^2>\mu^2)-R_{\rm mass}(\rho).
\end{align}

For a fixed coupling approximation, we get
\begin{align}\label{eq:mu2rec-fc}
  R_{\mu^2,\rm rec}^{{\rm (fixed)}}(z_1) & =
  \frac{\alpha_sC_R}{\pi}\Big\{
  \left[L_\mu^2/2-L_\mu L_\rho+2 L_\mu L_1+B_iL_\mu\right]\,\Theta(L_\mu<L_1)\nonumber\\
  & \phantom{= \frac{\alpha_sC_F}{\pi}}
  + \left[L_\mu^2 -L_\mu L_\rho  + L_\mu L_1+L_1^2/2+B_iL_1\right]\,\Theta(L_1<L_\mu<L_\rho)\nonumber\\
& \phantom{= \frac{\alpha_sC_F}{\pi}}
 +
 \Big[\frac{1}{2}(L_\rho+L_1+L_\mu+2B_i)(L_\mu+L_1-L_\rho)\Big]\,\Theta(L_\rho<L_\mu)\Big\}\nonumber\\
  & + \frac{\alpha_sC_A}{\pi}\left[(L_\mu-L_1)^2/2+B_g
    (L_\mu-L_1)\right]\,\Theta(L_\mu>L_1),
\end{align}
where the $C_R$ contribution is the same as for the recursive
$\tau_{21}$ cut and the $C_A$ contribution is the same as for the
non-recursive $\mu_{1/2}^2$ cut.

\subsection[Recursive $C_2$ cut]{Recursive \boldmath $C_2$ cut}\label{sec:bkg-ecfrec}

Again, the calculation unfolds as for the two recursive cases above
with a contribution from ``failed'' conditions for $\theta_2>\theta_1$
and a standard constraint for $\theta_2<\theta_1$. In the first case,
$e_2$ (resp. $e_3$) is set by emission $p_2$ (resp. $p_1$) and
$\theta_{12}\approx \theta_2$. 
In the second case, $e_2$ (resp. $e_3$) is set by emission $p_1$
(resp. $p_2$) and $\theta_{12}\approx \theta_1$, yielding
\begin{equation}
C_2 = \frac{z_1\theta_1^2}{z_2}\Theta(\theta_2>\theta_1) + \frac{z_2\theta_2^2}{z_1}\Theta(\theta_2<\theta_1).
\end{equation}
The Sudakov exponent will ultimately be given by
\begin{align}\label{eq:fullsud_C2rec}
R_{C,\rm rec}(z_1)
 & = \int_0^1\frac{d\theta_2^2}{\theta_2^2}\int_0^1dz_2 
   \,\frac{\alpha_s(z_2\theta_2)}{2\pi}\,P_i(z_2)
   \,\big[ \Theta(\theta_2^2>\theta_1^2)\,\Theta(z_2\theta_2^2>z_1\theta_1^2)\,\Theta(z_2>\rho/C)
   \nonumber\\
 & \phantom{= \int_0^1\frac{d\theta_2^2}{\theta_2^2}\int_0^1dz_2 
   \,\frac{\alpha_s(z_2\theta_2)}{2\pi}}
   + \Theta(\theta_2^2<\theta_1^2)\,\Theta(z_2\theta_2^2>z_1\theta_1^2)
   \nonumber\\
 & \phantom{= \int_0^1\frac{d\theta_2^2}{\theta_2^2}\int_0^1dz_2
   \,\frac{\alpha_s(z_2\theta_2)}{2\pi}}
   + \Theta(\theta_2^2<\theta_1^2)\,\Theta(z_2\theta_2^2<z_1\theta_1^2)\,\Theta(z_2\theta_2^2>\rho C/\theta_1^2) \big]
   \nonumber\\
 & + \int_0^{\theta_1^2}\frac{d\theta_{12}^2}{\theta_{12}^2}\int_0^1dz_2 
   \,\frac{\alpha_s(z_1z_2\theta_2)}{2\pi}\,P_g(z_2)
   \, \Theta(z_2\theta_{12}^2>C)-R_{\rm mass}(\rho).
\end{align}

For a fixed coupling approximation, we obtain
\begin{align}
R_{C,\rm rec}^{{\rm (fixed)}}(z_1)
& = \frac{\alpha_sC_R}{\pi}\Big\{
  \left[-L_e^2/2\right]\,\Theta(L_e<L_\rho-L_1)\\
& \phantom{=}
 + \left[(L_v+L_1-L_\rho)(L_v+2L_1-L_\rho+B_i)-L_e^2/2\right]
\,\Theta(0<L_\rho-L_e<L_1)\nonumber\\
& \phantom{=}
 + \left[
(L_e+2L_1-2L_\rho)(L_e+2L_1)/2+B_i(L_e-2L_\rho+2L_1)\right]\,\Theta(L_e>L_\rho)\Big\}\nonumber\\
& + \frac{\alpha_sC_A}{\pi}\left[(L_e+L_1-L_\rho)^2/2+B_g (L_e+L_1-L_\rho)\right]\,\Theta(L_e>L_\rho-L_1).
\nonumber
\end{align}

\subsection{Towards NLL accuracy}\label{sec:bkg-towards-nll}
In this article, as we have stated before, we are aiming to achieve
only a (modified) leading-logarithmic description of the shape
variables we study here. This level of approximation has already been
demonstrated to capture the main physical features of various jet
tagging and grooming tools (see
\eg Refs.~\cite{Dasgupta:2013ihk,Dasgupta:2013via} ).

Nevertheless it may ultimately prove important to extend the scope of our
current studies in various directions. One potential reason for this could be that here we
study tools that have some broad similarities \eg all of them place
constraints on subjet masses. In order to understand in more detail the
differences between these tools it would be helpful to increase the accuracy
of our analytical predictions, so that differences that may arise beyond
LL  effects are effectively highlighted. We would also expect
such differences to show up in the Monte Carlo event generator
studies, like those carried out below, since event generators would
partially capture many sources of subleading corrections. 

Secondly we do not study here the question of
optimal values of cuts on subjet variables, mainly confining ourselves
to the region with both $v_{\rm cut}$ and $\rho \ll 1$. To meaningfully explore the dependence
on $v_{\rm cut}$ and $\rho$ over a broader range of values of the
variables concerned, one may need to carefully investigate effects
beyond leading-logarithmic level including the role of hard
non-logarithmically enhanced contributions.

With such future developments in mind we discuss below several extra
ingredients that are required to reach NLL accuracy:
soft-and-large-angle contributions, multiple emissions, the two-loop
$\beta$ function for $\alpha_s$, finite $z_1$ corrections and
non-global logarithms~\cite{Dasgupta:2001sh}.

For the figures where we compare to Monte Carlo simulations, we will
include multiple emission effects (numerically important; see below
for their effect on the radiator function), two-loop running coupling
corrections (trivial to add, see
Appendix~\ref{app:res-building-blocks}) as well as finite $z_1$
corrections (important for the physics discussion; see
Appendix~\ref{app:finitez1}).

We have not included in our analytic results contributions which are
power-suppressed in the jet radius $R$. Although they would be
relevant for a full phenomenological prediction, and can be
substantial at the peak of the distributions (see e.g. Section 5
of~\cite{Dasgupta:2012hg}), these are expected to have little impact
when comparing the discriminative power of different jet
shapes. Moreover, they would be further reduced by the combination
with a grooming procedure which, as we argue in
Section~\ref{sec:grooming}, is the natural future direction of this
work.

\paragraph{Soft-and-large-angle radiation.} 
A source of single-logarithmic corrections comes from radiating
soft gluons at large angles. This would correspond to all the limits
beyond the strict collinear ordering that we have adopted until now
\ie it can come from either $\theta_1\sim R$, or $\theta_2\sim R$, or
$\theta_1\sim\theta_2$.

The first two regions would give single-logarithmic corrections
proportional to $R^2$. In the small-$R$ approximation we have adopted
so far, these would further be suppressed. 
At the same order of accuracy, one would also have to include
contributions coming from initial-state radiation and potential
colour-correlation with the recoiling partonic system
\cite{Dasgupta:2012hg}. Taking these into account would also add
single-logarithmic contributions to the mass distributions.
This significantly complicates the discussion, especially for signal
jets, where the mass would no longer be identical to the boosted
heavy-boson mass and we would have to impose a certain window around
the signal mass.
In practice, therefore, one usually applies these techniques together
with some grooming procedure which would drastically change this
discussion. Some first results have already been obtained in
\cite{Dasgupta:2015yua} for grooming techniques and we reserve for
future work the addition of radiation constraints to that
discussion. We will comment on that a bit further in
Section~\ref{sec:grooming}.

The situation for $\theta_1\sim\theta_2$ is a bit more involved and we
show in Appendix~\ref{app:soft-large-angle} that it would only
contribute to single-logarithmic corrections suppressed by
$\theta_1^2$.
These contributions are also at most proportional to $R^2$, although
since radiation constraints tend to take most of their discriminative
power from the large-angle region $\theta_2>\theta_1$, it makes sense
to consider a region $\theta_1\ll R$. In that case, the contribution
from the $\theta_1\sim\theta_2$ region would be even further
suppressed.

\paragraph{Multiple emissions.} Multiple gluon emissions also bring
single-logarithmic corrections to our results and we briefly discuss
below how to account for them for the non-recursive variants of the
shapes.

They correspond to cases where several gluon emissions, $(z_2,
\theta_2),\dots,(z_n,\theta_n)$, are only strongly ordered in angle and
give similar contributions to the shape $v$, \ie when
$v(z_2,\theta_2^2; z_1,\theta_1^2)\sim\dots\sim v(z_n,\theta_n^2;
z_1,\theta_1^2)$. This will come with a single-logarithmic correction
$\alpha_s^{n-1}L_v^{n-1}$ to the resummed exponent $R$. 

It is important to realise that we will keep working in the $v\ll 1$
limit and so neglect the contribution where all the $z_i\theta_i^2$,
$i\ge 2$, are of the same order as $z_1\theta_1^2$. This would also
give a single logarithmic correction of the form
$\alpha_s^nL_\rho^nf_n(v)$. Up to power corrections, we can take $f_n$
constant and this correction would therefore simply be equivalent to
the multiple-emission correction to the plain jet mass, cancelling
against the corresponding normalisation in the spectrum of
$v$.\footnote{These type of corrections may however be crucial in
  trying to obtain the spectrum of $v$ at finite $v$, a region of
  direct phenomenological relevance. We leave this for future work.}
So, from now on, we focus on the region where all the $z_i\theta_i^2$,
$i\ge 2$, are much smaller than $z_1\theta_1^2$ and compute the
corresponding correction to $R_v(z_1)$ for a fixed $z_1$.

The case of $N$-subjettiness and energy-correlation functions are
mostly straightforward. In the kinematical configurations under
consideration, the (optimal or gen-$k_t$) $N$-subjettiness axes will
still align with the jet axis and with the emission $(z_1, \theta_1)$
setting the mass. At a given $z_1$, both $\tau_{21}$ and $C_2$ will
therefore be additive and the correction to $R_v(z_1)$ will be $\gamma_E
R_v'(z_1)+\log[\Gamma(1+R_v'(z_1))]$ where $\gamma_E$ is the Euler
constant and $R_v'(z_1)$ is the derivative of $R_v(z_1)$ wrt $L_v$.

The situation is a bit more involved for the mass drop parameter. Had
we defined $\mu^2$ as $(m_{j_1}^2+m_{j_2}^2)/m^2$, $\mu^2$ would have
been additive and the similar conclusion as for $\tau_{21}$ and $C_2$
would have been reached.
Since $\mu^2$ is defined as a maximum over the two subjets rather than
a sum, we should instead use the fact that the condition
$\mu^2<\mu_{\rm cut}^2$ will be satisfied if both $m_{j_1}^2<\mu^2 m^2$
and $m_{j_2}^2<\mu^2 m^2$.

In practice, the emissions will either be clustered with the original
hard parton or with the emission setting the mass. How exactly the
particles in the jet are sifted in these two sets can depend
non-trivially on the details of the clustering. If we take as an
approximation, the assumption that particles behave independently,
they will be clustered with the hard parton or the emission setting
the mass according to which is geometrically closer, in a way similar
to the heavy-jet mass in $e^+e^-$ collisions~\cite{Catani:1991bd}.
If we split $R_{\mu_{1/2}^2}(z_1)$ in two contributions according to
whether the emissions are clustered with one or the other of the
subjets,
\begin{align}\label{eq:mu2_R0}
R_{\mu_{1/2}^2, 0}(z_1)
 & = \int_0^1\frac{d\theta_2^2}{\theta_2^2}\int_0^1dz_2 \,\frac{\alpha_s(z_2\theta_2)}{2\pi}\,P_i(z_2)
  \,\big[ \Theta(\theta_2^2<\theta_1^2) \,\Theta(\rho>z_2\theta_2^2>\rho\mu^2)\nonumber\\
 & \phantom{= \int_0^1\frac{d\theta_2^2}{\theta_2^2}\int_0^1dz_2 \,\frac{\alpha_s(z_2\theta_2)}{2\pi}P_i(z_2)}
 + \frac{1}{2} \Theta(\theta_2^2>\theta_1^2)\Theta(\rho>z_2\theta_2^2>\rho\mu^2)\big]
\end{align}
and
\begin{align}\label{eq:mu2_R1}
R_{\mu_{1/2}^2, 1}(z_1)
 & = \int_0^1\frac{d\theta_2^2}{\theta_2^2}\int_0^1dz_2 \,\frac{\alpha_s(z_2\theta_2)}{2\pi}\,P_i(z_2)
 \Theta(\theta_2^2>\theta_1^2) \frac{1}{2}\Theta(\rho>z_2\theta_2^2>\theta_1^2\mu^2)
\nonumber\\
 & + \int_0^{\theta_1^2}\frac{d\theta_{12}^2}{\theta_{12}^2}\int_0^1dz_2 \,\frac{\alpha_s(z_1z_2\theta_{12})}{2\pi}\,P_g(z_2)
  \,\Theta(z_1z_2\theta_{12}^2/\theta_1^2>\mu^2).
\end{align}
each of these two parts become additive and we obtain the following
correction to $R_{\mu_{1/2}^2}$
\begin{equation}\label{eq:resum_me_two}
  \gamma_E R_{\mu_{1/2}^2}'(z_1)+\log[\Gamma(1+R_{\mu_{1/2}^2,0}'(z_1))]+\log[\Gamma(1+R_{\mu_{1/2}^2,1}'(z_1))].
\end{equation}

This is however only an approximation and we leave a more precise
treatment for future work.
At this stage, it can also be seen as the fact that, compared to
$N$-subjettiness and energy-correlation functions, the mass-drop
parameter is more delicate to tackle analytically.

Before going to comparisons with Monte Carlo simulations, we can
observe that the two axes of $2$-subjettiness can be viewed as
partitioning the jet in two subjets, one with the jet constituents
closer to the hard parton, one with those closer to the emission
setting the mass. If instead of summing over all particles in the jet
we were summing independently over the contributions of each of the
two subjets and defining a modified $2$-subjettiness as the maximum of
these two contributions, the resummation of multiple emissions for
that observable would follow Eq.~(\ref{eq:resum_me_two}).
However, since $\Gamma(1+R_0')\Gamma(1+R_1')/\Gamma(1+R_0'+R_1')<1$ we
should expect this variant of $2$-subjettiness to perform worse than
its original definition. Conversely, defining the mass-drop parameter
as $(m_{j_1}^2+m_{j_2}^2)/m_{j}^2$ would not only make its analytic
behaviour simpler but could also translate into a slightly more
efficient tool.

\paragraph{Two-loop running coupling.} The inclusion of the two-loop
$\beta$ function is purely a technical complication. In the results
presented in Appendix~\ref{app:res-rc}, we have included their
effects.

\paragraph{Finite \boldmath $z_1$ corrections.} Finite $z_1$
corrections would typically give contributions to $R(z_1)$ like
$\alpha_s \log(1/v) \log(1/z_1)$ or
$\alpha_s \log(1/v) \log(1/(1-z_1))$. The first of these two terms,
integrated over the $1/z_1$ part of the splitting function
corresponding to the first emission, will give a double-logarithmic
contribution that we already have included. The second term, as well
as the first term integrated over the non-singular contributions to
the $P(z_1)$ splitting function will become important at NLL accuracy.
Indeed, after integration over $z_1$, they would give corrections
proportional to $\alpha_sL_v$ which contribute at the single-log
accuracy.
To properly include these corrections, it is sufficient to integrate
over the full $P(z_i)$ splitting function (rather than just including
the finite piece as a $B_i$ term) and to keep the full $z_1$
dependence when we calculate the shapes in order to get
single-logarithmic corrections to $R(z_1)$.

The corresponding results are presented in Appendix~\ref{app:finitez1}.
It is interesting to note that their calculation allows for a nice
physical discussion of similarities and differences between background
and signal jets.
Unless explicitly mentioned, these results will be used for the
figures in this paper.

\paragraph{Non-global logarithms.} Non-global logarithms are known to
be difficult contributions to handle, especially if we want to go
beyond the large-$N_c$ approximation, where a general treatment is
still lacking. We will not provide an explicit
calculation of their contribution in this paper. We note however that
it might be beneficial to apply grooming techniques such as SoftDrop
which are known to eliminate the contributions from non-global
logarithms. 
%

\subsection{Comparison with fixed-order Monte-Carlo}\label{sec:bkg-fomc}
As a partial cross-check of our results, the expressions obtained
above can be expanded in a series in $\alpha_s$ and compared to EVENT2
\cite{Catani:1996vz,event2} simulations. Here we compare the
(non-recursive) $\tau_{21}$, $\mu_{1/2}^2$ and $C_2$ distributions at order
$\alpha_s$.

Note that since we are using the $N$-subjettiness implementation from
FastJet contrib, we have to use $pp$ coordinates (transverse momentum,
rapidity and azimuth) rather than $e^+e^-$ ones (energy and polar
coordinates).\footnote{Alternatively, we could have used an $e^+e^-$
  implementation of the jet shapes (and clustering) together with
  unmodified $e^+e^-$ events. Such an implementation is already
  readily available in the fastjet-contrib implementation of Energy
  Correlation Functions. This would however give the same logarithms
  as in our $pp$ study so we decided to stay with a single coordinate
  system throughout this paper.} To maximise the efficiency and
provide quark jets with a monochromatic $p_t$, events are rotated so
that their original $2\to 2$ scattering gives 2 jets at
$y=0$.\footnote{Given the block structure of EVENT2 events, each event
  can be uniquely associated with a corresponding event with 2 partons
  in the final state. The latter can be used to define the event
  rotation. Another approach would be to rotate the event so as to
  align its thrust axis at $y=0$.}
After that rotation, jets are reconstructed with the standard ($pp$)
anti-$k_t$ algorithm \cite{antikt} with $R=0.4$.

On the analytic side, we take the fixed-order results\footnote{Running
  coupling corrections would only enter at order $\alpha_s^2$.},
expand (\ref{eq:distrib_generic_bkg}) to first order in $\alpha_s$,
and perform the $z_1$ integration.

For $N$-subjettiness, starting from (\ref{eq:tau21-fc}) we get
\begin{equation}\label{eq:bkg-tau21-fo}
\tau\,\frac{d\Sigma(\tau)}{d\tau}
 = \frac{\alpha_sC_F}{\pi} (L_\rho+L_\tau+B_q) + \frac{\alpha_sC_A}{\pi} (L_\tau+B_g).
\end{equation}
For the mass-drop parameter, we use (\ref{eq:mu2-fc}) and reach
\begin{align}\label{eq:bkg-mu2-fo}
\mu^2\,\frac{d\Sigma(\mu^2)}{d\mu^2}
& \overset{L_\mu<L_\rho}{=} \frac{1}{L_\rho+B_q}\Big[
    \frac{\alpha_sC_F}{4\pi} \big( 3 L_\rho^2+6 L_\rho L_\mu-L_\mu^2
                 +4 B_q (2 L_\rho+L_\mu) + 4 B_q^2 \big)\nonumber\\
& \phantom{\overset{L_\mu<L_\rho}{=} \frac{1}{L_\rho+B_q}}
   + \frac{\alpha_sC_A}{2\pi} \big( L_\mu^2+2 B_q L_\mu+2 B_g (L_\mu+B_q) \big) \Big]\nonumber\\
& \overset{L_\mu>L_\rho}{=} \frac{1}{L_\rho+B_q}\Big[
     \frac{\alpha_sC_F}{\pi} \big(L_\rho^2+L_\rho L_\mu+B_q (2 L_\rho+L_\mu) + B_q^2 \big)\nonumber\\
& \phantom{\overset{L_\mu<L_\rho}{=} \frac{1}{L_\rho+B_q}}
    +  \frac{\alpha_sC_A}{2\pi} \big( 2 L_\mu L_\rho-L_\rho^2+2 B_q L_\mu+2 B_g (L_\rho+B_q) \big) \Big].
\end{align}
Finally, for the energy correlation function, we start from
(\ref{eq:ecf-fc}) and obtain
\begin{align}\label{eq:bkg-ecf-fo}
C_2\,\frac{d\Sigma(C_2)}{dC_2}
& \overset{L_e<L_\rho}{=} \frac{1}{L_\rho+B_q}\Big[
    \frac{\alpha_sC_F}{2\pi} L_e ( 4 L_\rho-L_e+4B_q) 
  + \frac{\alpha_sC_A}{2\pi} L_e (L_e+2 B_g) \Big]\\
& \overset{L_e>L_\rho}{=} 
    \frac{\alpha_sC_F}{2\pi} \Big( 2 L_e+L_\rho+B_q \frac{L_\rho+2B_q}{L_\rho+B_q} \Big)
  + \frac{\alpha_sC_A}{2\pi} \Big( 2 L_e-L_\rho+2 B_g-B_q \frac{L_\rho}{L_\rho+B_q}\Big).  \nonumber
\end{align}

\begin{figure}[!t]
\includegraphics[width=0.33\textwidth]{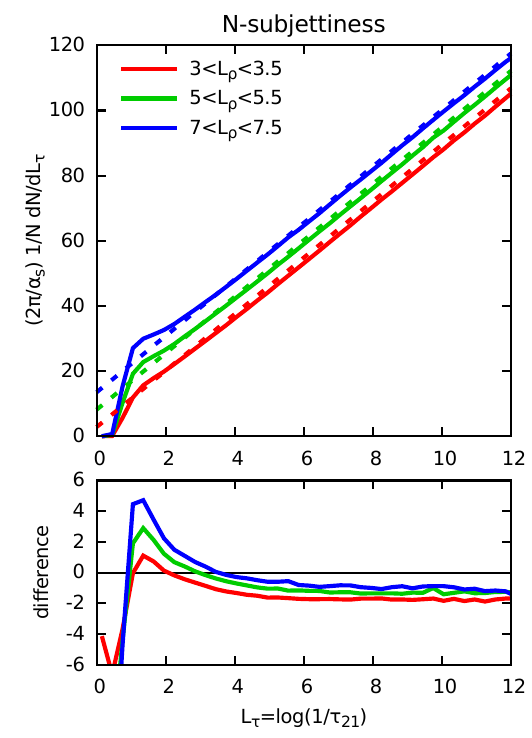}%
\includegraphics[width=0.33\textwidth]{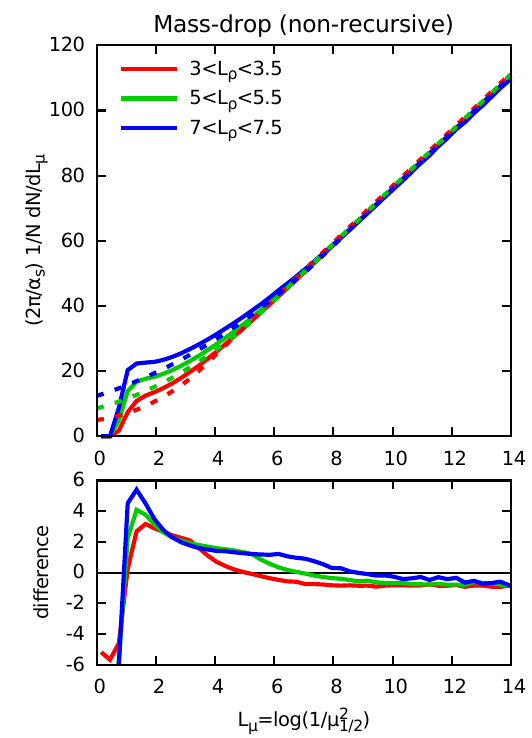}%
\includegraphics[width=0.33\textwidth]{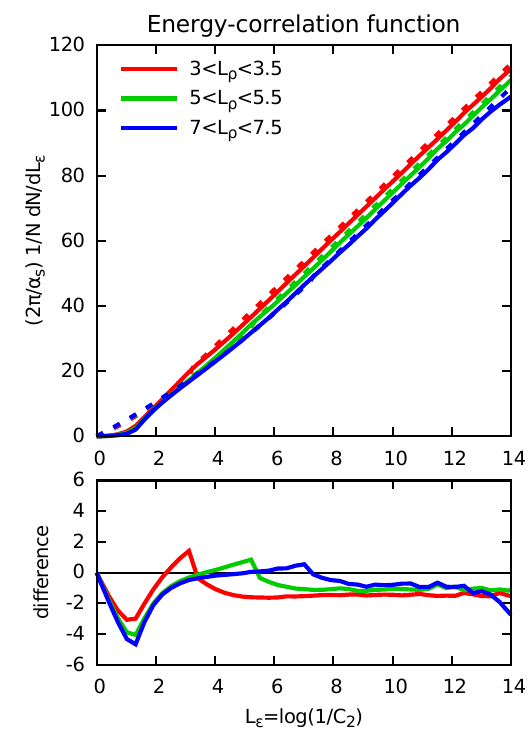}
\caption{Distributions for the (non-recursive) shapes at order
  $\alpha_s$ for a few specific bins in the jet mass. A constant
  factor $\alpha_s/(2\pi)$ has been factored out of the
  cross-section. The top row shows the distributions themselves, with
  solid lines corresponding to EVENT2 simulations and dashed lines to
  our analytic calculation. The bottom row show the difference between
  the two.}
\label{fig:bkg-fo-comparison}
\end{figure}

The comparison with EVENT2 is presented in
Fig.~\ref{fig:bkg-fo-comparison} where we have plotted the shape
distributions at order $\alpha_s$ together with our analytic
prediction. In these plots, a constant factor $\alpha_s/(2\pi)$ has
been factored out. 
From Fig.~\ref{fig:bkg-fo-comparison}, we see that this difference
goes at least to a constant at large $L_v$, meaning that we do control
the leading logarithmic behaviour.

\begin{figure}[!t]
\includegraphics[width=0.33\textwidth]{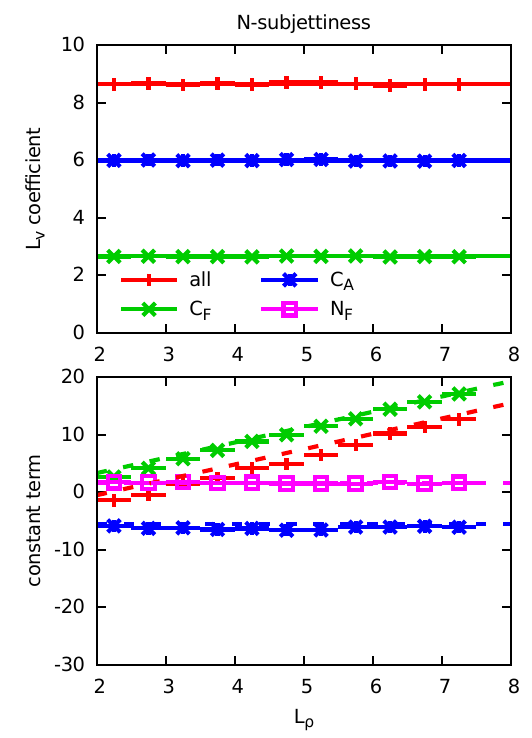}%
\includegraphics[width=0.33\textwidth]{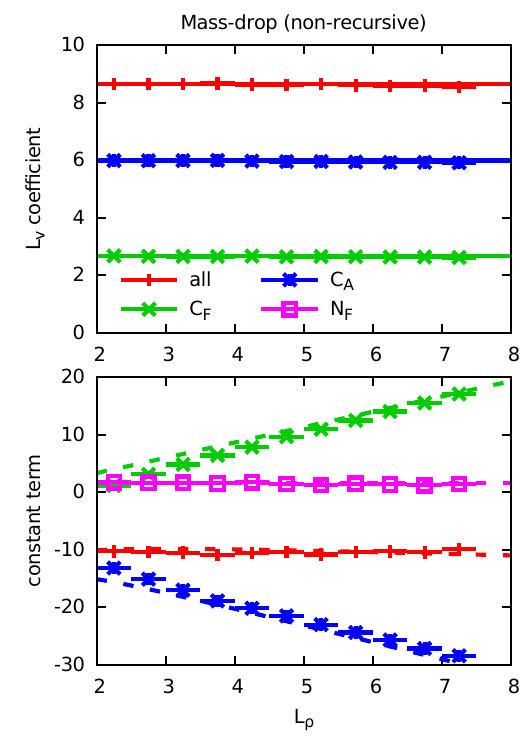}%
\includegraphics[width=0.33\textwidth]{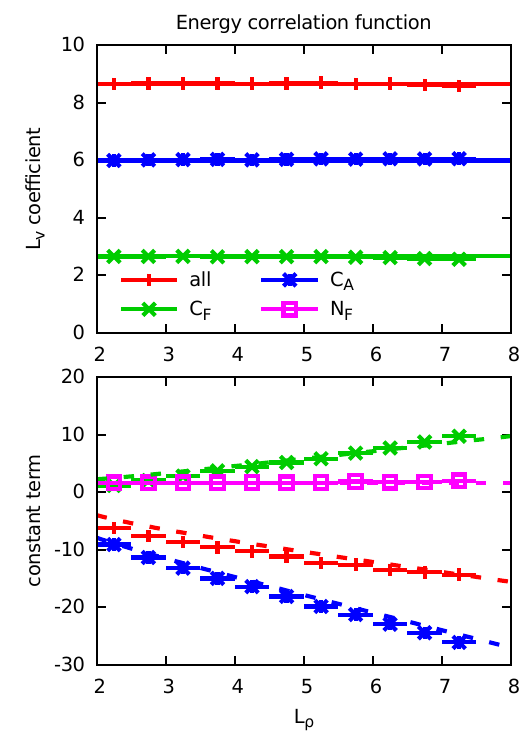}
\caption{Coefficients of the $L_v$ (top row) and constant (bottom row)
  terms extracted from the distributions in different bins of the jet
  mass. For each distribution, we have separated the results in the
  different colour channels. In all cases, a factor $\alpha_s/(2\pi)$
  has been factored out of the numbers that are shown.}
\label{fig:bkg-fo-coefficients}
\end{figure}

In principle, one can also wonder if the constant term can be obtained
from an analytic calculation, which is, strictly speaking, beyond
our leading-logarithmic accuracy.
For example, we have included in
equations~\eqref{eq:bkg-tau21-fo}-\eqref{eq:bkg-ecf-fo} corrections
coming from the hard part of the splitting function.
However, we have neglected large-angle contributions proportional to
$R^2$ and expected to be small for $R=0.4$, as well as possible finite
$z_1$ corrections. 
It is unclear from Fig.~\ref{fig:bkg-fo-comparison} whether or not
this fully accounts from the apparent constant value observed at large
$L_v$.
In this respect, it is also interesting to note that, contrary to the
jet mass where besides the logarithmic and constant terms we would
only have power corrections, the constant term in the $L_v$ expansion
has some corrections proportional to $1/L_\rho$, coming from the
normalisation of the shape distributions by the jet mass cross-section
(see Eq.~(\ref{eq:distrib_generic_bkg})). These terms can make the
convergence slower.

To extract more precise information, we have fitted, in each bin of
the jet mass, the coefficient of $L_v$ and the constant term.
This has been done in each colour channel and reported in
Fig.~\ref{fig:bkg-fo-coefficients}. Again, we see a good agreement
for the linear rise with $L_v$ as well as for the constant terms
proportional to $C_A$ and $N_f$. The slow convergence of the $C_F$
term is related to the above discussion.

More precise statements would require going to larger values of $L_v$
and $L_\rho$. This is difficult to explore due to limited machine
precision.

\subsection{Comparison with parton-shower Monte-Carlo}\label{sec:bkg-mc}

Our resummed analytic results can be directly compared to
parton-shower Monte Carlo event generators such as
Pythia~\cite{pythia} or Herwig~\cite{herwig}.
To do this, we have generated QCD dijet events in 14 TeV $pp$
collisions simulated with Pythia. We have selected anti-$k_t$(R=1)
jets with a transverse momentum of at least 3 TeV.

For our analytical predictions, we have used the results from
Appendix~\ref{app:finitez1}, which, unless explicitly mentioned
otherwise, include all the computed global NLL corrections discussed
in Section~\ref{sec:bkg-towards-nll}.
We have fixed $\alpha_s(M_z) =0.1185$ with $N_f=5$ and frozen the
coupling at $\mu_{fr}=1$~GeV.\footnote{Note that Pythia uses a
  different prescription for the strong coupling, with
  $\alpha_s(M_z) =0.1383$ and a 1-loop running. However, our analytic
  results use the 2-loop $\beta$ function. We show in
  Appendix~\ref{app:further-tests} that this does not affect our
  conclusions in any way.}

\begin{figure}[!t]
\includegraphics[width=0.48\textwidth]{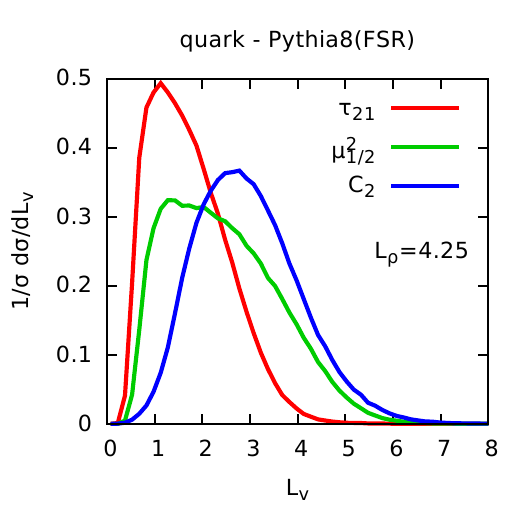}%
\hfill
\includegraphics[width=0.48\textwidth]{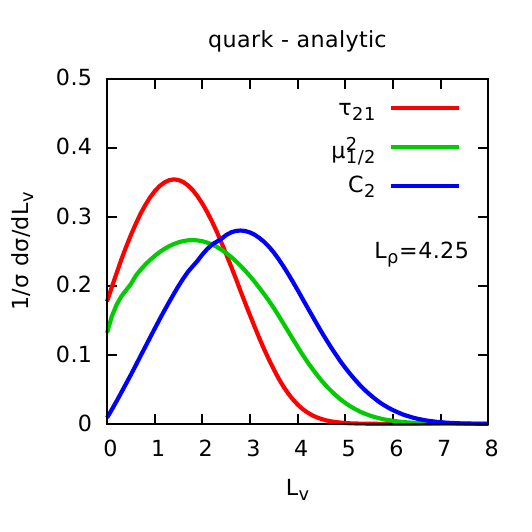}%
\caption{Distributions obtained from quark jets for each of the three
  shapes studies. Left: results obtained with Pythia including only
  final-state radiation (we used $p_{t,\rm jet}>3$~TeV, and
  $4<L_\rho<4.5$); right: results of our analytic calculations (for
  $p_t=3$~TeV and $L_\rho=4.25$).}\label{fig:MC-bkg-distribs}
\end{figure}

\begin{figure}[!ht]
\includegraphics[width=0.48\textwidth]{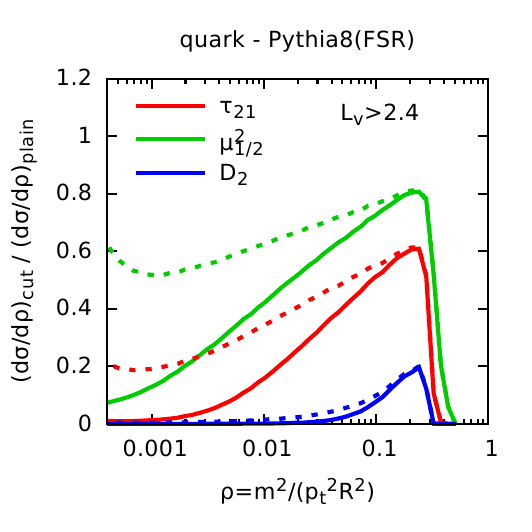}%
\hfill
\includegraphics[width=0.48\textwidth]{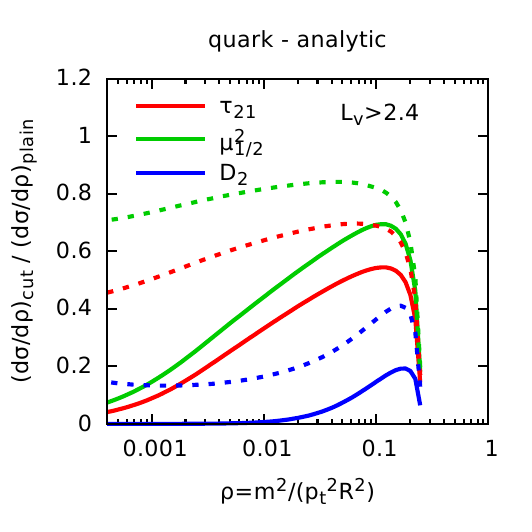}%
\caption{Ratio of the mass spectrum obtained with a cut on one of the
  shapes, divided by the plain jet mass spectrum. The solid lines are
  obtained imposing a fixed cut on the jet, while the dashed lines are
  obtained by imposing the cut recursively. Left: results obtained
  with Pythia including only final-state radiation (we used 
  $p_{t,\rm jet}>3$~TeV, and $L_v>2.4$ corresponding to $v<0.09$);
  right: results of our analytic calculations (for $p_t=3$~TeV). Note
  that multiple emissions are not included in these expressions since
  they have not been computed for the recursive
  versions.}\label{fig:MC-bkg-mass}
\end{figure}

In Fig.~\ref{fig:MC-bkg-distribs}, we compare the analytic results
obtained for the distribution of $N$-subjettiness, the mass-drop
parameter and the energy-correlation functions, at a given jet mass,
with the same distributions obtained with Pythia at parton-level,
including only final-state radiation.
First of all, if we look at the large $L_v$ region, where our analytic
description is valid, we see that it does reproduce nicely the Pythia
simulations.
However, at smaller $L_v$, Pythia tends to produce more peaked
distributions than what we obtain analytically.\footnote{Using the
  prescription from \cite{Jones:2003yv} we can replace $R(v)$ by
  $R(v/(1-v))$ and impose an endpoint, \eg at $v=1/2$, which would be
  the case for $N$-subjettiness at the order $\alpha_s$. That would
  produce distributions which look much closer to Pythia, although a
  more detailed resummation of subleading logarithms of $\rho$ (and
  $L_v$ when if becomes small), and potentially fixed-order
  corrections (e.g. for secondary emissions) would be needed to draw
  stronger conclusions.}
In any case, the main message that one has to take from this
comparison is that the generic ordering between the different shapes
is well captured by our analytic calculations.

Instead of plotting the distributions themselves, we can instead look
at the mass distributions. This has the advantage that we can also
consider the recursive versions of the cuts on the shapes.
In Fig.~\ref{fig:MC-bkg-mass}, we plotted the ratio of the mass
distribution obtained after a given cut, $L_v>2.4$, applied
recursively (dashed lines) or not (solid lines) on our three shapes,
divided by the jet mass distribution without applying any cut.
Globally, our analytic calculations tends to reproduce the main
features of the Monte Carlo simulations, although they show longer
tails at small masses.
Note that for these plots, we have used $D_2$ instead of $C_2$ since,
compared to the latter, the former peaks at values of $L_v$ closer to
the other two shapes.
Furthermore, since we have not computed multiple-emission corrections
for the recursive versions of the shape constraints, we have also left
aside the multiple-emission corrections to the non-recursive versions
for the analytic results plotted in Fig.~\ref{fig:MC-bkg-mass}. It is
interesting to notice that including the multiple-emission corrections
for the non-recursive shapes tends to reduce the tails towards small
mass, bringing more resemblance to the Pythia results. We could expect
a similar behaviour for the corresponding recursive versions.

\begin{figure}[!ht]
\includegraphics[width=0.32\textwidth]{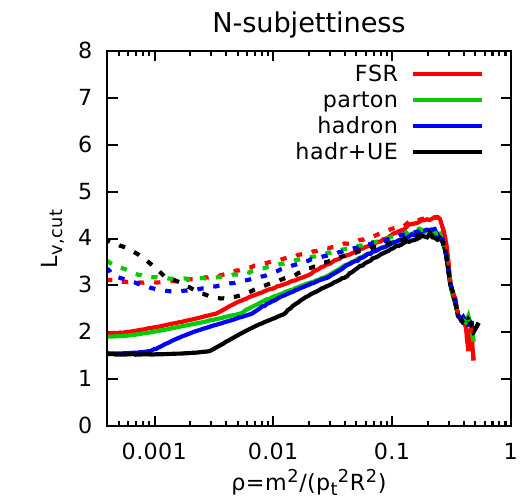}%
\hfill
\includegraphics[width=0.32\textwidth]{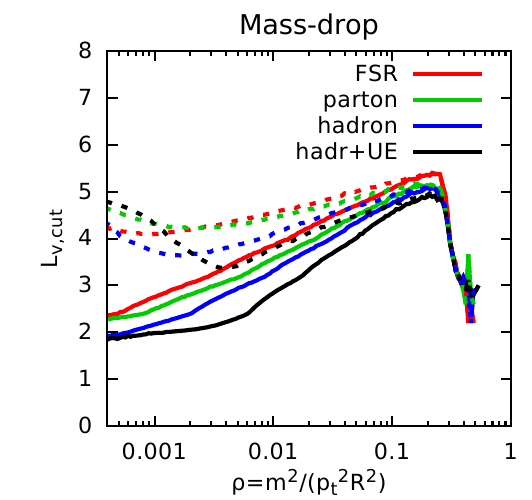}%
\hfill
\includegraphics[width=0.32\textwidth]{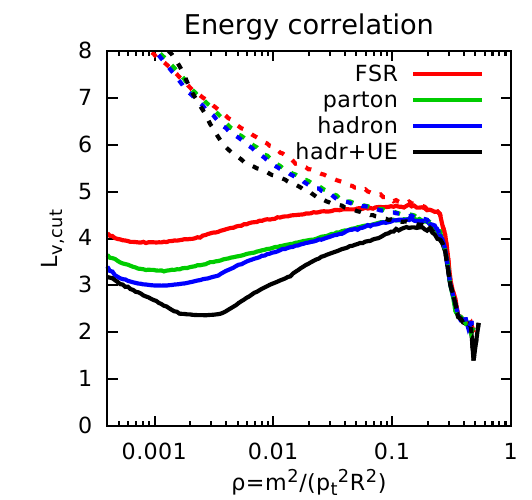}%
\caption{As a function of the jet mass, value of the cut on a given
  shape, $\log(1/v_{\rm cut})$ which would correspond to a 25\%
  tagging rate. Results correspond to dijet events obtained with
  Pythia with $p_{t,\rm jet}>3$~TeV. The various curves correspond to
  different levels of the simulations. The three plots, from left to
  right, correspond to $N$-subjettiness, the mass-drop parameter and
  the energy-correlation function.}\label{fig:MC-bkg-npeffects}
\end{figure}

Finally, we want to investigate how the three shapes we have
considered are affected by initial-state radiation (ISR) and
non-perturbative effects such as hadronisation and the Underlying
Event (UE). 
To get an insight about the importance of these effects, we have
looked, for each jet mass, at the cut on $L_v$ that has to be applied
to obtain a 25\% tagging rate compared to the plain jet mass. 
This is plotted in Fig.~\ref{fig:MC-bkg-npeffects} where we see that,
as expected, the cuts are quite sensitive to ISR and the UE, with
hadronisation effects remaining relatively small.

We attribute this behaviour to the sensitivity of the shapes to soft
and large-angle radiation.
We also see that the energy correlation function tends to be more
sensitive to these effects than $N$-subjettiness and the mass-drop
parameter.

These conclusions however have to be taken with a bit of care since
the mass of the jet itself will also be subject to the
non-perturbative effects.
In practice, one would rarely use such a cut without some additional
grooming of the jet, limiting the non-perturbative effects at least on
the reconstruction of the jet mass. We will come back to this point
later, in Section~\ref{sec:grooming}.

\section{Calculations for the signal}\label{sec:sig}

We now turn to the case of signal jets, \ie jets coming from
boosted colourless objects that decay into a $q\bar q$ pair
(or a pair of gluons), like a $W$, $Z$ or Higgs boson, or a photon.

As already briefly discussed in Sec.~\ref{sec:generic}, the splitting
of such a boosted object $X$ into a $q\bar q$ pair differs from a QCD
gluon emission in the sense that it does not diverge as $1/z$ at small
transverse-momentum fraction. 
This means that, although we are still in the regime $\rho\ll 1$ and
we shall still consider the limit of small $v$ for all jet shapes $v$
we study in this paper, now $L_1=\log(1/z_1)$ is no longer large.
As for the case of QCD jets, we shall write the results as a function
of $z_1$, see eq.~(\ref{eq:distrib_generic_sig}), but now we will keep
the correction in $z_1$ and $1-z_1$. These finite $z_1$ corrections
would generate single-logarithmic terms under the form of
contributions with one logarithm of $z_1$ or $1-z_1$ and one logarithm
of $\rho$ or $v$. It is illustrative to expand out results in series
of $\log(1/\rho)$ and $\log(1/v)$  to see explicitly how these terms
appear. We shall do this in this Section and use a fixed-coupling
approximation to better highlight the physics behind our
calculation. In Appendices~\ref{app:sig-rc} and \ref{app:finitez1}, we
give the results with a running coupling. In that case, we found it
easier to keep the $z_1$ dependence without making an explicit series
expansion, knowing that both results are equivalent at
single-logarithmic accuracy.

Besides the careful inclusion of the $z_1$ and $1-z_1$ dependence, the
calculation follows the same logic as what has been done above and
mostly consists of two copies of the contribution from ``secondary
emissions'' in the QCD case, one for each of the decay products of the
boosted colourless object. The contributions from each parton will
just differ by the replacement $z_1\leftrightarrow (1-z_1)$.
For simplicity, we still use $L_1 = \log(1/z_1)$ and additionally
introduce $L_-=\log(1/(1-z_1))$.

Finally, as was already seen to be the case for the secondary emission
contributions for QCD jets, the results presented in this section
apply invariantly for the recursive or non-recursive versions of the
shapes.

\subsection[$\tau_{21}$ cut]{\boldmath $\tau_{21}$ cut}
\label{sec:sig-tau}

Following the same construction as in Section~\ref{sec:bkg-tau}, we find
that for an emission off the parton carrying a momentum $(1-z_1)p_t$, we
have
\begin{equation}\label{eq:sig-tau-value}
\tau_{21} = \frac{z_2\theta_2^2}{z_1\theta_1^2}.
\end{equation}

This leads to
\begin{align}\label{eq:sig-tau}
R_{\tau}(z_1)
 & = \int_0^{\theta_1^2}\frac{d\theta_2^2}{\theta_2^2}\int_0^1dz_2 \,\frac{\alpha_s(z_2\theta_2)}{2\pi}\,P_q(z_2)
  \,\Theta(z_2\theta_2^2/\theta_1^2>z_1\tau_{21}) + [z_1\leftrightarrow (1-z_1)],
\end{align}
where $\theta_1^2=\rho/[z_1(1-z_1)]$.

For a fixed coupling approximation, and keeping only the first
non-trivial terms in $L_1$ and $L_-$,  terms we find
\begin{equation}\label{eq:sig-tau-fc}
R_{\tau}^{{\rm (fixed)}}(z_1)
 = \frac{\alpha_sC_R}{\pi}\left[L_\tau^2+(L_1+L_-+2B_i)L_\tau\right].
\end{equation}

\subsection[$\mu^2$ cut]{\boldmath $\mu^2$ cut}\label{sec:sig-mu2}

As for the case of QCD jets discussed in Section~\ref{sec:bkg-mu2},
expressions for $\mu^2$ differ from the $N$-subjettiness ones due to
the fact that the $p_t$ normalisations are different.

For an emission off the parton carrying a momentum $(1-z_1)p_t$, we have
\begin{equation}\label{eq:sig-mu2-value}
\mu^2_{1/2} = \frac{(1-z_1)z_2\theta_2^2}{z_1\theta_1^2}.
\end{equation}

This leads to
\begin{align}\label{eq:sig-mu2}
R_{\mu^2_{1/2}}(z_1)
 & = \int_0^{\theta_1^2}\frac{d\theta_2^2}{\theta_2^2}\int_0^1dz_2 \,\frac{\alpha_s(z_2\theta_2)}{2\pi}\,P_q(z_2)
  \,\Theta(z_2\theta_2^2/\theta_1^2>z_1/(1-z_1)\mu^2_{1/2}) + [z_1\leftrightarrow (1-z_1)]
\end{align}

Note that formally the $\Theta$ constraint above will result in the
condition $\Theta(\mu^2<(1-z_1)/z_1)$ but this will only lead to power
corrections in $\mu^2$ and can hence be neglected. 

For a fixed coupling approximation the extra contributions from the
two legs thus cancel, giving
\begin{equation}\label{eq:sig-mu2-fc}
R_{\mu^2_{1/2}}^{{\rm (fixed)}}(z_1)
 = \frac{\alpha_sC_R}{\pi}\left[L_\mu^2+2B_iL_\mu\right].
\end{equation}

Note that in the case of the signal, the calculation for $\mu_0^2$
would lead to the same result. However, other effects like soft and
large-angle gluon emissions that we have neglected here would appear
at the same order and lead to an infrared divergence for $\mu_0^2$.

\subsection[$C_2$ cut]{\boldmath $C_2$ cut}\label{sec:sig-ecf}

This time for emissions off the parton carrying a momentum
$(1-z_1)p_t$, we find
\begin{equation}\label{eq:sig-ecf-value}
C_2 = \frac{\rho}{z_1^2(1-z_1)}\,z_2\,\frac{\theta_2^2}{\theta_1^2}.
\end{equation}

This leads to
\begin{align}\label{eq:sig-ecf}
R_{C_2}(z_1)
 & = \int_0^{\theta_1^2}\frac{d\theta_2^2}{\theta_2^2}\int_0^1dz_2 \,\frac{\alpha_s(z_2\theta_2)}{2\pi}\,P_q(z_2)
  \,\Theta\Big(\frac{z_2\theta_2^2}{\theta_1^2}>\frac{z_1^2(1-z_1)C_2}{\rho}\Big)
 + [z_1\leftrightarrow (1-z_1)]
\end{align}

For a fixed coupling approximation, we get
\begin{equation}\label{eq:sig-ecf-fc}
R_{C_2}^{{\rm (fixed)}}(z_1)
 = \frac{\alpha_sC_R}{\pi}\left[(L_e-L_\rho)^2+(3L_1+3L_-+2B_i)(L_e-L_\rho)\right]\,\Theta(L_e>L_\rho).
\end{equation}
Again, formally the extra factor $z_1^2(1-z_1)$ will enter in the
$\Theta(L_e>L_\rho)$ condition but its effect is only power
corrections and then can be neglected.

\subsection[Integration over the $z_1$ splitting]{\boldmath Integration over the $z_1$ splitting}\label{sec:sig-split}

For most of the splitting relevant for phenomenological studies, the
splitting function in terms of $z_1$ is expressed as $z_1^k(1-z_1)^k$
or as a linear combination of such terms (typically, only $k=0$ and
$k=1$ are needed for $W/Z/H$ or photon signals).

Introducing $B_2(x)=B(x,x)=\Gamma^2(x)/\Gamma(2x)$, the integration
over $z_1$ can be performed in the fixed-coupling approximation, using
\begin{equation}\label{eq:sig-integ-splitting}
\int_0^1 dz_1 \, z_1^k(1-z_1)^k
\exp\left(-\frac{\alpha_sC_R}{\pi}\,p\,L_v\,(L_1+L_-)\right)
 = B_2\left(1+k+\frac{\alpha_sC_R}{\pi}\,p L_v\right),
\end{equation}
with $p$ a number varying from one shape to another.

\subsection{Comparison with fixed-order Monte-Carlo}\label{sec:sig-fomc}

Similarly to what was presented in Section~\ref{sec:bkg-fomc} for QCD
jets, we can compare our results with EVENT2 simulations.
In this case, we boost the event along the $z$ axis and rotate it to
obtain boosted photons decaying to a jet at $y=0$.\footnote{It appears
  that the exact outcome depends on the value used for the EVENT2
  parameter {\tt metype}, referring to the matrix elements. Set to 1,
  our default here, we recover the expected situation of a boosted
  photon. Set to 0, it behaves like a boosted scalar particle, \ie
  with a $z$-independent splitting function.}

\begin{figure}[!t]
\includegraphics[width=0.33\textwidth]{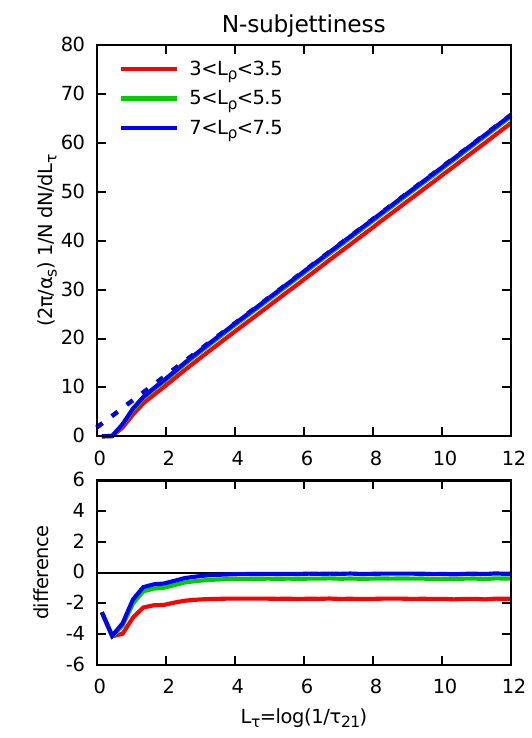}%
\includegraphics[width=0.33\textwidth]{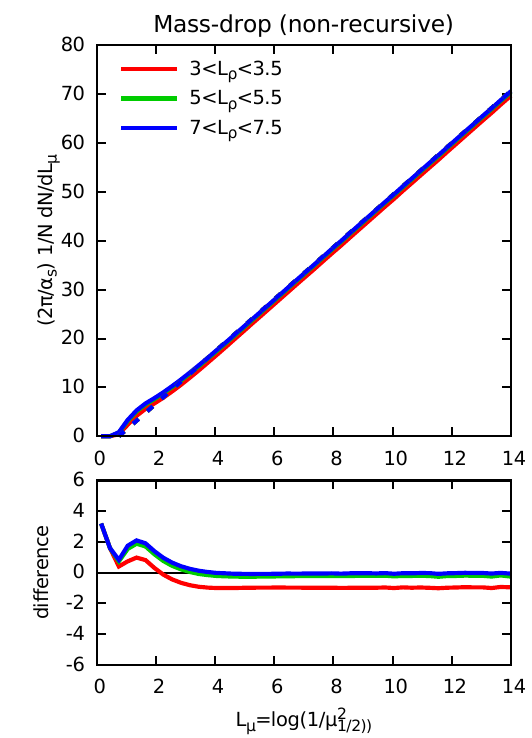}%
\includegraphics[width=0.33\textwidth]{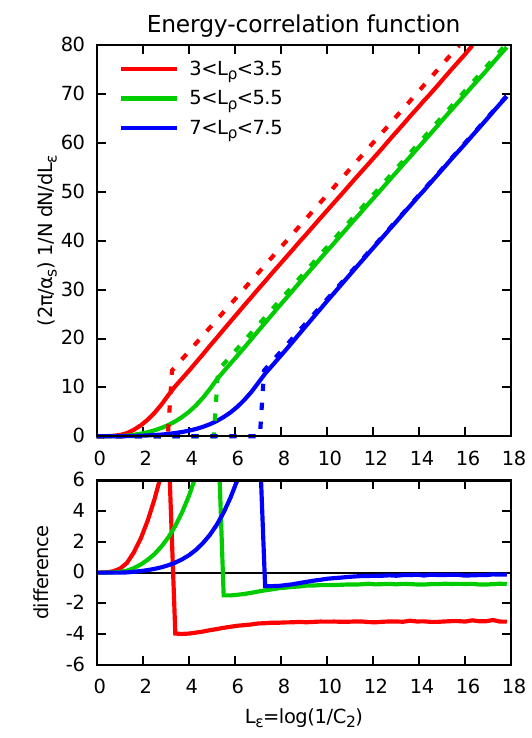}
\caption{Distributions for the (non-recursive) shapes at order
  $\alpha_s$ for a few specific bins in the jet mass for the hadronic
  decay of a $Z$ boson. A constant factor $\alpha_s/(2\pi)$ has been
  factored out of the cross-section. The top row shows the
  distributions themselves, with solid lines corresponding to EVENT2
  simulations and dashed lines to our analytic calculation. The bottom
  row shows the difference between the two.}\label{fig:sig-fo-comparison}
\end{figure}

The expansion of the above results to first order in $\alpha_s$ gives,
after integration over $z_1$
\begin{align}
\tau\frac{d\Sigma(\tau)}{d\tau}
& = \frac{\alpha_sC_F}{\pi} (2L_\tau+2B_q+a_\gamma),
\label{eq:gamma-fo-tau21}\\
\mu^2\,\frac{d\Sigma(\mu^2)}{d\mu^2}
& = \frac{\alpha_sC_F}{\pi} (2L_\mu+2B_q),
\label{eq:gamma-fo-mu2}\\
C_2\,\frac{d\Sigma(C_2)}{dC_2}
 & = \frac{\alpha_sC_F}{\pi} \big(2(L_e-L_\rho)+2B_q+3a_\gamma\big)\Theta(L_e>L_\rho)\label{eq:gamma-fo-C2}.
\end{align}
In the above expressions, $a_\gamma=\frac{3}{2} a_0-\frac{1}{2} a_1 = \frac{13}{6}$ with
$a_0=2$ and $a_1=\frac{5}{3}$.

The comparison of these analytic results with EVENT2 simulations
is presented in Fig.~\ref{fig:sig-fo-comparison} and shows a good
agreement. It is also interesting to notice that the convergence seems
faster than it was for QCD jets, probably due to the fact that here
the jet mass is fixed.

\subsection{Comparison with parton-shower Monte-Carlo}\label{sec:sig-mc}

\begin{figure}[!t]
\includegraphics[width=0.48\textwidth]{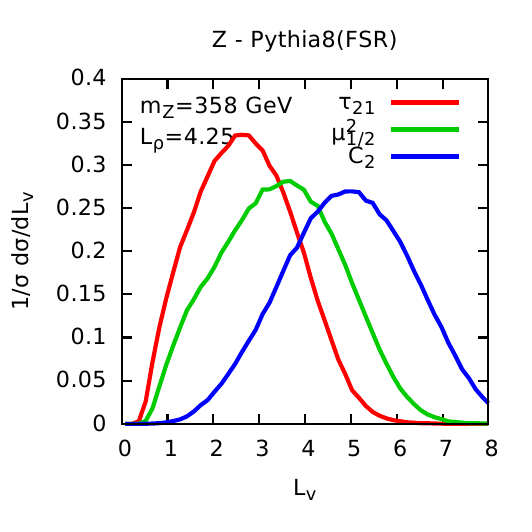}%
\hfill
\includegraphics[width=0.48\textwidth]{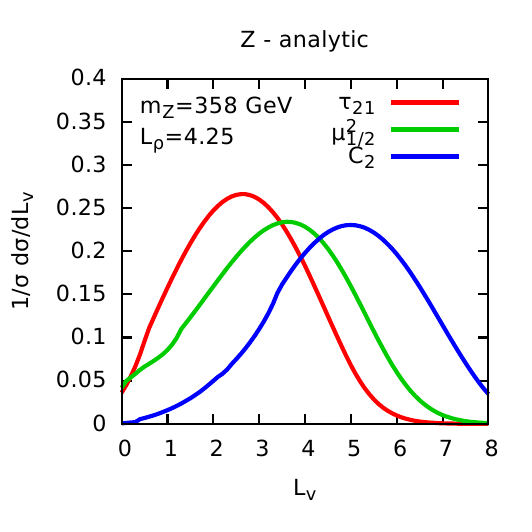}%
\caption{Distributions obtained from $Z\to q\bar q$ jets for each of
  the three shapes studies. Left: results obtained with Pythia
  including only final-state radiation (for $4<L_\rho<4.5$); right:
  results of our analytic calculations (for
  $L_\rho=4.25$).}\label{fig:MC-sig-distribs}
\end{figure}

\begin{figure}[!t]
\includegraphics[width=0.48\textwidth]{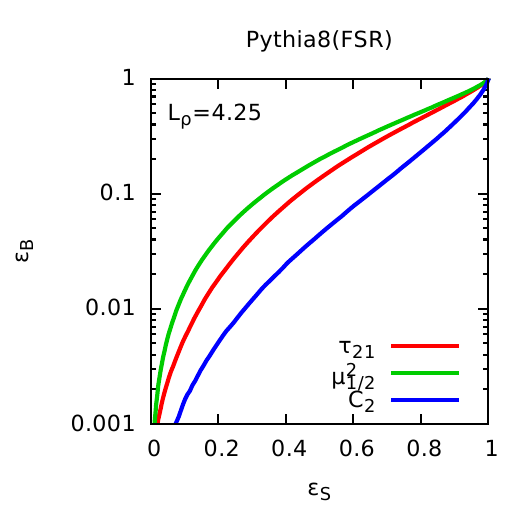}%
\hfill
\includegraphics[width=0.48\textwidth]{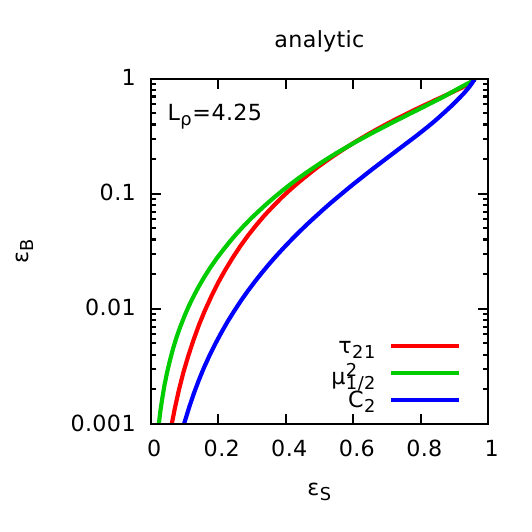}%
\caption{ROC curves showing the background fake rate as a function of
  the signal efficiency obtained from $Z\to q\bar q$ jets for each of
  the three shapes studies. Left: results obtained with Pythia
  including only final-state radiation (for $4<L_\rho<4.5$); right:
  results of our analytic calculations (for
  $L_\rho=4.25$).}\label{fig:MC-ROC}
\end{figure}

As for the case of the QCD background jets, we want to compare our
analytic calculations to parton-shower Monte Carlo simulations. 
This time, we used Pythia to generate $ZZ$ events with both $Z$ bosons
decaying to hadrons. To match the jet selection of
Section~\ref{sec:bkg-mc} in the case of QCD jets, we have selected
anti-$k_t$($R=1$) jets with $p_t\ge 3$~TeV and artificially varied the
mass of the $Z$ boson to scan over the $\rho$ range.

The distributions obtained for the shapes are plotted on
Fig.~\ref{fig:MC-sig-distribs} for $Z$ bosons decaying
hadronically. As for the case of QCD jets, we see a good overall
description of the features of the distributions and of the
differences between the three shapes, particularly in the large $L_v$
region which is targeted by our calculation.

Based on the results for both the signal and the QCD background, we
have plotted a set of ROC curves on Fig.~\ref{fig:MC-ROC} obtained by
varying the cut on the three shapes for a given value of the jet
mass.
Note that here, the signal and background efficiencies are normalised
to the sample of jets that are within the mass window under
investigation.
The main result here is that a cut on the energy correlation function
is more efficient at rejecting the QCD background than a cut on
$N$-subjettiness, itself performing a bit better than a cut on the
mass-drop parameter. This behaviour is clearly seen in both the Pythia
simulations and our analytic calculations.\footnote{We show in
  Appendix~\ref{app:further-tests} that this remains valid for less
  boosted jets, e.g. with $p_t=500$~GeV.}
We leave a detailed discussion of this comparison for
Section~\ref{sec:discussion}.

\section{Non-perturbative effects and combination with grooming}\label{sec:grooming}

\begin{figure}[!htbp]
\centering
\includegraphics[width=0.32\textwidth]{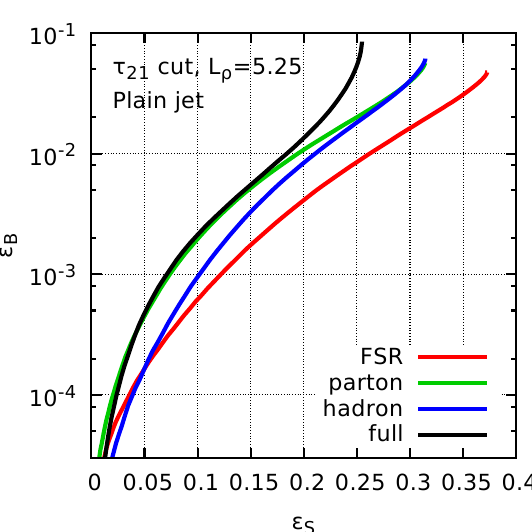}
\hfill
\includegraphics[width=0.32\textwidth]{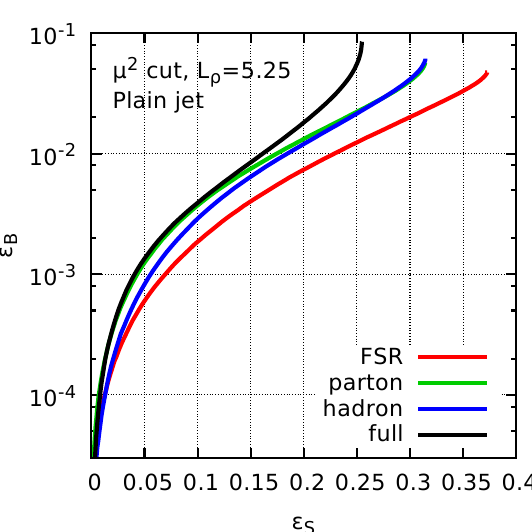}
\hfill
\includegraphics[width=0.32\textwidth]{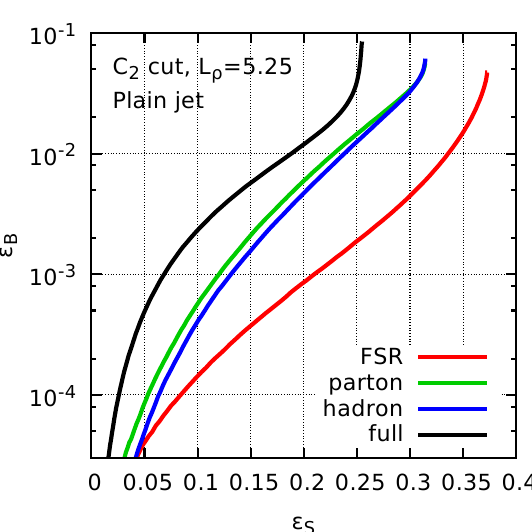}
\includegraphics[width=0.32\textwidth]{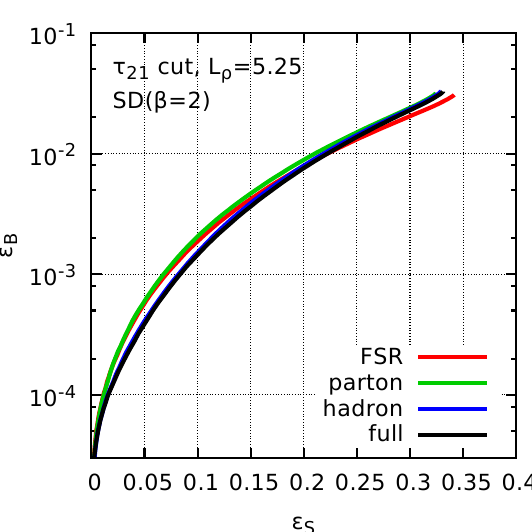}
\hfill
\includegraphics[width=0.32\textwidth]{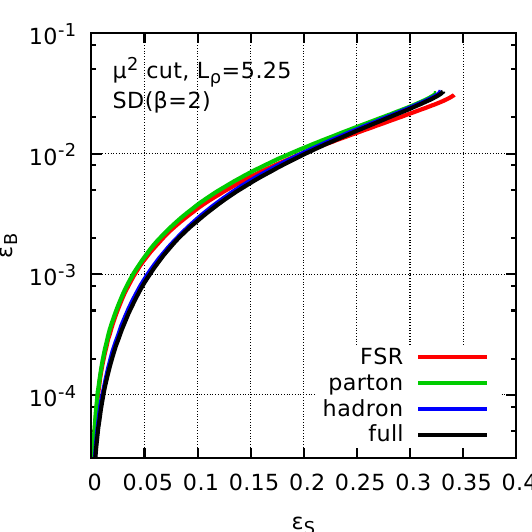}
\hfill
\includegraphics[width=0.32\textwidth]{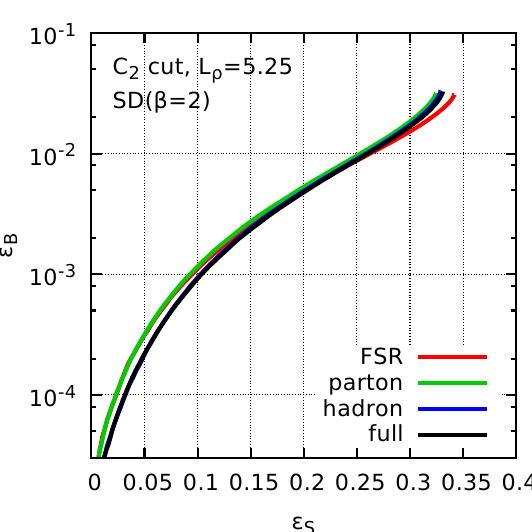}
\caption{Effects of the initial-state radiation (green), hadronisation
  (blue) and Underlying Event (black) on the ROC curves, compared to 
  pure final-state radiation (red). In all cases, we impose that 
  $5<\log(p_t^2R^2/m^2)<5.5$.
  The left, central and right columns correspond to $\tau_{21}$, 
  $\mu_{1/2}^2$ and $C_2$, respectively. For the top row, the mass
  and shape constraints are imposed on the plain, ungroomed, jet. 
  For the plots on the bottom row, we have first applied a SoftDrop
  procedure with $\beta=2$ and $z_{\rm cut}=0.1$ before imposing the
  mass and shape constraints.}\label{fig:with-softdrop}
\end{figure}

We have already seen in Section~\ref{sec:bkg-mc} and in
Figure~\ref{fig:MC-bkg-npeffects} that initial-state radiation and
non-perturbative effects can have a large impact on the shapes we have
studied.
One difficulty in trying to assess these effects is that they do not
only affect the different shapes we are interested in but also the jet
mass and hence our selection of a sample of jets with a mass lying
within a given window.

To make a physically meaningful comparison, we have to adapt our
normalisation of the background and signal efficiencies compared to
what we used to produce Figure~\ref{fig:MC-ROC}. Instead, we shall now
compute the efficiencies as the fraction of the jets passing the
initial $p_t$ cut which satisfy both the constraint on the mass and
the constraint on the shape.
In such a case, as the cut on the shape increases, the signal and
background efficiencies progressively increase to ultimately reach an
endpoint, common to all shapes, where just the cut on the mass is
effective.

As before, we work with anti-$k_t$ jets with $R=1$ and impose a $p_t$
cut of 3 TeV. For the signal, we used a massive $Z'$ boson with a mass
of 217~GeV and impose the constraint on the mass that
$5<\log(p_t^2R^2/m^2)<5.5$.\footnote{Working with the nominal $Z$ mass
  would bring us yet closer to the non-perturbative region and
  increase even further the effects observed here.}
Here the background is taken as quark-only to match with the results
presented in the previous sections. 

The top row of Fig.~\ref{fig:with-softdrop} show the ROC curves
obtained for our three shapes starting from events including only
final-state radiation effects at parton level (in red) and adding
successively initial-state radiation (in green), hadronisation effects
(in blue) and the Underlying Event (in black).
We clearly see large deviations from what we observe for pure FSR
results, noticeably when adding initial-state radiation and the
Underlying Event. Concentrating on the endpoint of these curves, where
the cut on the shapes has no effect, we see that these effects are
already present when applying the initial mass cut.

In practice, when working with large-$R$ jets, one usually first
applies a grooming procedure in order to obtain, at the very least, a
good resolution on the jet mass.
The bottom row of Fig.~\ref{fig:with-softdrop} shows the same plot as
on the top row, now obtained by first grooming the jet with the
SoftDrop procedure \cite{Larkoski:2014wba}, using $z_{\rm cut}=0.1$
and $\beta=2$, before imposing the cut on the mass and on the shapes.
Although this reduces the performance observed on events with pure
final-state radiation, this has two positive effects: (i) it
stabilises remarkably the ROC curves against initial-state radiation
and non-perturbative effects, and (ii) at full parton level it even
gives better performance than without the grooming procedure.
Again, the ordering between the three shapes remains the same, albeit
with strongly reduced differences compared to the plain jet case.

\section{Discussion and conclusions}\label{sec:discussion}

In this paper, we have provided a first-principles comparison of the
performance of three common jet-shapes --- $N$-subjettiness, the
mass-drop parameter and Energy-Correlation Functions --- used to
discriminate boosted two-prong decays from QCD jets. 
In order to ensure infrared safety, we have defined the mass-drop
parameter based on the subjets obtained via a clustering with the
generalised $k_t$ algorithm with the extra parameter $p$ set to $1/2$.
Similarly, for $N$-subjettiness, we find that using the exclusive
gen-$k_t$($p=1/2$) algorithm is an efficient alternative to the
more complicated optimal axes.
The usage of the gen-$k_t$ algorithm is closely connected to the fact
that it respects the ordering in mass, which is helpful in our
situation where we work at a fixed jet mass and study shapes that have
a mass-like behaviour.

The main observation from our analytical results and simulations
involving only final-state radiation is that there appears to be a
clear ordering in the discriminating power of the shapes we have
studied: the energy-correlation function ratio is more powerful than
the $N$-subjettiness ratio which, in turn, is more powerful than a
cut on the $\mu^2$ parameter.

Our results indicate a Sudakov suppression of both the signal and the
background for $v\ll 1$. This suppression is however more powerful for
the background for two major reasons. 
Recall that, since we work at a fixed jet mass, both the QCD jets and
the signal jets can be seen as two-pronged objects.\footnote{Strictly
  speaking, this is only true in the strongly-ordered limit, relevant
  in the small $v$ context considered in this paper (up to NLL in
  $L_v$). For more generic situations, one would also have to consider
  multi-pronged QCD jets.}
A cut on the shape thus constrains additional radiation from that
system.
Given that, discrimination power comes from constraints on radiation
at angles smaller and larger than the opening angle between the two
prongs.
For large angles, the cut on the shape only affects the background due
to the colour-singlet nature of the signal.
At small angles, the radiation from each of the two prongs is
proportional to their colour factors, which tend to be larger for QCD
jets, involving gluons in their two-prong decay, than for resonances
mostly decaying to quarks.\footnote{This argument would be reversed for
  resonances decaying to gluons.}
Since we know from experience with quark-gluon discrimination that
exploiting differences in colour factors only lead to moderate
discrimination power
\cite{Larkoski:2013eya,Gallicchio:2011xq,Larkoski:2014pca,Aad:2014gea},
we expect that the large-angle effect would be the main source of
difference in tagging two-body decays.

The ordering in discrimination power between the different shapes can
also be understood from that viewpoint. 
Say we work at a given signal efficiency. The corresponding cut on the
shape would determine the constraints on small-angle radiation for both
the signal and the background (up to colour-factor effects discussed
above).
Once this is fixed, one has to look at the constraint put on the
large-angle radiation for QCD jets.
In that region, it is clear from our results, that the radiation veto
imposed by a cut on $C_2$ is more constraining than that imposed by a
cut on $\tau_{21}$, itself more constraining than a cut on $\mu^2$.
This can be deduced from Fig.~\ref{fig:phasespace}: fixing the signal
efficiency amounts to fix the rejected region at small angle and once
this is held equal for all three shapes, the vetoed region at large
angle shows a clear ordering between $C_2$, $\tau_{21}$ and
$\mu^2$.\footnote{Strictly speaking, this is only true at a fixed
  value of $z_1$ but the integration over $z_1$ will not significantly
  affect the argument.}

\begin{figure}[!t]
\includegraphics[width=0.48\textwidth]{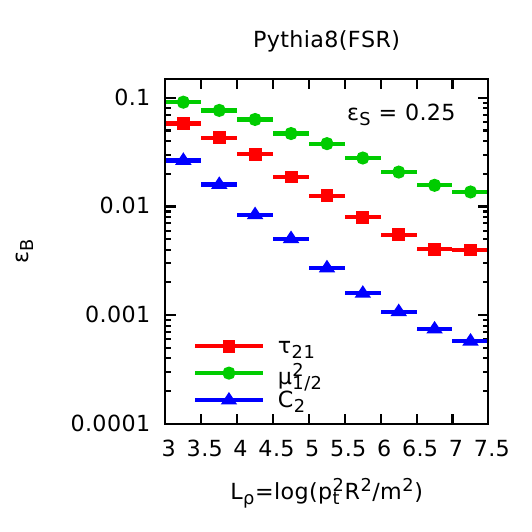}%
\hfill
\includegraphics[width=0.48\textwidth]{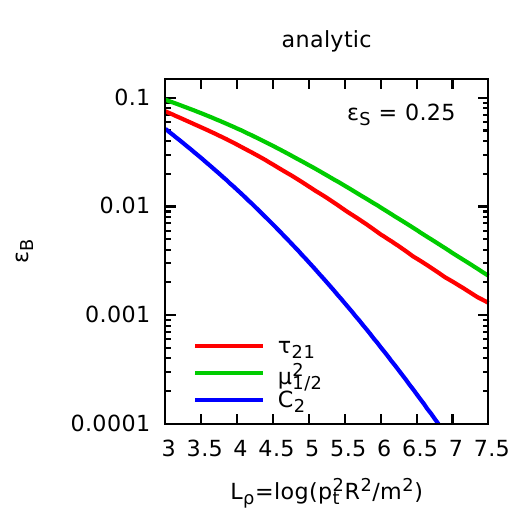}%
\caption{Background fake rate for a 25\% signal efficiency as a
  function of the jet mass. As above, we used $R=1$ and
  $p_{t,\rm jet}>3$~TeV for the Pythia simulation (left plot) and
  $p_t=3$~TeV, for the analytic calculation (right
  plot).}\label{fig:MC-Lrho-dep}
\end{figure}

This statement can be made more quantitative from our analytic
results.
First, the difference between $\tau_{21}$ and $\mu^2$ mostly comes
from the large-angle region where gluon emissions are clustered with
the gluon setting the mass. The extra $z_1$ factor in the expression
for $\mu^2$ compared to $\tau_{21}$, see
Eq.~(\ref{eq:tau21-value-primary}) v. (\ref{eq:mu2value}), results in
a smaller vetoed region for $\mu^2$. Parametrically, this region
scales like $\alpha_s\log(1/\theta_1^2)\log(1/v)\propto
\alpha_s\log(1/\rho)\log(1/v)$. This can be deduced algebraically from
our results by fixing the signal efficiency and computing the
background for the corresponding cut (with additional $\alpha_s
\log^2(1/v)$ terms also coming from the small-angle region).
In the case of $C_2$, the constraint at large angle now becomes
proportional to $\theta_2^4$, see Eq.~(\ref{eq:ecf-value-primary}),
and this translates into an additional vetoed region compared to
$\tau_{21}$ which is proportional to
$\alpha_s\log^2(1/\theta_1^2)\propto\alpha_s\log^2(1/\rho)$. 
In conclusion, we expect the ordering between the shapes to be more
visible when increasing the boost of the jet. This difference should
also grow faster with $p_t/m$ when comparing $C_2$ and $\tau_{21}$
than for $\tau_{21}$ and $\mu^2$.
This is indeed what is observed from both pure-FSR Monte-Carlo studies
and from our analytic calculations, as seen in
Fig.~\ref{fig:MC-Lrho-dep}, where we have plotted the background
rejection rate for a 25\% signal efficiency as a function of
$\log(1/\rho)=\log(p_t^2R^2/m^2)$.\footnote{We used the same samples
  as in Sections \ref{sec:bkg-mc} and \ref{sec:sig-mc}, using a 3 TeV
  cut on the jet $p_t$ and varying its mass.}

Note that our explanation of the differences between $C_2$ and
$\tau_{21}$ is consistent with a similar observation made in
\cite{Larkoski:2013eya} but our more detailed analytic treatment
allows for more quantitative understanding.

The next important observation is that, without grooming, the shapes
are significantly affected by ISR and non-perturbative effects, UE
in particular. These model-dependent effects can be substantial enough
to wash out or even invert the differences between the shapes observed
from pure FSR and analytic studies (see \eg the top row of
Fig.~\ref{fig:with-softdrop}).
This is due to the impact of these effects on both the mass resolution
for the jet --- mostly for signal jets --- and the sensitivity of the
shapes themselves.
Since ISR and UE mostly affect the soft-and-large-angle region, we expect
$C_2$ to be more affected than $\tau_{21}$, itself more affected than
$\mu^2$ (see the discussion above) and this is indeed what we observe
from Monte Carlo studies.

Furthermore, we have seen that applying a grooming procedure on the
jet before computing its mass and values of the shapes largely
improves the robustness against ISR and non-perturbative effects, also
restoring the ordering between the shapes observed with pure FSR.
Again, this can be interpreted as grooming cutting away a part of the
soft-and-large-angle region. 
This increased robustness however comes at a price in that reducing
the soft-and-large-angle region using grooming also reduces the
discriminating power of the shape cuts.
In practice, there will be a trade-off between sheer efficiency and
robustness against model-dependent effects.
We reserve the detailed study of an optimal combination of a shape cut
with a proper grooming procedure for future work.

In addition, note that working at a fixed jet mass ensures that our
results are infrared-and collinear safe because it fixes automatically
the value of $\tau_1$ and $e_2$. If we were to impose a cut on the
shapes without fixing the jet mass, our results would still be finite
after integration of (\ref{eq:mass_generic_bkg_full}) over $\rho$
because the infrared region is killed by the plain mass Sudakov. This
is an example of Sudakov-safe observables
\cite{Larkoski:2013paa,Larkoski:2015lea}.
It is interesting to note that, after integration over the jet mass,
we recover a distribution that can be expressed as a series in
$\sqrt{\alpha_s}\log(1/v)$, similar to what was obtained for ratios of
angularities in \cite{Larkoski:2013paa}.

The arguments above can be applied when comparing the recursive and
non-recursive versions of the shapes: the recursive versions have a
smaller vetoed region at large angle while retaining the same
small-angle region as their corresponding non-recursive version. Thus,
although the recursive versions have the advantage of being less
sensitive to ISR and non-perturbative effects, they have a smaller
discriminating power. A combination of a non-recursive cut on the
shape with a proper grooming of the jet is expected to perform better
while at the same time limiting non-perturbative effects.

Another key aspect of our results is that a cut on the shapes leads to
an exponential suppression of the signal efficiency. This has to be
contrasted with two-prong taggers like the mass-drop tagger, trimming
or pruning which would only give a linear suppression
\cite{Dasgupta:2015yua}. This means that although it initially seems
natural to work in the small $v$ limit, in practice one will not be
able to take the cut on $v$ too small. Computing corrections for
finite $v$ could then become relevant for this discussion.

Finally, there are several other developments that can be made based
on this study. In this paper, we have focused on a subset of jet
shapes sensitive to the mass of the subjets. It would be interesting
to extend this study to more generic jet shapes, \eg studying the
$\beta$ dependence of energy-correlation-function ratios and
$N$-subjettiness ratios.
On the more formal side, we could also refine our calculations to
include effects such as the initial-state radiation and finite jet
radius contributions as well as attaining full NLL accuracy,
optionally matched to a fixed-order calculation.

\section*{Acknowledgements}
This work has benefited from fruitful interactions with several
colleagues. In particular, we wish to thank Gavin Salam for
stimulating conversations and collaboration in the early stages of
this work. 
We also want to thank Jesse Thaler, for several interactions about
$N$-subjettiness and making some features available in its code, as
well as Andrew Larkoski.
MD wishes to thank the IPhT and the French ANR for hospitality and
financial support.
MD's work is supported in part by the Lancaster-Manchester-Sheffield
Consortium for Fundamental Physics under STFC grant ST/L000520/1.
This work was also in part supported by the ERC advanced grant
Higgs@LHC, by the French Agence Nationale de la Recherche, under grant
ANR-10-CEXC-009-01, and by the Paris-Saclay IDEX.

\appendix

\section{Results with running coupling: QCD background}\label{app:res-rc}

Results including running-coupling corrections can be
straightforwardly obtained from the expressions before integration
over $z_2$ and $\theta_2$ given in Section \ref{sec:bkg}. The running
of the coupling is expressed wrt its value
$\alpha_s\equiv\alpha_s(p_tR)$ taken at the physical scale of the
problem, $p_tR$, using the CMW scheme as appropriate for
resummations~\cite{Catani:1990rr,Dokshitzer:1995ev}. We also freeze
the coupling at a scale $\mu_{\rm fr}$, giving
\begin{equation}\label{eq:running-alphas}
\alpha_s(k_t) =
\frac{\alpha_s}{D}
-\alpha_s^2\frac{\beta_1}{\beta_0} \frac{\log(D)}{D^2}
+\alpha_s^2\frac{K}{2\pi} \frac{1}{D^2},
\text{ with }D=1+2\alpha_s\beta_0\log\left(\frac{\max(k_t,\mu_{\rm fr})}{p_tR}\right)
\end{equation}
with
\begin{equation}
\beta_0=\frac{11C_A-2n_f}{12\pi},\quad
\beta_1=\frac{17C_A^2-5C_An_f-3C_Fn_f}{24\pi^2},\quad
K=\left(\frac{67}{18}-\frac{\pi^2}{6}\right)C_A-\frac{5}{9}n_f.
\end{equation}

To keep the notations concise, we introduce
$\lambda_x=2\alpha_s\beta_0L_x$ where $L_x$ denotes any symbol we have
introduced in (\ref{eq:deflogs}) and $L_{\rm fr}=\log(p_tR/\mu_{\rm fr})
=\log(1/\tilde\mu_{\rm fr})$. 

\subsection{Basic building blocks}\label{app:res-building-blocks}

\begin{figure}
\centerline{%
\includegraphics[width=0.43\textwidth]{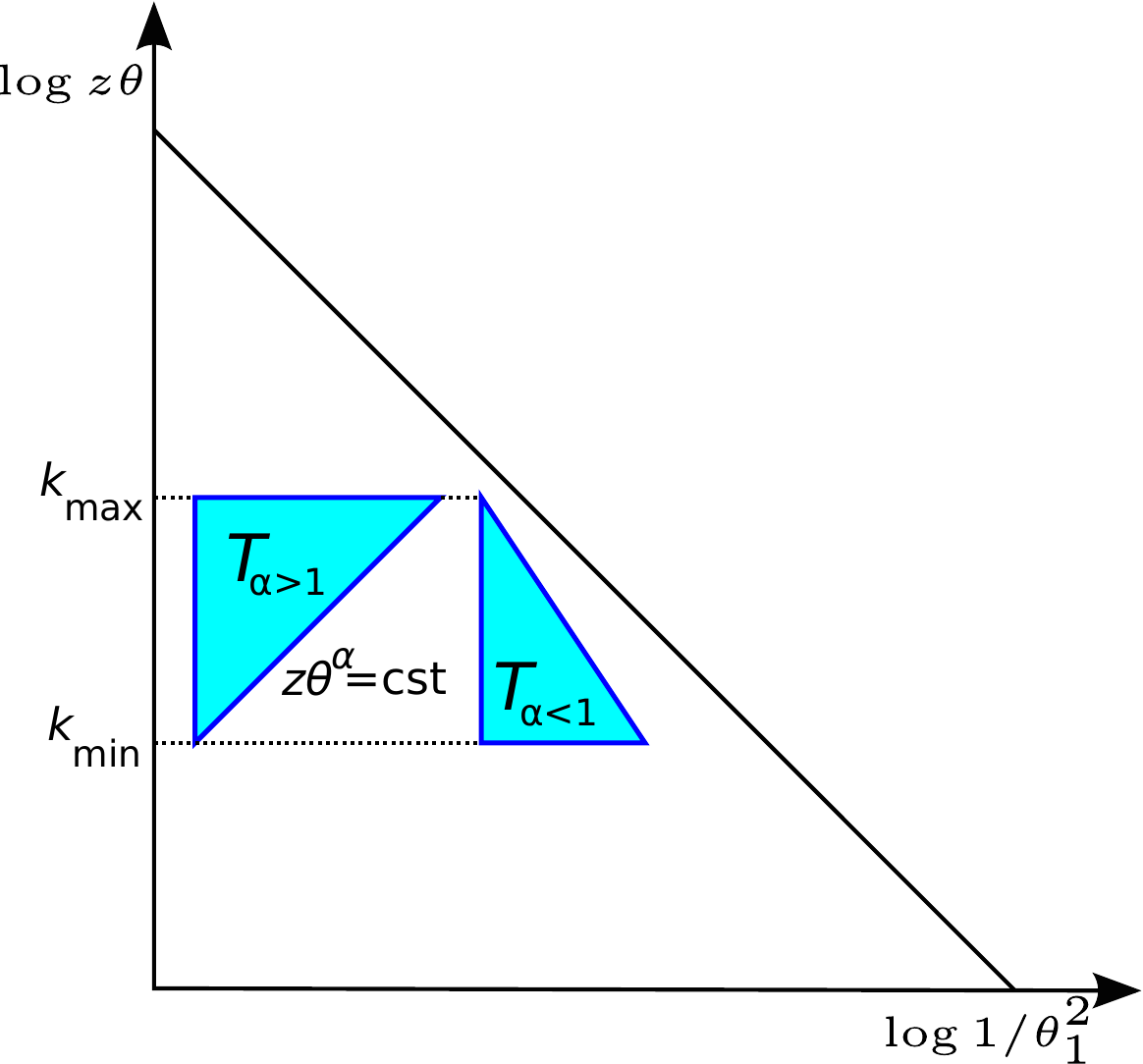}\hspace*{0.5cm}%
\includegraphics[width=0.43\textwidth]{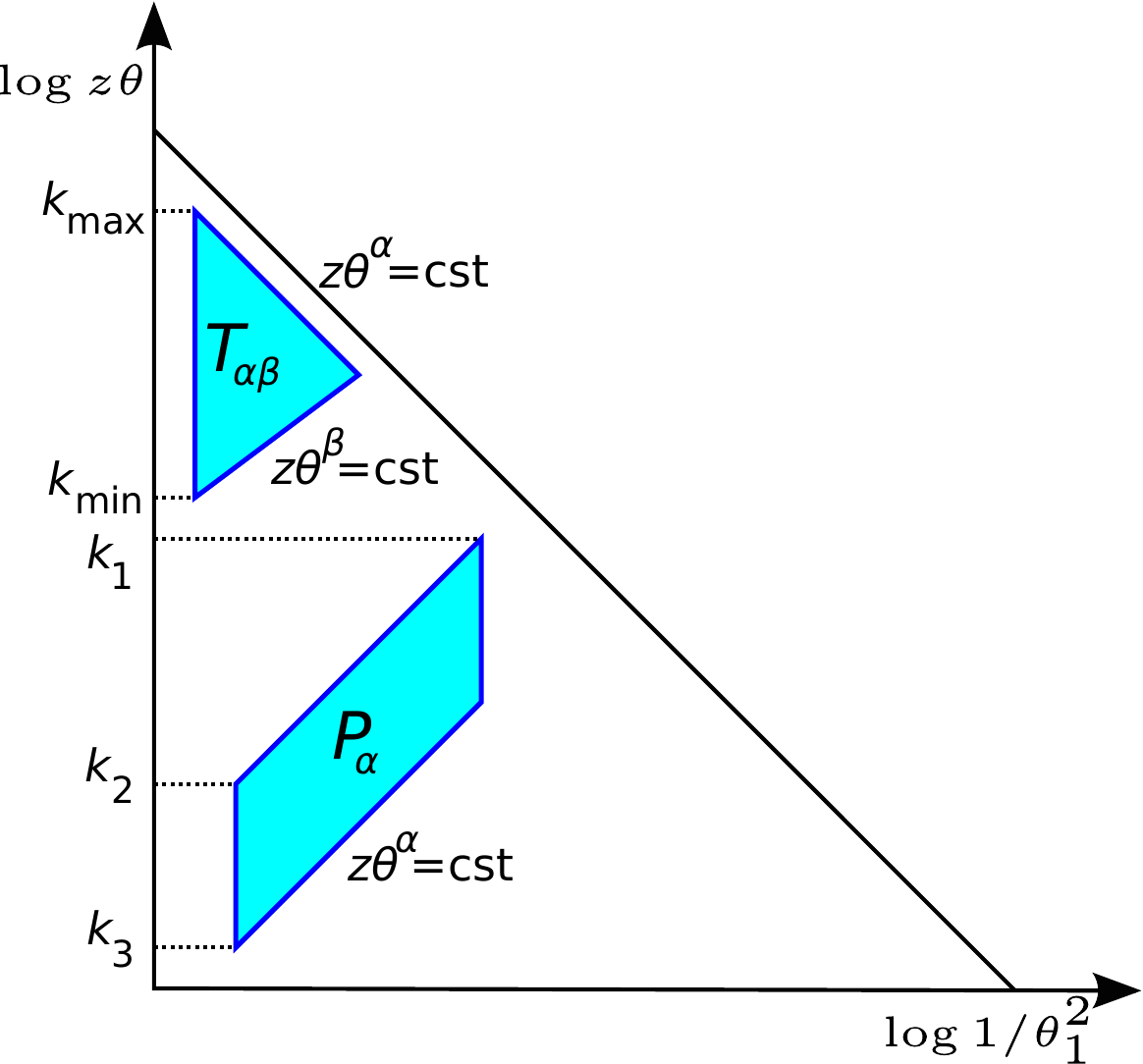}}
\caption{Left: representation of the basic building block used to
  present our results. It appears in two different forms whether we
  have $\alpha<1$ or $\alpha>1$. Right: two additional fundamental
  objects built from $T_\alpha$.}\label{fig:basic-building-blocks}
\end{figure}

It is helpful to introduce a few building blocks that will greatly help
in writing the several results below in a short and understandable way.

The most basic building block we shall use is the integral over a
``triangle'' bounded by a maximal angle, a constant
$k_t\propto z\theta$ line (upper or lower bound) and a constant
generic line of constant $z\theta^\alpha$, as represented on
Fig.~\ref{fig:basic-building-blocks}.
Expressed as a function of the minimal and maximal $k_t$ scales
of this triangle, this triangle can be written as
\begin{align}
&T_{\alpha}(k_{\max}, k_{\min}; C_R, B_i) \nonumber\\
&\quad\overset{\alpha<1}{=} \int
   \frac{d\theta^2}{\theta^2}\,dz\,P(z)\,\frac{\alpha(z\theta)}{2\pi}\,\Theta(\theta<k_{\max})\,\Theta(z\theta>k_{\min})\,\theta(z\theta^\alpha<k_{\max}^\alpha)\\
&\quad\overset{\alpha>1}{=} \int
   \frac{d\theta^2}{\theta^2}\,dz\,P(z)\,\frac{\alpha(z\theta)}{2\pi}\,\Theta(\theta<1)\,\Theta(z\theta<k_{\max})\,\theta(z\theta^\alpha>k_{\min})
\end{align}

The exact expressions for these integrals depend on the positions of
$k_{\min}$ and $k_{\max}$ compared to $\tilde\mu_{\rm fr}$. For
$k_{\min}>\tilde\mu_{\rm fr}$ we find, introducing
$L_{\min}=\log(1/k_{\min})$, $\lambda_{\min}=2\alpha_s\beta_0L_{\min}$
and similar quantities associated with $k_{\max}$,
\begin{align}
&T_{\alpha}(k_{\max}, k_{\min}; C_R, B_i)\\
&\;\;\overset{\alpha<1}{=}
  \frac{C_R}{2\pi\alpha_s\beta_0^2}\frac{1}{1-\alpha}\Big\{\Big[
  (1\!-\!\lambda_{\max}+2\alpha_s\beta_0B_i\Theta(\alpha=0))
  \log\Big(\frac{1\!-\!\lambda_{\max}}{1\!-\!\lambda_{\min}}\Big)
  +\lambda_{\max}-\lambda_{\min}\Big]\nonumber\\
&\qquad -\frac{\alpha_s\beta_1}{\beta_0}\Big[
   \frac{1}{2}\log^2(1\!-\!\lambda_{\min})
  -\frac{1}{2}\log^2(1\!-\!\lambda_{\max})
  +\frac{1\!-\!\lambda_{\max}}{1\!-\!\lambda_{\min}}\log(1\!-\!\lambda_{\min})
  -\log(1\!-\!\lambda_{\max})\nonumber\\
& \phantom{\qquad -\frac{\alpha_s\beta_1}{\beta_0}\;}
  + \frac{\lambda_{\min}-\lambda_{\max}}{1\!-\!\lambda_{\min}} \Big]
 +\frac{\alpha_sK}{2\pi}\Big[
  \log\Big(\frac{1\!-\!\lambda_{\min}}{1\!-\!\lambda_{\max}}\Big)
  +\frac{\lambda_{\min}-\lambda_{\max}}{1\!-\!\lambda_{\min}}
  \Big]\Big\}\nonumber\\
&\;\;\overset{\alpha>1}{=}
  \frac{C_R}{2\pi\alpha_s\beta_0^2}\frac{1}{\alpha-1}\Big\{\Big[
  (1\!-\!\lambda_{\min})
  \log\Big(\frac{1\!-\!\lambda_{\min}}{1\!-\!\lambda_{\max}}\Big)
  +\lambda_{\min}-\lambda_{\max}\Big] \nonumber\\
&\qquad -\frac{\alpha_s\beta_1}{\beta_0}\Big[
   \frac{1}{2}\log^2(1\!-\!\lambda_{\max})
  -\frac{1}{2}\log^2(1\!-\!\lambda_{\min})
  +\frac{1\!-\!\lambda_{\min}}{1\!-\!\lambda_{\max}}\log(1\!-\!\lambda_{\max})
  -\log(1\!-\!\lambda_{\min})\nonumber\\
& \phantom{\qquad -\frac{\alpha_s\beta_1}{\beta_0}\;}
  + \frac{\lambda_{\max}-\lambda_{\min}}{1\!-\!\lambda_{\max}} \Big]
 +\frac{\alpha_sK}{2\pi}\Big[
  \log\Big(\frac{1\!-\!\lambda_{\max}}{1\!-\!\lambda_{\min}}\Big)
  +\frac{\lambda_{\max}-\lambda_{\min}}{1\!-\!\lambda_{\max}}
  \Big]\Big\},\nonumber
\end{align}
where the $B_i$ term for $\alpha<1$  only has to be included if the
``triangle'' upper edge corresponds to $z=1$.
For $\kmin<\tilde\mu_{\rm fr}$ but $\kmax>\tilde\mu_{\rm fr}$, one
obtains
\begin{align}
&T_{\alpha}(k_{\max}, k_{\min}; C_R, B_i)\\
&\;\;\overset{\alpha<1}{=}
  \frac{C_R}{2\pi\alpha_s\beta_0^2}\frac{1}{1-\alpha}\Big\{\Big[
  (1\!-\!\lambda_{\max}+2\alpha_s\beta_0B_i\Theta(\alpha=0))
  \log\Big(\frac{1\!-\!\lambda_{\max}}{1\!-\!\lambda_{\rm fr}}\Big)
  +\lambda_{\max}-\lambda_{\rm fr}\Big]\nonumber\\
&\qquad -\frac{\alpha_s\beta_1}{\beta_0}\Big[
   \frac{1}{2}\log^2(1\!-\!\lambda_{\rm fr})
  -\frac{1}{2}\log^2(1\!-\!\lambda_{\max})
  +\frac{1\!-\!\lambda_{\max}}{1\!-\!\lambda_{\rm fr}}\log(1\!-\!\lambda_{\rm fr})
  -\log(1\!-\!\lambda_{\max})\nonumber\\
& \phantom{\qquad -\frac{\alpha_s\beta_1}{\beta_0}\;}
  + \frac{\lambda_{\rm fr}-\lambda_{\max}}{1\!-\!\lambda_{\rm fr}} \Big]
 +\frac{\alpha_sK}{2\pi}\Big[
  \log\Big(\frac{1\!-\!\lambda_{\rm fr}}{1\!-\!\lambda_{\max}}\Big)
  +\frac{\lambda_{\rm fr}-\lambda_{\max}}{1\!-\!\lambda_{\rm fr}}
  \Big]\Big\}\nonumber\\
&\qquad +\frac{C_R}{\pi(1-\alpha)}(L_{\min}-L_{\rm fr})
  \big[\alpha_s(\tilde\mu_{\rm fr})(L_{\min}+L_{\rm fr}-2L_{\max})+2\alpha_{s,\text{1-loop}}(\tilde\mu_{\rm fr})B_i\Theta(\alpha=0)\big]\nonumber\\
&\;\;\overset{\alpha>1}{=}
  \frac{C_R}{2\pi\alpha_s\beta_0^2}\frac{1}{\alpha-1}\Big\{\Big[
  (1\!-\!\lambda_{\min})\log\Big(\frac{1\!-\!\lambda_{\rm fr}}{1\!-\!\lambda_{\max}}\Big)
  +\lambda_{\rm fr}-\lambda_{\max}\Big]\nonumber\\
&\qquad -\frac{\alpha_s\beta_1}{\beta_0}\Big[
   \frac{1}{2}\log^2(1\!-\!\lambda_{\max})
  -\frac{1}{2}\log^2(1\!-\!\lambda_{\rm fr})
  +\frac{1\!-\!\lambda_{\min}}{1\!-\!\lambda_{\max}}\log(1\!-\!\lambda_{\max})
  -\frac{1\!-\!\lambda_{\min}}{1\!-\!\lambda_{\rm fr}}\log(1\!-\!\lambda_{\rm fr})\nonumber\\
& \phantom{\qquad -\frac{\alpha_s\beta_1}{\beta_0}\;}
  + \frac{(\lambda_{\max}\!-\!\lambda_{\rm fr})(1\!-\!\lambda_{\min})}{(1\!-\!\lambda_{\max})(1\!-\!\lambda_{\rm fr})} \Big]
 +\frac{\alpha_sK}{2\pi}\Big[
  \log\Big(\frac{1\!-\!\lambda_{\max}}{1\!-\!\lambda_{\rm fr}}\Big)
  +\frac{(\lambda_{\max}\!-\!\lambda_{\rm fr})(1\!-\!\lambda_{\min})}{(1\!-\!\lambda_{\max})(1\!-\!\lambda_{\rm fr})}
  \Big]\Big\},\nonumber\\
&\qquad +\frac{\alpha_s(\tilde\mu_{\rm fr})C_R}{\pi(\alpha-1)}(L_{\min}-L_{\rm fr})^2. \nonumber
\end{align}
In that expression, we have introduced
$\alpha_{s,\text{1-loop}}(k_t)=\alpha_s/(1-2\alpha_s\beta_0\log(p_tR/k_t))$,
the running-coupling at 1-loop, which multiplies the contributions
proportional to $B_i$ in the frozen region. This reflects the fact
that contributions proportional to $\beta_1B_i$ and $K B_i$, coming
from the 2-loop corrections to the running of $\alpha_s$ are
subleading. They are not included, neither in the frozen region, nor
in the running-coupling region. 

And, finally, for $\kmax<\tilde\mu_{\rm fr}$, one gets
\begin{align}
&T_{\alpha}(k_{\max}, k_{\min}; C_R, B_i)\\
&\quad = \frac{C_R}{\pi|1-\alpha|}(L_{\min}-L_{\max})
  \big[\alpha_s(\tilde\mu_{\rm fr})(L_{\min}-L_{\max})+2\alpha_{s,\text{1-loop}}(\tilde\mu_{\rm fr})B_i\Theta(\alpha=0)\big].\nonumber
\end{align}

From this fundamental building block, we can build two derived objects
which will be used to describe all the expressions we have below. The
first one is again a triangle bound by a maximal angle, a maximal
$z\theta^\alpha$ line and a minimal $z\theta^\beta$ line, see the
right plot of Fig.~\ref{fig:basic-building-blocks}. This can be seen
as a superposition of two of the above triangles.
Again, we can express this new object as a function of the minimal and
maximal $k_t$ scales on the maximal-angle side of the triangle, and,
assuming $\alpha<\beta$, we get
\begin{align}
T_{\alpha\beta}(\kmax,\kmin;C_R,B_i)
& \overset{\alpha<\beta<1}{=}
   T_\alpha(\kmax,k_{\rm med};C_R,B_i)-T_\beta(\kmin,k_{\rm med};C_R,B_i)\\
& \overset{\alpha<1<\beta}{=}
   T_\alpha(\kmax,k_{\rm med};C_R,B_i)+T_\beta(k_{\rm med},\kmin;C_R,B_i)\\
& \overset{1<\alpha<\beta}{=}
   T_\beta(k_{\rm med},\kmin;C_R,B_i)-T_\alpha(k_{\rm med},\kmax;C_R,B_i),
\end{align}
with $k_{\rm
  med}=\kmax^{\frac{\beta-1}{\beta-\alpha}}\kmin^{\frac{1-\alpha}{\beta-\alpha}}$.\footnote{$T_{0\beta}(\kmax=1,\kmin)$
  is related to the radiator given in Appendix~A.1 of Ref.~\cite{Banfi:2004yd}.}

The last object we shall use is a ``parallelogram'' bounded by a
minimal and a maximal angle and two parallel lines of constant
$z\theta^\alpha$, assuming here $\alpha>1$, see again the right plot
of Fig.~\ref{fig:basic-building-blocks}. This is expressed as a
function of the maximal $k_t$ scale $k_1$ (at the minimal angle) and
the maximal and minimal $k_t$ scales, $k_2$ and $k_3$ at the maximal
angle.
We can view this as a function of three of our basic triangles
\begin{equation}
P_\alpha(k_1,k_2,k_3;C_R) =
T_\alpha(k_1,k_3;C_R,0)-T_\alpha(k_1,k_2;C_R,0)-T_\alpha(k_1,k_4;C_R,0)
\end{equation}
with $k_4 = k_1k_3/k_2$.

Note that we will often substitute the $k_t$ scale with their
logarithm, $\log(1/k_t)$ and it is worth keeping in mind that the
maximal $k_t$ would correspond to the minimal $\log(1/k_t)$.

\subsection{Results for the QCD background}\label{app:bkg-rc}

Now that we have building blocks corresponding to the
integration of Sudakov factors over basic phase-space regions, we can
use them to find simple expressions for the Sudakov factors
corresponding to the shapes we are studying.

The phase-space regions will correspond exactly to the regions
we have already used for the fixed-coupling calculation given in the
main text, so we just list the results here.
 
\paragraph{\boldmath $N$-subjettiness.} This is the most simple result
because the phase-space just corresponds to a triangle for the primary
emissions and another one for the secondary emissions:
\begin{align}
R_{\tau}(z_1)
& = T_{02}(0,L_\rho+L_v;C_R,B_i)-T_{02}(0,L_\rho;C_R,B_i)\nonumber\\
& + T_{02}\Big(\frac{L_\rho+L_1}{2},\frac{L_\rho+L_1}{2}+L_v;C_A,B_g\Big),
\label{eq:bkg-tau-rc}
\end{align}
where the negative term subtracts the Sudakov factor for the plain jet
mass which has been factored out in our expressions.

\paragraph{Mass-drop (non-recursive).} Here we split the result in a
part, $R_0$ clustered with the main parton and a part, $R_1$,
clustered with the emission setting the mass.
\begin{align}
R_{\mu_{1/2}^2,0}(z_1)
& = T_{02}\Big(\frac{L_\rho-L_1}{2},\frac{L_\rho+L_1}{2}+L_v;C_R,B_i\Big)
  - T_{02}\Big(\frac{L_\rho-L_1}{2},\frac{L_\rho+L_1}{2};C_R,B_i\Big)\nonumber\\
& + \frac{1}{2} P_2\Big(\frac{L_\rho+L_1}{2}, L_\rho, L_\rho+L_v; C_R, B_i\Big)\nonumber\\
R_{\mu_{1/2}^2,1}(z_1)
& = \Big[ \frac{1}{2} P_2\Big(\frac{L_\rho+L_1}{2}, L_\rho, L_\rho-L_1+L_v; C_R, B_i\Big)\nonumber\\
& \phantom{=\Big[} + T_{02}\Big(\frac{L_\rho+L_1}{2},\frac{L_\rho-L_1}{2}+L_v;C_A,B_g\Big)
  \Big]\,\Theta(L_v>L_1)
\label{eq:bkg-mu2-rc}
\end{align}
The total Sudakov $R_{\mu_{1/2}^2}$ is the sum of these two contributions.

\paragraph{Energy correlation function.} For $C_2$ we have to
disentangle two cases depending on whether we have a contribution from
emissions at small angles of not:
\begin{align}
R_{C_2}(z_1)
  & \overset{L_v<L_\rho-L_1}{=}
    T_{24}(L_\rho,L_\rho+L_v;C_R,B_i)\nonumber\\
  & \overset{L_v>L_\rho-L_1}{=}
    T_{02}(0,L_\rho-L_1+L_v;C_R,B_i)-T_{02}(0,L_\rho;C_R,B_i)\label{eq:bkg-ecf-rc}\\
  & \phantom{\overset{L_v>L_\rho-L}{=}}
   + T_{24}(L_1+L_v,L_\rho+L_v;C_R,B_i)
   + T_{02}\Big(\frac{L_\rho+L_1}{2},\frac{3L_1-L_\rho}{2}+L_v;C_A,B_g\Big)\nonumber
\end{align}
This expression can be trivially expressed as a result for $D_2$
replacing $L_v$ by $L_v-L_\rho$.

\paragraph{\boldmath Recursive $N$-subjettiness.} Here, the phase-space
constraints can take three different forms. Remember also that we do
subtract the Sudakov factor corresponding to the plain jet mass.
\begin{align}
R_{\tau,\rm rec}(z_1)
& \overset{L_v<L_1}{\quad\;\;=\quad\;\;}
   T_{02}\Big(\frac{L_\rho-L_1}{2},\frac{L_\rho+L_1}{2}+L_v;C_R,B_i\Big)
 - T_{02}\Big(\frac{L_\rho-L_1}{2},\frac{L_\rho+L_1}{2};C_R,B_i\Big)\nonumber\\
& \phantom{\overset{L_v>L}{\quad\;\;=\quad\;\;}}
 - P_2\Big(\frac{L_\rho-L_1}{2}-L_v,L_\rho-L_v,L_\rho;C_R,B_i\Big)\nonumber\\
& \phantom{\overset{L_v>L}{\quad\;\;=\quad\;\;}}
 + T_{02}\Big(\frac{L_\rho+L_1}{2},\frac{L_\rho+L_1}{2}+L_v;C_A,B_g\Big)\nonumber\\
& \overset{L_1<L_v<L_\rho}{\quad\;\;=\quad\;\;}
   T_{02}\Big(\frac{L_\rho-L_1}{2},\frac{L_\rho+L_1}{2}+L_v;C_R,B_i\Big)
 + T_{02}(0,L_v-L_\rho;C_R,B_i)\nonumber\\
& \phantom{\overset{L_v>L}{\quad\;\;=\quad\;\;}}
 - T_{02}(0,L_\rho;C_R,B_i)
 + T_{02}\Big(\frac{L_\rho+L_1}{2},\frac{L_\rho+L_1}{2}+L_v;C_A,B_g\Big)\nonumber\\
& \overset{L_v>L_\rho}{\quad\;\;=\quad\;\;}
   T_{02}\Big(\frac{L_\rho-L_1}{2},\frac{L_\rho+L_1}{2}+L_v;C_R,B_i\Big)
 - T_{02}(0,L_\rho;C_R,B_i)\nonumber\\
& \phantom{\overset{L_v>L}{\quad\;\;=\quad\;\;}}
 + T_{02}\Big(\frac{L_\rho+L_1}{2},\frac{L_\rho+L_1}{2}+L_v;C_A,B_g\Big)
\label{eq:bkg-taurec-rc}
\end{align}

\paragraph{Recursive Mass-drop.} The expression is the same as for the
recursive $N$-subjettiness cut,Eq.~(\ref{eq:bkg-taurec-rc}), except
that the second argument of the $C_A$ term should be
$\frac{L_\rho-L_1}{2}+L_v$ instead of $\frac{L_\rho+L_1}{2}+L_v$ and
that term comes with a $\Theta(L_v>L_1)$.

\paragraph{Recursive energy correlation function.} Again, we have
three different situations
\begin{align}
R_{C_2,\rm rec}(z_1)
& \overset{L_v<L_\rho-L_1}{\quad\;\;=\quad\;\;}
 - T_{02}(L_\rho-L_v,L_\rho;C_R,B_i\Big)\nonumber\\
& \overset{L_v<L_\rho}{\quad\;\;=\quad\;\;}
   T_{02}\Big(\frac{L_\rho-L_1}{2},\frac{3L_1-L_\rho}{2}+L_v;C_R,B_i\Big)
 - T_{02}\Big(\frac{L_\rho-L_1}{2},\frac{L_\rho+L_1}{2};C_R,B_i\Big)\nonumber\\
& \phantom{\overset{L_v>L}{\quad\;\;=\quad\;\;}}
 - T_{02}(L_\rho-L_v,L_\rho;C_R,B_i)
 + T_{02}\Big(\frac{3L_\rho-L_1}{2}-L_v,\frac{L_\rho+L_1}{2};C_R,B_i\Big)\nonumber\\
& \phantom{\overset{L_v>L}{\quad\;\;=\quad\;\;}}
 + T_{02}\Big(\frac{L_\rho+L_1}{2},\frac{3L_1-L_\rho}{2}+L_v;C_A,B_g\Big)\nonumber\\
& \overset{L_v>L_\rho}{\quad\;\;=\quad\;\;}
   T_{02}\Big(\frac{L_\rho-L_1}{2},\frac{3L_1-L_\rho}{2}+L_v;C_R,B_i\Big)
 - T_{02}(0,L_\rho;C_R,B_i)\nonumber\\
& \phantom{\overset{L_v>L}{\quad\;\;=\quad\;\;}}
 + T_{02}\Big(\frac{L_\rho+L_1}{2},\frac{3L_1-L_\rho}{2}+L_v;C_A,B_g\Big)
\label{eq:bkg-ecfrec-rc}
\end{align}
This expression can be trivially expressed as a result for $D_2$
replacing $L_v$ by $L_v-L_\rho$.

\subsection{Results for the signal}\label{app:sig-rc}

As previously, it is fairly straightforward to use the ``triangular''
building blocks to express our findings.
Note also that, compared to the results presented for fixed-coupling
in the main text, we have not expanded our results to first order in
$z_1$ and $1-z_1$. This would only lead to more complicated
expressions without changing the formal accuracy of our results.
Remember also that for the case of signal jets and at NLL (and
small-$R$) accuracy, the results are the same for the recursive and
non-recursive versions of the shapes.

\paragraph{\boldmath $N$-subjettiness (recursive or non-recursive).} From the
expression in eq.~(\ref{eq:sig-tau}) it is easy to find
\begin{align}\label{eq:sig-tau-rc}
R_{\tau}(z_1)
 &= T_{02}\Big(\frac{L_\rho+L_--L_1}{2},\frac{L_\rho+L_-+L_1}{2}+L_v;C_R,B_i\Big)\\
 &+ T_{02}\Big(\frac{L_\rho+L_1-L_-}{2},\frac{L_\rho+L_1+L_-}{2}+L_v;C_R,B_i\Big)\nonumber
\end{align}

\paragraph{Mass-drop (recursive or non-recursive).} As for the
fixed-coupling case, the only difference between $N$-subjettiness and
a $\mu_{1/2}^2$ cut lies in the $z_1$ and $1-z_1$ corrections. We find
\begin{align}\label{eq:sig-mu2-rc}
R_{\mu_{1/2}^2}(z_1)
 &= T_{02}\Big(\frac{L_\rho+L_--L_1}{2},\frac{L_\rho-L_-+L_1}{2}+L_v;C_R,B_i\Big)\,\Theta(L_v>L_--L_1)\\
 &+ T_{02}\Big(\frac{L_\rho+L_1-L_-}{2},\frac{L_\rho-L_1+L_-}{2}+L_v;C_R,B_i\Big)\,\Theta(L_v>L_1-L_-)\nonumber
\end{align}

\paragraph{Energy correlation function (recursive or non-recursive).}
Again, the expression for $C_2$ looks very similar, except for the
logarithms involving $z_1$. We find
\begin{align}\label{eq:sig-ecf-rc}
R_{C_2}(z_1)
 &= T_{02}\Big(\frac{L_\rho+L_--L_1}{2},\frac{3L_-+3L_1-L_\rho}{2}+L_v;C_R,B_i\Big)\,\Theta(L_v>L_\rho-L_--2L_1)\\
 &+ T_{02}\Big(\frac{L_\rho+L_1-L_-}{2},\frac{3L_1+3L_--L_\rho}{2}+L_v;C_R,B_i\Big)\,\Theta(L_v>L_\rho-L_1-2L_-)\nonumber
\end{align}
This expression can be trivially expressed as a result for $D_2$ 
replacing $L_v$ by $L_v-L_\rho$.

\subsection[Including finite $z_1$ corrections: QCD (background) and
signal jets]{\boldmath Including finite $z_1$ corrections: QCD
  (background) and signal jets}\label{app:finitez1}

We have argued in Section~\ref{sec:bkg-towards-nll} that if we wish to
achieve NLL accuracy it is mandatory to include all finite $z_1$ and
$1-z_1$ factors in our expressions for the shapes, with $z_1$ the
fraction of the jet transverse momentum carried by the emission that
dominates the mass of the jet. The main reason behind that is that 
they can be raised to powers of order $\alpha_s \log(1/v)$ which would
give single-logarithmic corrections after integration over $z_1$.

\begin{figure}
  \centering
  \includegraphics[width=0.32\textwidth]{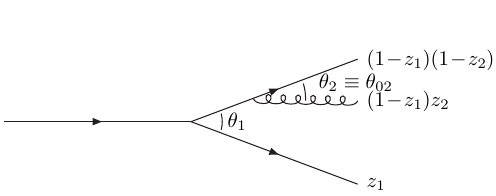}%
  \hfill%
  \includegraphics[width=0.32\textwidth]{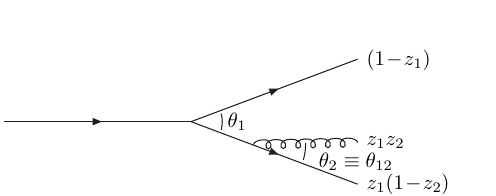}%
  \hfill%
  \includegraphics[width=0.32\textwidth]{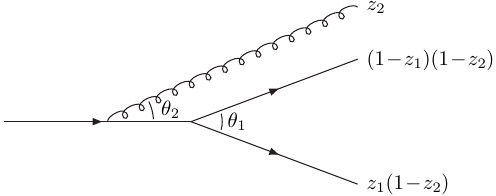}%
  \caption{Three topologies potentially contributing to the emission
    of the gluon dominating the value of the shape, starting with a
    massive two-pronged object. Left: small-angle emission from the
    prong carrying a fraction $1-z_1$ of the jet $p_t$ (``prong 1''),
    centre: small-angle emission from the prong carrying a fraction
    $z_1$ of the jet $p_t$ (``prong 2''), right: large-angle emission
    from the parent object (``parent'').}\label{figs:emissions}
\end{figure}

In this Section, our main goal is to discuss these extra source of
NLL terms. As a fringe benefit of this discussion, we will at the same
time provide a unified description of the signal and background
distributions, allowing for interesting interpretations of the results
obtained in this paper.

If we want to properly include the finite $z_1$ corrections we first
need to carefully identify the origin of the gluon emissions. In the
collinear limit, sufficient to capture all the finite $z_1$
corrections, colour coherence indicates that we can encounter three
situations, represented in Fig.~\ref{figs:emissions}. The first two
situations correspond to gluon emissions at small angle
$\theta_2\ll\theta_1$ from the splitting of either the hardest or the
softest of the two prongs (carrying respectively a fraction $1-z_1$
and $z_1$ of the jet transverse momentum). These are the first two
plots of Fig.~\ref{figs:emissions} and will be referred to as the
``prong 1'' and ``prong 2'' topologies respectively for the $1-z_1$
and $z_1$ case. The third option corresponds to gluons emitted at
large angle $\theta_2\gg\theta_1$ from the parent parton in the jet.
This is represented on the rightmost plot of Fig.~\ref{figs:emissions}
and will be called the ``parent'' topology in what follows.
In that approach, the distribution for QCD jets will receive
contributions from all three topologies --- the first and third
weighted by $C_R$ and the second, corresponding to secondary
emissions, weighted by $C_A$ --- while signal jets coming from the
decay of colour-neutral bosons would only receive contributions from
the first two topologies, both weighted by $C_R$.

For each of the three topologies, one then has to find the expression
for the shape in the soft and collinear limit for the gluon
emission,\footnote{Meaning in particular that one can discard the
  $1-z_2$ factors.}  and impose that the first emission
$(z_1,\theta_1)$ dominates the mass. 
The Sudakov factors for a given mass $\rho$, splitting momentum
fraction $z_1$ and shape cut $v$, would then take the following form
for each topology:
\begin{align}
R_{\rm prong1}
  & = \int_0^{\theta_1^2} \frac{d\theta_2^2}{\theta_2^2}
      \int_0^1 dz_2 P_{\rm prong1}(z_2)\,
      \frac{\alpha_s}{2\pi}\,
      \Theta(v_{\rm prong1}(z_1,\rho;z_2,\theta_2)>v)
      \Theta((1-z_1)^2z_2\theta_2^2<\rho),\nonumber\\
R_{\rm prong2}
  & = \int_0^{\theta_1^2} \frac{d\theta_{12}^2}{\theta_{12}^2}
      \int_0^1 dz_2 P_{\rm prong2}(z_2)\,
      \frac{\alpha_s}{2\pi}\,
      \Theta(v_{\rm prong2}(z_1,\rho;z_2,\theta_{12})>v)
      \Theta(z_1^2z_2\theta_{12}^2<\rho),\nonumber\\
R_{\rm parent}
  & = \int_{\theta_1^2}^1 \frac{d\theta_2^2}{\theta_2^2}
      \int_0^1 dz_2 P_{\rm parent}(z_2)\,
      \frac{\alpha_s}{2\pi}\,
      \Theta(v_{\rm parent}(z_1,\rho;z_2,\theta_2)>v)
      \Theta(z_2\theta_2^2<\rho),
\end{align}
where the splitting function would be the one of a quark, a gluon, or
simply 0 for emissions from a colour-neutral object, and
$\rho=z_1(1-z_1)\theta_1^2$.

In practice, the two ``prong'' contributions are the same as the ones
we have computed in the case of signal jets, up to the constraint that
the $(z_1,\theta_1)$ emission dominates the mass. This last term is
irrelevant for signal jets as it would only contribute to a
constant. For QCD jets it is however crucial to impose it for the
emissions from the hard prong since, there, the $z_1\ll 1$ region can
give rise to large logarithms.

Strictly speaking, the finite $z_1$ corrections should only be kept in
the expression for the shapes and the mass constraint in the emission
from the soft prong is subleading for both the signal and the
background. However, keeping these contributions makes the expressions
more symmetric.

To fully specify our results, we just have to find the expressions
of the three shapes we consider in each of the three topologies
above. Following the same considerations as in the main text, it is
easy to obtain
\begin{align}
\tau_{21,\rm prong1} & = \frac{z_2}{z_1}\frac{\theta_2^2}{\theta_1^2} &
\tau_{21,\rm prong2} & = \frac{z_2}{1-z_1}\frac{\theta_{12}^2}{\theta_1^2}&
\tau_{21,\rm parent} & = \frac{z_2\theta_2^2}{\rho}\\
\mu^2_{\rm prong1} & = \frac{(1-z_1)z_2}{z_1}\frac{\theta_2^2}{\theta_1^2} &
\mu^2_{\rm prong2} & = \frac{z_1z_2}{1-z_1}\frac{\theta_{12}^2}{\theta_1^2}&
\mu^2_{\rm parent} & \overset{\theta_{2}>\theta_{12}}{=} \frac{z_1z_2\theta_2^2}{\rho}\\
&&
&&
                 & \overset{\theta_{12}>\theta_{2}}{=} \frac{(1-z_1)z_2\theta_2^2}{\rho}
\nonumber\\
C_{2,\rm prong1} & = \frac{z_2}{z_1}\theta_2^2 &
C_{2,\rm prong2} & = \frac{z_2}{1-z_1}\theta_{12}^2&
C_{2,\rm parent} & = \frac{z_2\theta_2^4}{\rho},
\end{align}
For parent emissions, we again had to separate two cases for
the mass-drop parameter corresponding to the clustering of the second
emission with one of the two prongs, with $\theta_{2}$ being the
angle wrt ``prong 1'' and $\theta_{12}$ the angle to ``prong 2''.

With these expressions and the building blocks introduced in
Appendix~\ref{app:bkg-rc}, we can compute the Sudakov form factors. It
is convenient to introduce $C_{R,1}$, $C_{R,2}$ and $C_{R,p}$
respectively as the colour factors associated with the ``prong 1'',
``prong 2'' and ``parent'' topologies. Similarly, we denote $B_{1}$,
$B_{2}$ the hard-splitting coefficient associated with the two
``prong'' configurations, realising that the large-angle
topology will not receive a hard-splitting correction. Note that in
the case of a boson decay, we can simply set $C_{R,p}=0$.

The results for the emissions collinear to the $1-z_1$ branch (``prong
1'') are as follows:
\begin{align}
R_{\tau,\rm prong1}(z_1)
 = &\Big[T_{02}\Big(\frac{L_\rho-L_1+L_-}{2},\frac{L_\rho+L_1+L_-}{2}+L_v;C_{R,1},B_1\Big)\\
  &- T_{02}\Big(\frac{L_\rho-L_1+L_-}{2},\frac{L_\rho+L_1-L_-}{2};C_{R,1},B_1\Big)\Theta(L_1>L_-)\Big] \nonumber\\
  & \Theta(L_v+L_1>0)\,\Theta(L_v+L_->0)\nonumber\\
R_{\mu^2,\rm prong1}(z_1)
 = & \Big[T_{02}\Big(\frac{L_\rho-L_1+L_-}{2},\frac{L_\rho+L_1-L_-}{2}+L_v;C_{R,1},B_1\Big)\\
  &- T_{02}\Big(\frac{L_\rho-L_1+L_-}{2},\frac{L_\rho+L_1-L_-}{2};C_{R,1},B_1\Big)\Theta(L_1>L_-)\Big] \nonumber\\
  & \Theta(L_v>L_--L_1)\,\Theta(L_v>0)\nonumber\\
R_{C_2,\rm prong1}(z_1)
 =& \Big[T_{02}\Big(\frac{L_\rho-L_1+L_-}{2},\frac{3L_-+3L_1-L_\rho}{2}+L_v;C_R,B_i\Big)\\
  &- T_{02}\Big(\frac{L_\rho-L_1+L_-}{2},\frac{L_\rho+L_1-L_-}{2};C_{R,1},B_1\Big)\Theta(L_1>L_-)\Big] \nonumber\\
  & \Theta(L_v>L_\rho-L_--2L_1)\,\Theta(L_v>L_\rho-L_1-2L_-),\nonumber
\end{align}
where the last two $\Theta$ constraints come from the fact that the
first term has to be positive and larger than the second term.
Note that the second term in each of these three expressions is the
same and come from the kinematic constraint than the second emission $(z_2,\theta_2)$
does not dominate the mass.

The results for the ``prong 2'' topology have not been given
explicitly but can be directly obtained from the ``prong 1'' topology
by inverting $L_1$ and $L_-$ which corresponds to inverting $z_1$ and
$1-z_1$.

For the emissions from the parent object, we find in a similar way
\begin{align}
R_{\tau,\rm parent}(z_1)
 = & P_2\Big(\frac{L_\rho+L_1+L_-}{2},L_\rho,L_\rho+L_v;C_{R,p},0\Big)\,\Theta(L_v>0)\\
R_{\mu^2,\rm parent}(z_1)
 = & \frac{1}{2} P_2\Big(\frac{L_\rho+L_1+L_-}{2},L_\rho,L_\rho-L_-+L_v;C_{R,p},0\Big)\,\Theta(L_v>L_-)\nonumber\\
 + & \frac{1}{2} P_2\Big(\frac{L_\rho+L_1+L_-}{2},L_\rho,L_\rho-L_1+L_v;C_{R,p},0\Big)\,\Theta(L_v>L_1)\\
R_{C_2,\rm parent}(z_1)
 = & \Big[P_2\Big(\frac{L_\rho+L_1+L_-}{2},L_\rho,L_1+L_-+L_v;C_{R,p},0\Big)\nonumber\\
   & \phantom{\Big[}+T_{24}(L_1+L_-+L_v,L_\rho+L_v;C_{R,p},0)\Big]\,\Theta(L_v>L_\rho-L_1-L_-)\nonumber\\
   & +T_{24}(L_\rho,L_\rho+L_v;C_{R,p},0)\,\Theta(0<L_v<L_\rho-L_1-L_-)
\end{align}

\section{Details for the computation the shape value}\label{app:technical-details}

In this Appendix we give all the technical details related to the
calculation of the leading-logarithmic expressions for each of the
shapes we consider. 

\subsection[$N$-subjettiness calculation and axes
  choice]{\boldmath $N$-subjettiness calculation and axes
  choice}\label{app:tau21-details}

We need to justify the result in
Eq.~(\ref{eq:tau21-value-primary}). For $N$-subjettiness with
$\beta=2$, we do not have to worry about recoil effects and we can
focus on $E$-scheme recombinations, which uses 4-momentum sum of the
particles.

We consider a hard parton($p_0$) accompanied by two emissions, $p_1$
and $p_2$, of transverse momentum fraction $z_1$ and $z_2$
respectively emitted at angles $\theta_1$ and $\theta_2$. We work in
the strongly-ordered limit where we can assume that the mass (and
$\tau_1$) are dominated by the first emission:
$\rho=\tau_1\approx z_1\theta_1^2$, neglecting a subleading
$(1-z_1)$ power correction, with the axis defining $\tau_1$ aligned
with the jet axis.

For $\tau_2$, three different situations are possible:
\begin{itemize}
\item one axis coincides with $p_0$, the other with $p_1+p_2$, 
  giving $\tau_2^{(0,12)}=z_1z_2/(z_1+z_2)\theta_{12}^2$,
\item one axis coincides with $p_1$, the other with $p_0+p_2$, 
  giving $\tau_2^{(1,02)}=z_2\theta_2^2$,
\item one axis coincides with $p_2$, the other with $p_0+p_1$, 
  giving $\tau_2^{(2,01)}=z_1\theta_1^2$,
\end{itemize}
where we have again neglected subleading large-$z_i$ contributions,
and $\theta_{12}$ is the angle between the first and second emissions.

Since the emission $p_1$ dominates the mass, we have
$\tau_2^{(2,01)}\gg \tau_2^{(1,02)}$. The ordering between
$\tau_2^{(0,12)}$ and $\tau_2^{(1,02)}$ is less clear.
When $\theta_2\gg\theta_1$, $z_2\theta_2^2\ll z_1\theta_1^2$ imposes
$z_2\ll z_1$; we can then approximate $\theta_{12}\approx\theta_2$ and
get $\tau_2^{(0,12)}\approx z_2\theta_2^2$, \ie both choices
$\tau_2^{(0,12)}$ and $\tau_2^{(1,02)}$ are equivalent.
In the opposite case, when $\theta_2\ll\theta_1$,
$\theta_{12}\approx\theta_1$ and $\tau_2^{(0,12)}\approx
z_1z_2/(z_1+z_2)\theta_1^2$. For $z_1\ll z_2$, we get
$\tau_2^{(0,12)}\approx z_1\theta_1^2\gg z_2\theta_2^2$, while for
$z_1\gg z_2$, we get $\tau_2^{(0,12)}\approx z_2\theta_1^2\gg
z_2\theta_2^2$.

Note that if we target single logarithmic accuracy, we should also
worry about the situation where $\theta_2\approx\theta_1$. In that
case, $z_2\ll z_1$ and $\tau_2^{(0,12)}\approx z_2\theta_{12}^2$. This
would give at most a constant-factor correction to $\tau_{21}$ and
hence only contribute at a NNLL compared to the approximation
$\tau_2\approx z_2\theta_2^2$.

Which of the three options is used depends on the specific choice of
axes we use to define $\tau_{21}$:
\begin{itemize}
\item the {\it optimal axes} should minimise $\tau_2$ and hence give
  $\tau_2=z_2\theta_2^2$.
\item for the {\it $k_t$ axes}, we should therefore find the minimum
  of $d^{(k_t)}_{01}=z_1 \theta_1$, $d^{(k_t)}_{02}=z_2 \theta_2$, and
  $d^{(k_t)}_{12}={\rm min}(z_1,z_2) \theta_{12}$. In that case, we
  also will find $\tau_2\approx z_2\theta_2^2$ {\it except} in a
  region $z_2\theta_2^2\ll z_1\theta_1^2$,
  $z_2\theta_2\gg z_1\theta_1$, \ie the region where the emission
  $p_2$ has smaller mass but larger $k_t$ than the emission $p_1$, and
  where we get $\tau_2\approx z_1\theta_1^2$.
\item for the {\it gen-$k_t(1/2)$ axes}, we should find the pair that
  minimises the distance
  $d^{(1/2)}_{ij}={\rm min}(z_i,z_j)\theta_{ij}^2$. In this case, the
  minimum will always be $d_{02}$ or $d_{12}$ and yield
  $\tau_2=z_2\theta_2^2$.
\end{itemize}

In the end, the case of {\it $k_t$ axes} is clearly more complex.
In what follows we shall therefore focus on the two other axes
choices.
Based on considerations similar to the ones above, one can show that
the gen-$k_t(1/2)$ axes will agree with the minimal axes up to NNLL
corrections (mostly occuring when two angles become comparable or when
there is a hard splitting). In practice, computing the optimal axes
can be an expensive step and we can view the gen-$k_t(1/2)$ option as
a simpler alternative reproducing essentially the same
performance.

Concentrating on optimal axes or gen-$k_t(1/2)$ axes, we recover
\eqref{eq:tau21-value-primary}.

\subsection{Details of the mass-drop calculation}\label{app:mu2-details}

We now move to the mass-drop parameter and the result quoted in
Eq.~(\ref{eq:mu2value}).

Again, we consider a the leading parton $p_0$ and two emissions
$p_1(\theta_1,z_1)$ and $p_2(\theta_2,z_2)$ with
$z_1\theta_1^2\gg z_2\theta_2^2$. In order to find the two subjets, we
need to find the minimal distance amongst the gen-$k_t$($1/2$)
distances $d_{01}$, $d_{02}$ and $d_{12}$ which gives the two subjets
and $\mu_{1/2}^2$ will be given by the mass of the two particles which
have been clustered divided by the total mass of the jet.
The smallest distance is either $d_{02}=z_2\theta_2^2$ or
$d_{12}={\rm min}(z_1,z_2)\theta_{12}^2$.
For $\theta_2\ll\theta_1$, $\theta_{12}\approx \theta_1$ and
$d_{12}\ge z_2\theta_1^2\gg z_2\theta_2^2$, so that the hard subjet
mass is $z_2\theta_2^2$.
The opposite case, $\theta_2\gg\theta_1$ (implying $z_2\ll z_1$), is
more subtle: one has to compare the pairwise clustering distances
$d_{02}=z_2\theta_2^2$ with $d_{12}=z_2\theta_{12}^2$, where we have
used $\theta_{12}\approx \theta_2$. If we remember that each emission
comes with an additional angle, $\varphi_i$ around the jet axis, the
minimum depends on $\varphi_2-\varphi_1$. In half the cases this will
cluster 0 and 1 and giving a subjet mass $z_2\theta_2^2$, in the other
half, it will cluster 1 and 2, giving a subjet mass of $z_1
z_2\theta_2^2$. 
Similar considerations allow one to show that the secondary emissions
also have an extra factor $z_1$ compared to the $N$-subjettiness case.

\section{Infrared (un)safety of Cambridge/Aachen
  de-clustering}\label{app:irc-safety}

In this Appendix, we provide a few additional details regarding the
infrared unsafety of the $\mu^2$ parameter with Cambridge/Aachen
de-clustering. To avoid any possible confusion, we must stress that
the discussion below only applies to the non-recursive version of the
$\mu^2$ parameter and that the recursive aplication of a $\mu_p^2$ cut
is infrared-safe for any $p$.

That said, let us consider a jet with three particles: a hard parton,
a first emission with momentum fraction $z_1$ at an angle $\theta_1$
and a second emission with momentum fraction $z_2$ at an angle
$\theta_2$, with $z_1\theta_1^2>z_2\theta_2^2$ and
$\theta_2\ll\theta_1$. This corresponds to the leading-order
(${\cal O}(\alpha_s^2)$) configuration for a jet with
$m^2=(z_1 \theta_1^2+z_2\theta_2^2)p_t^2$ and with a generic
$\mu^2=z_2\theta_2^2/(z_1 \theta_1^2+z_2\theta_2^2)$ (using
Cambridge/Aachen de-clustering).
At the next order of the perturbation theory, one would have to
include real emissions of gluons with momentum fraction $z_3$ and
angle $\theta_3$ as well as the corresponding virtual corrections and
the soft divergence $z_3\to 0$ is supposed to cancel between the real
and virtual contributions. However, for $\theta_3\gg \theta_1$ and
$z_3\to 0$, the virtual contribution would give
$\mu^2_{\rm virt}=z_2\theta_2^2/(z_1 \theta_1^2+z_2\theta_2^2)$ as for
the 2-particle configuration, but the real emissions would give
$\mu^2_{\rm real}=1$ because of the Cambridge/Aachen
de-clustering. This would lead to an infrared unsafety at
$\mu^2_{\rm virt}$.
This situtation can happen at any value of $\mu$, depending on the
original three-particle configuration.

Although we have not made an explicit calculation, one might expect
that the Sudakov $R_{\mu_p^2}$ would receive a contribution
proportional to $(\alpha_s/p) \log^2(1/\theta_1^2)$, with
$\theta_1^2=\rho/z_1$, which diverges in the limit $p\to 0$.

\section{Soft and large-angle emissions}\label{app:soft-large-angle}

In all the calculations we have performed so far, we have included
hard collinear splittings which correspond to the terms proportional
to $B_i$ and $B_g$ in our results. At the same order we could also
have single-logarithmic contributions coming from soft and large-angle
emissions. In practice, keeping the same notations as above, this
means working in the approximation $z_2\ll z_1$ without assuming any
specific ordering between $\theta_1$ and $\theta_2$.

This can affect the calculations above at various levels: either
through changes in the approximation used for the shape, where so far
we have assumed a strong ordering, or through modifications of the
matrix element for soft gluons at large angles.\footnote{In this
  discussion, we neglect additional effects from non-global
  logarithms. Since they will be impacted by grooming, we defer their
  study to a forthcoming study.}

Let us first discuss the first effect. Since the expressions we have
used so far are correct when $\theta_2\ll\theta_2$ or when
$\theta_2\gg\theta_1$ we only have to worry about the region
$\theta_2\sim\theta_1$.

For $N$-subjettiness and the energy correlation functions, the correct
expression in that region will only differ from the asymptotic one
used so far by a constant, not enhanced by any
parametrically large quantities. As a consequence, if we compute the
difference to what has already been included, the integration
over $z_2$ will at most bring a constant. Then, the angular
integration over $\theta_2\sim\theta_1$ will also at most bring a
constant giving an overall NNLL subleading correction, as already
briefly discussed in Section~\ref{sec:bkg-tau}.

The situation is potentially a bit more tricky for $\mu^2$ since the
expression at $\theta_2\sim\theta_1$ can vary between
$z_2\theta_2^2/\rho$ and $z_2\theta_2^2/\theta_1^2$ potentially
introducing a correction enhanced by $\log(1/z_1)$. Not making any
assumption about angular ordering, Eq.~(\ref{eq:mu2}) becomes
\begin{align}
R_{\mu_{1/2}^2}(z_1)
 & = \int_0^1\frac{d\theta_2^2}{\theta_2^2}\int_0^1\frac{dz_2}{z_2} \,\frac{\alpha_sC_R}{\pi}
  \,\Theta(z_2\theta_2^2<\rho)\,\bigg\{ \Theta(\theta_2^2<\theta_1^2/4) \,\Theta(z_2\theta_2^2>\rho\mu^2)\\
 & \qquad\qquad\qquad+ \Theta(\theta_2^2>\theta_1^2/4)\int_0^{2\pi}\frac{d\phi}{2\pi}
   \,\Big[\Theta(\theta_{12}^2>\theta_2^2)\,\Theta(z_2\theta_2^2>\rho\mu^2)\nonumber\\
 & \qquad\qquad\qquad\phantom{+\Theta(\theta_2^2>\theta_1^2/4)\int_0^{2\pi}\frac{d\phi}{2\pi}}
       +\Theta(\theta_{12}^2<\theta_2^2)\,\Theta(z_2\theta_2^2>\theta_1^2\mu^2)
    \Big]\bigg\}\nonumber
\end{align}
where we have only considered primary emissions, worked with a fixed
coupling approximation, and noticed that, for the sake of our
calculation, we can safely replace $P(z_2)$ by $2C_R/z_2$. The angle
$\phi$ that we have introduced is the angle between the two emissions,
measured from the jet axis. This means that we have
$\theta_{12}^2=\theta_1^2+\theta_2^2-2\theta_1\theta_2\cos(\phi)$.
The calculation of the above integral is a bit tedious but, in the
end, we find that all single-logarithmic terms cancel, leaving the
same result as what we have obtained in Section~\ref{sec:bkg-mu2}.

We are therefore left with potential single logarithms coming from the
matrix element for the emission of soft and large-angle gluons.
Taking the case of a quark jet, we therefore have to compute the
following generic expression:
\begin{equation}
R = \int \frac{dz_2}{z_2}\,d^2\theta_2\,\frac{\alpha_s}{\pi^2}
 \Big[\frac{C_A}{2}\frac{1}{\theta_{12}^2}
      +\frac{C_A}{2}\frac{\theta_{01}^2}{\theta_{02}^2\theta_{12}^2}
      +\Big(C_F-\frac{C_A}{2}\Big)\frac{1}{\theta_{02}^2}\Big] 
 \,\Theta(z_2\theta_2^2<\rho)\,\Theta(v(z_i,\theta_i)>v).
\end{equation}

If we focus on the single-logarithmic contribution, we can subtract
the double-logarithmic piece,
$C_F/\theta_{02}^2+C_A/\theta_{12}^2\,\Theta(\theta_{12}<\theta_{01})$,
and set $v(z_i,\theta_i)=(z_2\theta_2^2)/(z_1\theta_1^2)$ in what
remains so that the $z_2$ integration yields a $\log(1/v)$. This gives
\begin{equation}
R_{\rm SL} = \frac{\alpha_s C_A}{2\pi^2}\log(1/v)
  \int d^2\theta_2\,
  \frac{\theta_{01}^2}{\theta_{02}^2\theta_{12}^2}
      +\frac{2}{\theta_{12}^2}\,\Theta(\theta_{12}^2<\theta_{01}^2),
\end{equation}
where we have used the fact that
$\int d^2\theta_2/\theta_{02}^2=\int d^2\theta_2/\theta_{12}^2$. Up to
subleading corrections, we can extend the $\theta_2$ integration to
infinity and show, \eg using dimensional regularisation, that
it vanishes. In the end, there are therefore no soft and large-angle
single-logarithmic corrections to what we have computed earlier in
the text.

\section{Further comparisons}\label{app:further-tests}

In this last appendix, we provide a few additional comparisons between
our analytic predictions and Monte-Carlo simulations. 

\begin{figure}[!ht]
\includegraphics[width=0.48\textwidth]{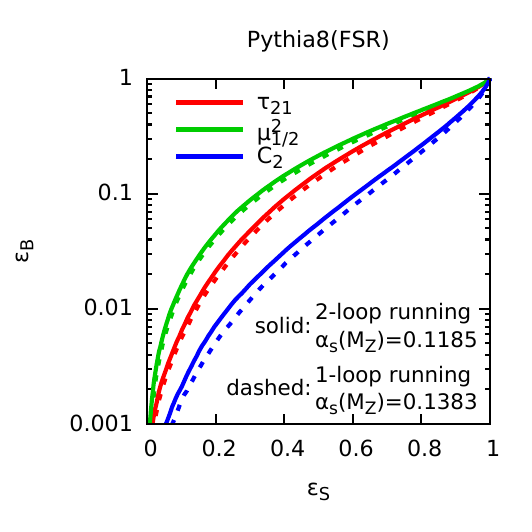}%
\hfill
\includegraphics[width=0.48\textwidth]{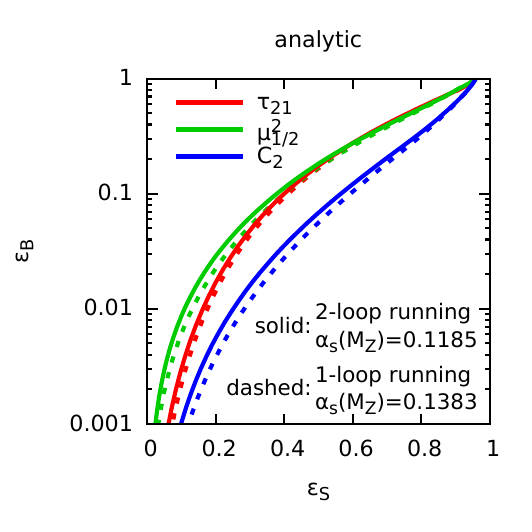}%
\caption{Similar plot as in Fig.~\ref{fig:MC-ROC} where we show Pythia
  results (Left) and analytic calculations (right) of the signal and
  background efficiencies for two different running-coupling
  prescriptions: a one-loop running with $\alpha_s(M_Z)=0.1383$
  (dashed, our default for Pythia in the main text) and a two-loop
  running with $\alpha_s(M_Z)=0.1185$ (solid, uor default for analytic
  results in the main text).}\label{fig:MC-ROC-2loops}
\end{figure}

\begin{figure}[h]
\includegraphics[width=0.48\textwidth]{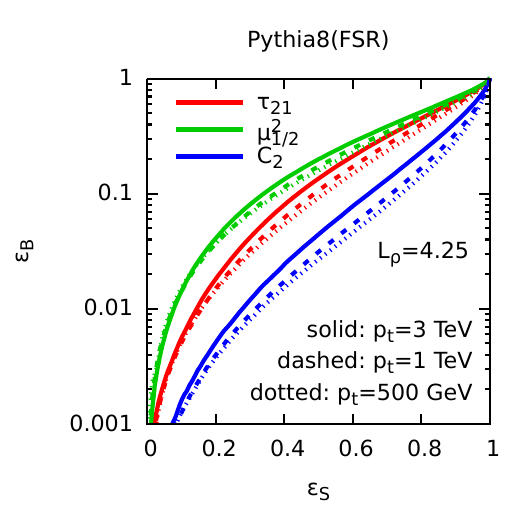}%
\hfill
\includegraphics[width=0.48\textwidth]{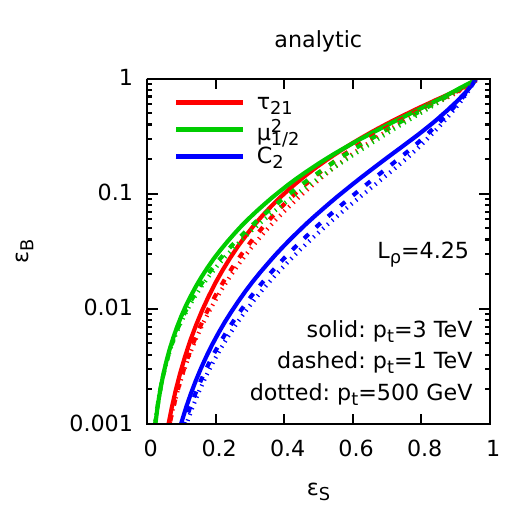}%
\caption{Similar plot as in Fig.~\ref{fig:MC-ROC} where we show Pythia
  results (Left) and analytic calculations (right) of the signal and
  background efficiencies for two different running-coupling
  prescriptions obtained for different jet transverse momenta, keeping
  $L_\rho$ fixed to 4.25 (or, in the 4-4.5 range for Pythia
  simulations).}\label{fig:MC-ROC-ptdep}
\end{figure}

\paragraph{One-loop v. two-loop running coupling.} First, in
Sections~\ref{sec:bkg-mc} and \ref{sec:sig-mc}, we have used a
one-loop running of $\alpha_s$, with $\alpha_s(M_Z)=0.1383$, for
Pythia simulations, and compared that to analytic calculations
including two-loop corrections and using $\alpha_s(M_Z)=0.1185$.
In the case of our analytic calculation, this choice is motivated by
the fact that two-loop corrections are easily included and we then
used the world-average value~\cite{pdg} at the $Z$-boson mass.
For the Pythia simulation, we simply kept the default which is a
one-loop running.

We could also have run Pythia with a two-loop running of the coupling
and impose $\alpha_s(M_Z)=0.1185$. We did not do that in the main text
because that can only safely be done with a retuning of other
parameters in Pythia (mostly for the non-perturbative effects). It is
however interesting to check that this difference in the treatment of
the running of the strong coupling does not come with large effects.
The result is presented in Fig.~\ref{fig:MC-ROC-2loops}, where we see
that this is indeed a small effect which does not alter in any way the
conclusions of this paper. We also see from that figure that the size
of the effect is similar in Monte-Carlo simulations and in our
analytic predictions.

Note also that another interesting check of our results is to compare
our fixed-order results with Pythia simulations also done with a fixed
coupling. Although we do not show explicit plots here, this comparison
shows similar features as the ones observed with a running-coupling
prescription.

\paragraph{Dependence on the jet transverse momentum.}
Throughout this paper, we have shown results for jets with a large
transverse momentum of 3~TeV. Here, we briefly show that our
calculations remain valid for less boosted jets, closer to those used
in today's phenomenological analyses. 

In Fig.~\ref{fig:MC-ROC-ptdep}, we show ROC curves obtained from
Pythia simulations and our analytic calculations, for three different
jet transverse momenta: 3~TeV, 1~TeV and 500~GeV. For this comparison,
we have kept the ratio $m/p_t$ fixed, i.e. considered a mass of 358,
120 and 60~GeV respectively for each of the three $p_t$ scales.
We see that the dependence on the jet $p_t$ is mild, which is expected
since the result only depend on $p_t$ through the $p_tR$ scale
entering in $\alpha_s$. Our conclusions are therefore also valid for
jets of more moderate transverse momentum. Note that the small
differences observed in Pythia simulations between different jet $p_t$
are well reproduced by our analytic calculation.

\end{document}